\documentclass[twocolumn, twocolappendix]{aastex63}

\usepackage{amsmath}

\def\aliii     {\ensuremath{\text{Al\,\textsc{iii}}}}

\def\siiii     {\ensuremath{\text{Si\,\textsc{iii]}}}}

\def\siiv     {\ensuremath{\text{Si\,\textsc{iv}}}}

\def\civ     {\ensuremath{\text{C\,\textsc{iv}}}}

\def\mgii     {\ensuremath{\text{Mg\,\textsc{ii}}}}

\def\heii     {\ensuremath{\text{He\,\textsc{ii}}}}

\def\feii     {\ensuremath{\text{Fe\,\textsc{ii}}}}

\def\feiii    {\ensuremath{\text{Fe\,\textsc{iii}}}}

\def\cii     {\ensuremath{\text{[C\,\textsc{ii]}}}}

\def\ciii     {\ensuremath{\text{C\,\textsc{iii]}}}}

\def\ciilong    {\ensuremath{\text{[C\,\textsc{ii]}}_{158\,\mu\text{m}}}}

\def\heii    {\ensuremath{\text{He\,\textsc{ii}}}}

\def\oiiibr    {\ensuremath{\text{O\,\textsc{iii]}}}}

\shorttitle{The X-SHOOTER/ALMA sample I}
\shortauthors{Schindler et al.}
\graphicspath{{./}{}}

\newcommand{\Nsamp}{38}

\defcitealias{Vestergaard2001}{VW01}
\defcitealias{Tsuzuki2006}{T06}

\begin{document}

\title{The X-SHOOTER/ALMA sample of Quasars in the Epoch of Reionization. \\ I. NIR spectral modeling, iron enrichment and broad emission line properties}

\correspondingauthor{Jan-Torge Schindler}
\email{schindler@mpia.de}
\author[0000-0002-4544-8242]{Jan-Torge Schindler}
\affiliation{Max Planck Institut f\"ur Astronomie, K\"onigstuhl 17, D-69117, Heidelberg, Germany}

\author[0000-0002-6822-2254]{Emanuele Paolo Farina}
\affiliation{Max Planck Institut f\"ur Astrophysik, Karl--Schwarzschild--Stra{\ss}e 1, D-85748, Garching bei M\"unchen, Germany}

\author[0000-0002-2931-7824]{Eduardo Ba{\~n}ados} 
\affiliation{Max Planck Institut f\"ur Astronomie, K\"onigstuhl 17, D-69117, Heidelberg, Germany}

\author[0000-0003-2895-6218]{Anna-Christina Eilers}\thanks{NASA Hubble Fellow}
\affiliation{MIT Kavli Institute for Astrophysics and Space Research, 77 Massachusetts Ave., Cambridge, MA 02139, USA}

\author[0000-0002-7054-4332]{Joseph F. Hennawi}
\affiliation{Department of Physics, University of California, Santa Barbara, CA 93106-9530, USA}
\affiliation{Max Planck Institut f\"ur Astronomie, K\"onigstuhl 17, D-69117, Heidelberg, Germany}

\author[0000-0003-2984-6803]{Masafusa Onoue}
\affiliation{Max Planck Institut f\"ur Astronomie, K\"onigstuhl 17, D-69117, Heidelberg, Germany}

\author[0000-0001-9024-8322]{Bram P. Venemans}
\affiliation{Max Planck Institut f\"ur Astronomie, K\"onigstuhl 17, D-69117, Heidelberg, Germany}

\author[0000-0003-4793-7880]{Fabian Walter}
\affiliation{Max Planck Institut f\"ur Astronomie, K\"onigstuhl 17, D-69117, Heidelberg, Germany}

\author[0000-0002-7633-431X]{Feige Wang}
\altaffiliation{NHFP Hubble Fellow}
\affiliation{Steward Observatory, University of Arizona, 933 N Cherry Ave, Tucson, AZ 85721, USA}

\author[0000-0003-0821-3644]{Frederick B. Davies}
\affiliation{Lawrence Berkeley National Laboratory, 1 Cyclotron Rd, Berkeley, CA 94720, USA} 

\author[0000-0002-2662-8803]{Roberto Decarli}
\affiliation{INAF --- Osservatorio di Astrofisica e Scienza dello Spazio, via Gobetti 93/3, I-40129, Bologna, Italy}

\author[0000-0003-3242-7052]{Gisella De Rosa}
\affiliation{Space Telescope Science Institute, 3700 San Martin Dr, MD 21218, Baltimore, US}

\author[0000-0002-0174-3362]{Alyssa Drake}
\affiliation{Max Planck Institut f\"ur Astronomie, K\"onigstuhl 17, D-69117, Heidelberg, Germany}

\author[0000-0003-3310-0131]{Xiaohui Fan}
\affiliation{Steward Observatory, University of Arizona, 933 N Cherry Ave, Tucson, AZ 85721, USA}

\author[0000-0002-5941-5214]{Chiara Mazzucchelli}
\affiliation{European Southern Observatory, Alonso de C\'ordova 3107, Vitacura, Regi\'on Metropolitana, Chile}

\author[0000-0003-4996-9069]{Hans-Walter Rix}
\affiliation{Max Planck Institut f\"ur Astronomie, K\"onigstuhl 17, D-69117, Heidelberg, Germany}

\author[0000-0003-0960-3580]{G\'abor Worseck}
\affiliation{Institut f\"ur Physik und Astronomie, Universit\"at Potsdam, Karl-Liebknecht-Str.\ 24/25, D-14476 Potsdam, Germany}

\author[0000-0001-5287-4242]{Jinyi Yang}
\affiliation{Steward Observatory, University of Arizona, 933 N Cherry Ave, Tucson, AZ 85721, USA}



\begin{abstract}



We present X-SHOOTER near-infrared spectroscopy of a large sample of \Nsamp{} luminous ($M_{1450}=-29.0$ to $-24.4$) quasars at $5.78<z<7.54$, which have complementary \ciilong{} observations from ALMA. This X-SHOOTER/ALMA sample provides us with the most comprehensive view of reionization-era quasars to date, allowing us to connect the quasar properties with those of its host galaxy. 
In this work we introduce the sample, discuss data reduction and spectral fitting, and present an analysis of the broad emission line properties.
The measured \feii{}/\mgii{} flux ratio suggests that the broad line regions of all quasars in the sample are already enriched in iron.
We also find the \mgii{} line to be on average blueshifted with respect to the \cii{} redshift with a median of $-391\,\rm{km}\,\rm{s}^{-1}$. A significant correlation between the \mgii{}-\ciilong{} and \civ{}-\ciilong{} velocity shifts indicates a common physical origin.
Furthermore, we frequently detect large \civ{}-\mgii{} emission line velocity blueshifts in our sample with a median value of $-1848\,\rm{km}\,\rm{s}^{-1}$. 
While we find all other broad emission line properties not to be evolving with redshift, the median \civ{}-\mgii{} blueshift is much larger than found in low-redshift, luminosity-matched quasars ($-800\,\rm{km}\,\rm{s}^{-1}$). Dividing our sample into two redshift bins, we confirm an increase of the average \civ{}-\mgii{} blueshift with increasing redshift. 
Future observations of the rest-frame optical spectrum with the James Webb Space Telescope will be instrumental in further constraining the possible evolution of quasar properties in the epoch of reionization.

\end{abstract}



\section{Introduction}\label{sec:intro}


Quasars are the most luminous non-transient light sources in the universe. They are galaxies in which mass accretion onto a supermassive black hole (SMBH) dominates UV and optical emission, quasars can be discovered well into the epoch of reionization \citep[$z>6$,][]{Fan2006}. In this last major phase transition of the universe neutral hydrogen is being ionized by UV emission of the first generation of galaxies and accreting SMBHs.
High-redshift quasars at $z>6$ not only provide a window into the formation and early growth of SMBHs, they also facilitate the study of massive high-redshift galaxy evolution, probe the onset of black hole host galaxy co-evolution, and shed light on the process of reionization. 

The advent of wide-area photometric surveys has increased the number of known quasars at $z>6$ to $\sim200$ by today \citep[e.g.,][]{Fan2001c, Banados2016, Matsuoka2019b, Reed2019, YangJinyi2019, WangFeige2019}.
Above $z=7$ only seven quasars are known to date \citep{Mortlock2011, WangFeige2018, Matsuoka2018b, Matsuoka2019a, YangJinyi2019, YangJinyi2020} with ULAS~J1342+0928 at $z=7.54$ \citep{Banados2018} being the most distant quasar known.

Rest-frame UV and optical spectra of quasars have been key to identifying the origin of the emission as mass accretion onto a SMBH \citep{LyndenBell1969}. We now understand that the broad emission lines ($\rm{FWHM} \gtrsim 1000\,\rm{km}\,\rm{s}^{-1}$) seen in the spectra originate from mostly virialized gas orbiting the central SMBH at sub-parsec scales, the so-called broad line region (BLR).  
Narrow emission lines ($\rm{FWHM} \lesssim 500\,\rm{km}\,\rm{s}^{-1}$) often seen in addition to the broad lines emanate from gas at kilo-parsec scales, the narrow line region (NLR).
The kinematics of the BLR imprinted on the broad emission lines allow us to estimate the SMBH mass and further understand the dynamics of the accretion process \citep{Peterson1993, Peterson2004}. 

At $z>6$ the rest-frame UV spectrum is shifted into the optical/NIR wavelength range, and the region blueward of $1216\,$\AA, including parts of the the Ly$\alpha$ line, is strongly absorbed by the intergalactic medium due to the resonant nature of Ly$\alpha$ photons in neutral hydrogen \citep[e.g.][]{Michel-Dansac2020}.
Therefore, near-infrared (NIR) spectroscopy is necessary to fully characterize the quasar's spectrum and exploit the information provided by the broad and narrow emission lines. 
Many high-redshift quasars have thus been followed up either individually or in small ($N<10$) samples \citep[e.g.][]{Kurk2007, Jiang2007, DeRosa2014, Onoue2019}. However, the discovery of hundreds of quasars above $z\approx6$ has paved the way for studies of increasingly larger samples \citep{DeRosa2011, Mazzucchelli2017, Becker2019}, enabling first insights into the population properties of high-redshift quasars. The largest study at $z \gtrsim 5.7$ to date \citep{Shen2019a} presents near-infrared spectra and measured properties for a total of 50  quasars. 


Studies of $z>6$ quasars have revealed large SMBH masses, $\sim10^{8}-10^{10}\,M_{\odot}$ \citep[e.g.][]{Wu2015, Onoue2019}, only 1\,Gyr after the Big Bang, setting strong constraints on models of black hole formation and evolution \citep[for a review see][]{Volonteri2012}. The majority of $z>6$ quasars are found to have high accretion rates as characterized by their high Eddington luminosity ratios of $L_{\rm{bol}}/L_{\rm{Edd}}\ge0.1$.
Interestingly, general properties (spectral shape, maximum SMBH mass, BLR metallicity, \feii{}/\mgii{} flux ratio) of quasars at $z>6$ show no or only a weak evolution with redshift \citep[e.g.][]{Jiang2007, DeRosa2011, DeRosa2014, Mazzucchelli2017, Shen2019a}. 
The only exception seems to be the \civ{}-\mgii{} velocity shift. It was already known that a large fraction of $z\gtrsim6$ quasars exhibit highly blueshifted \civ{} emission  compared to their \mgii{} redshift \citep[e.g.][]{DeRosa2014, Mazzucchelli2017, Reed2019}, indicative of an outflowing component in the \civ{} emission line \citep[e.g.][]{Gaskell1982}. A comparison across (luminosity-matched) quasar samples at different redshifts \citep{Meyer2019c} has highlighted that large \civ{} blueshifts are much more common in $z>6.5$ quasars than at lower redshifts.


On the other hand, it is currently unclear whether this evolution is an intrinsic change or induced by selection effects. Quasars at $z{>}6$ are predominantly selected by the strong Lyman-$\alpha$ break in their spectrum. In addition, available photometry limits $z{>}6$ quasar searches to the bright end ($M_{1450} \le -25.5$) of the quasar distribution \citep{WangFeige2019}. Only the Canada-France High-z Quasar Survey \citep{Willott2010a} and the recent efforts of the Subaru High-z Exploration of Low-luminosity Quasars (SHELLQs) project \citep[e.g.][]{Matsuoka2016, Matsuoka2019b} have provided a first look at the fainter $z>6$ quasar population. These lower luminosity quasars show on average less massive SMBHs $\sim10^{7}-10^{9}\,M_{\odot}$ black holes \citep[e.g.][]{Willott2017, Onoue2019} compared to their luminous counterparts. Unfortunately, only a handful of NIR spectroscopic measurements exist to date for low luminosity $z>6$ quasars.


Investigations of high-redshift quasars are often complemented with studies of the host galaxy gas via rotational transitions of the carbon monoxide (CO) molecule or the fine structure line of singly ionized carbon \cii{} at $158\,\mu\rm{m}$, which enters the 1.2\,mm atmospheric window for quasars at $z\gtrsim6$. 
Millimeter observations so far provide the only direct probes for the host galaxy in high-redshift quasars.
As the \cii{} line is the main coolant of the cool ($<1000\,\rm{K}$) interstellar material, it is a very bright line easily detectable at cosmological distances. 
The \cii{} line and the underlying far-infrared (FIR) dust continuum emission, allow measurements of precise \cii{} redshifts, estimates of the dynamical masses, and star formation rates. 
Since the first \cii{} line detection at $z>0.1$ in the host galaxy of J1148+5251, a quasar at $z=6.4$ \citep{Maiolino2005}, the \cii{} line has become a widely used diagnostic for high-redshift quasar hosts \citep[e.g.][]{Walter2009, Venemans2012, Wang2013, Willott2013, Willott2015, Banados2015, Venemans2016, Mazzucchelli2017, Izumi2018, Izumi2019, Eilers2020a, Venemans2020}. 


We here present the analysis of the NIR spectra of \Nsamp{} quasars, capitalizing on new and archival VLT/X-SHOOTER data.
All quasars in our sample have also been targeted and observed at millimeter (mm) wavelengths to detect the \cii{} emission. Successful detection of 34 of our \Nsamp{} quasars \citep{Decarli2018, Eilers2020a, Venemans2020} thus complement our sample with precise systemic redshifts and additional information on the cold ISM and dust emission of the host galaxy.  
The combined information on the quasar and its host provide us with a comprehensive view on the full quasar phenomenon, unique to the X-SHOOTER/ALMA sample.
In this paper we present the X-SHOOTER NIR spectral analysis of the quasar sample and an in-depth discussion of the quasars' rest-frame UV properties.
A companion paper (Farina et al., in prep.) will present the SMBH masses and discuss them in context with their host galaxies.
That paper will also put the sample in context with VLT/MUSE observations \citep[REQUIEM][]{Farina2019}, which probe the immediate environment in Lyman-$\alpha$ emission \citep[see also][]{Drake2019}.
Data reduction of the optical quasar spectra taken by the X-SHOOTER visual arm (VIS) is on-going and will be presented in a future publication.

In Section\,\ref{sec:sample} we give an overview of the quasar sample and describe the data reduction. We lay out our spectral fitting methodology in detail in Section\,\ref{sec:fitting} and describe the analysis of the fits in Section\,\ref{sec:fitanalysis}. Section\,\ref{sec:irondiscussion} is devoted to a discussion on the biases inherent in adopting different iron pseudo-continuum templates. We analyze the iron enrichment of the broad line region in Section\,\ref{sec:results_feIImgIIratios} and examine the properties of the broad \civ{} and \mgii{} lines in Section\,\ref{sec:results_BELS}. Our findings are summarized in Section\,\ref{sec:conclusion}.

Throughout this work we adopt a standard flat $\Lambda$CDM cosmology with $\rm{H}_0=70\,\rm{km}\,\rm{s}^{-1}\,\rm{Mpc}^{-1}$, $\Omega_{\rm M}=0.3$, and \mbox{$\Omega_\Lambda=0.7$} in broad agreement with the results of the Planck mission \citep{PlanckCollaboration2016}. All magnitudes are reported in the AB photometric system. 

\section{The X-SHOOTER/ALMA sample}\label{sec:sample}
\begin{figure}
    \centering
    \includegraphics[width=0.45\textwidth]{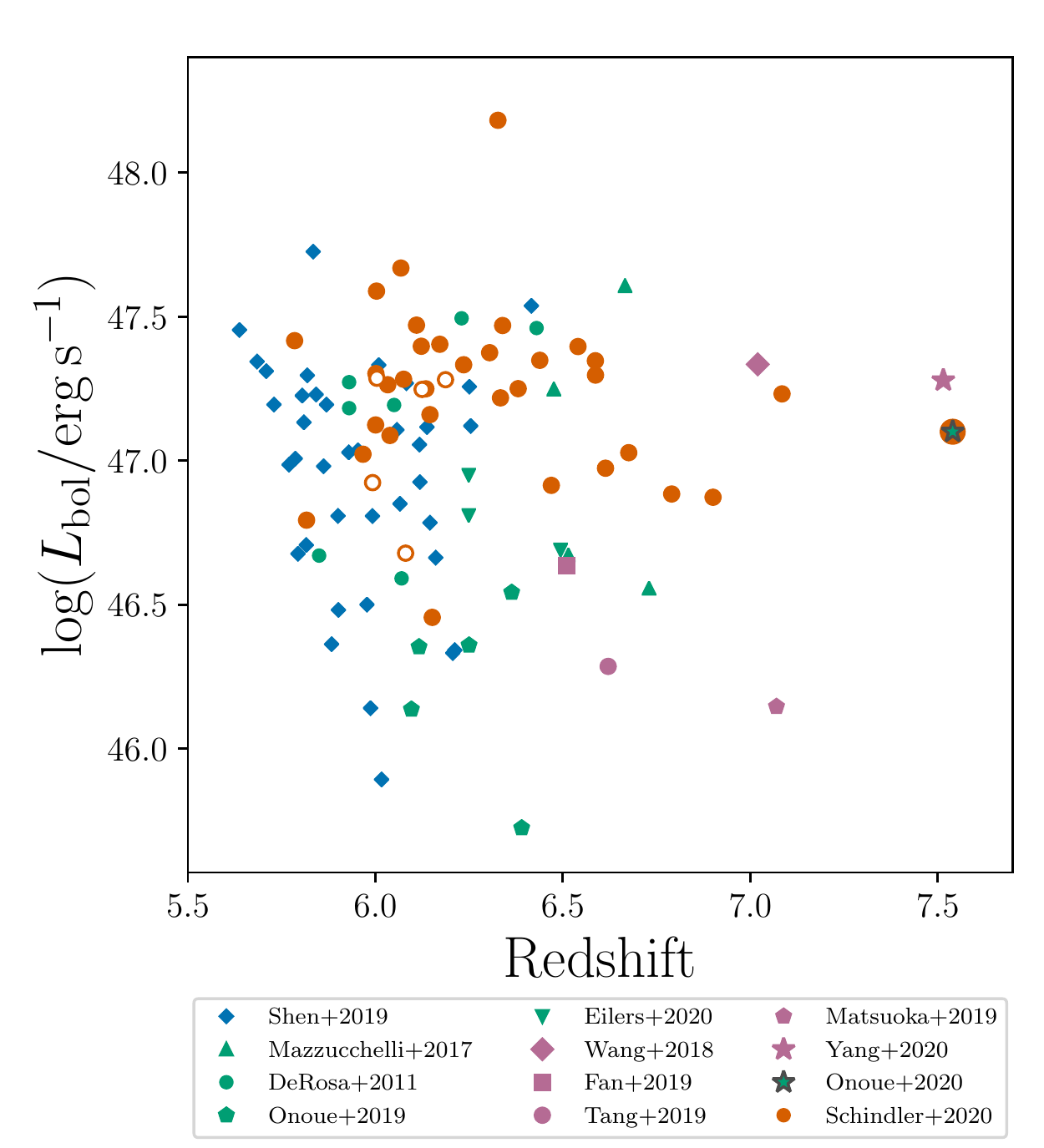}
    \caption{Quasars at $z>5.5$ with available near-infrared spectroscopy as a function of their bolometric luminosity and redshift. Quasars of this study are highlighted as orange circles. Filled orange circles refer to objects with successful continuum fits, whereas open orange circles refer to the five objects, where we could not fit the continuum shape with a power law.  Quasars from other studies are represented with blue and green symbols according to the legend.}
    \label{fig:lbol_z}
\end{figure}

\begin{figure*}
\centering
\includegraphics[width=0.95\textwidth]{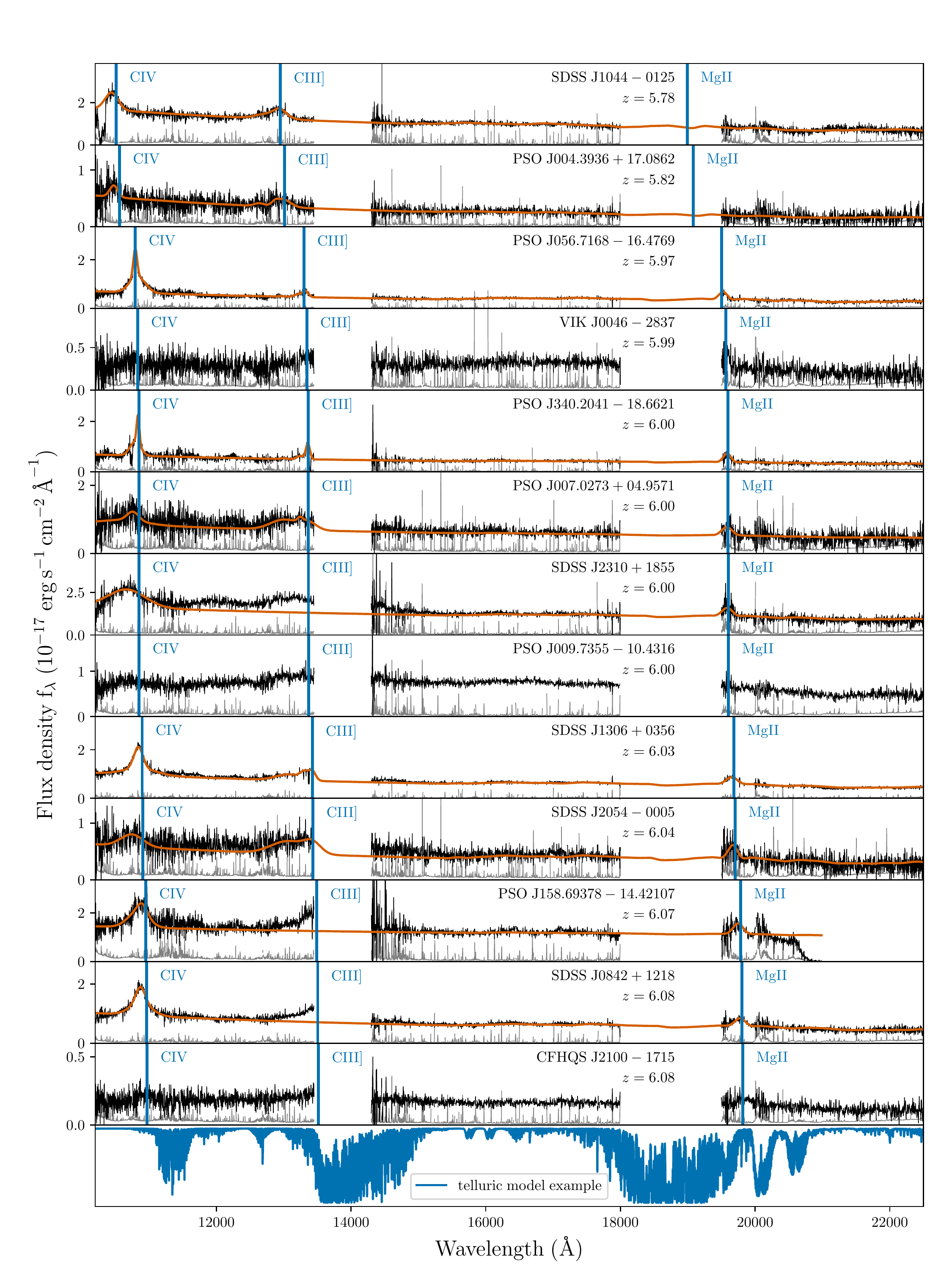}
\caption{We display the near-infrared X-SHOOTER spectra of all \Nsamp{} quasars in our sample. The spectra have been binned by 4 pixels and we show the flux uncertainty in grey. Model fits are over-plotted in orange for all cases where fitting the continuum with a power-law model was possible. We also highlight the positions of the broad \civ{}, C\,III] and \mgii{} lines according to the systemic redshift. We have removed wavelength ranges of strong telluric absorption as highlighted by the telluric model example in the bottom panel. \label{fig:sample}}
\end{figure*}

\begin{figure*}
\centering
\includegraphics[width=0.95\textwidth]{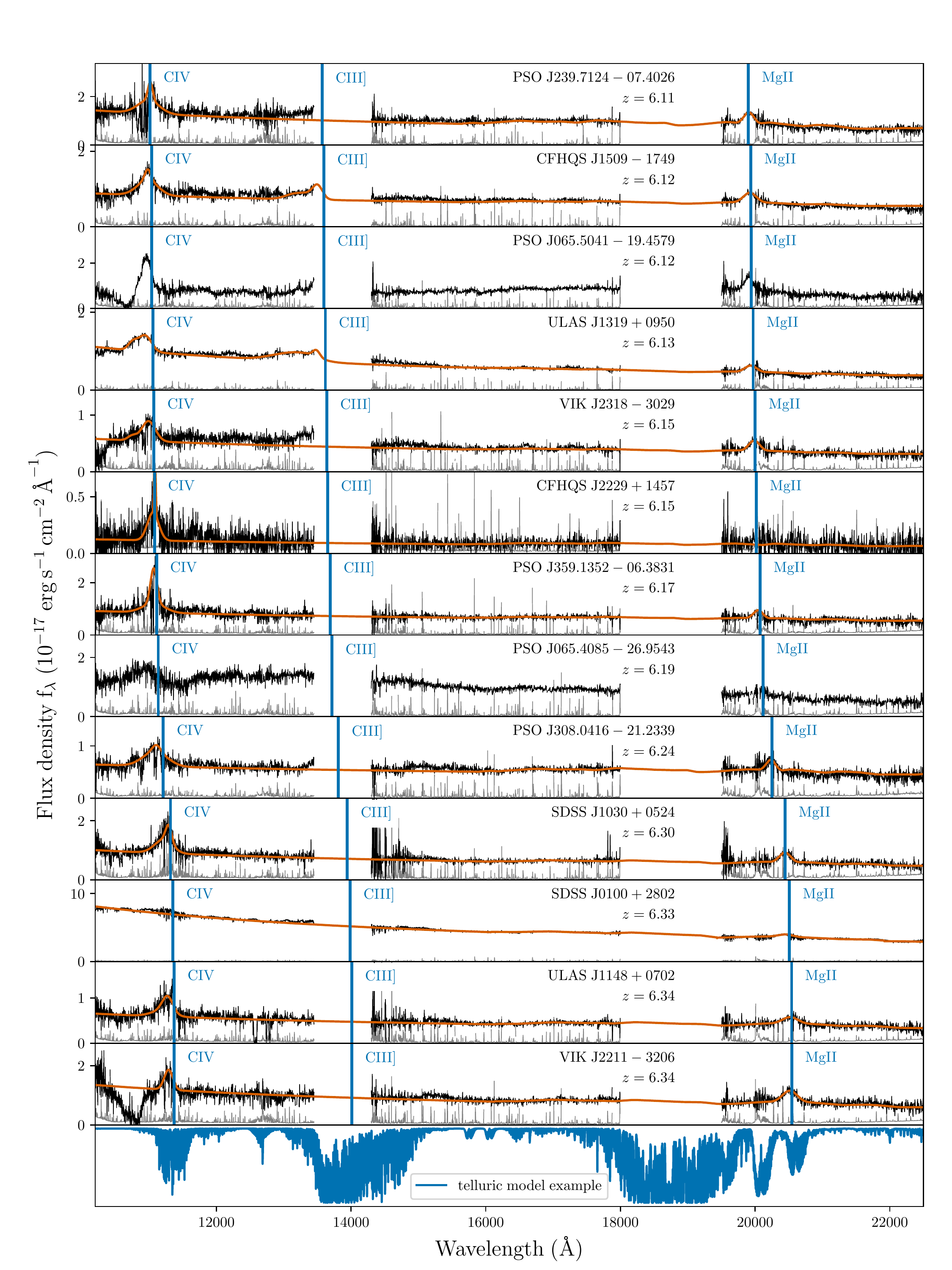}
\caption{Same as Figure\,\ref{fig:sample}\label{fig:sample2}}
\end{figure*}

\begin{figure*}
\centering
\includegraphics[width=0.95\textwidth]{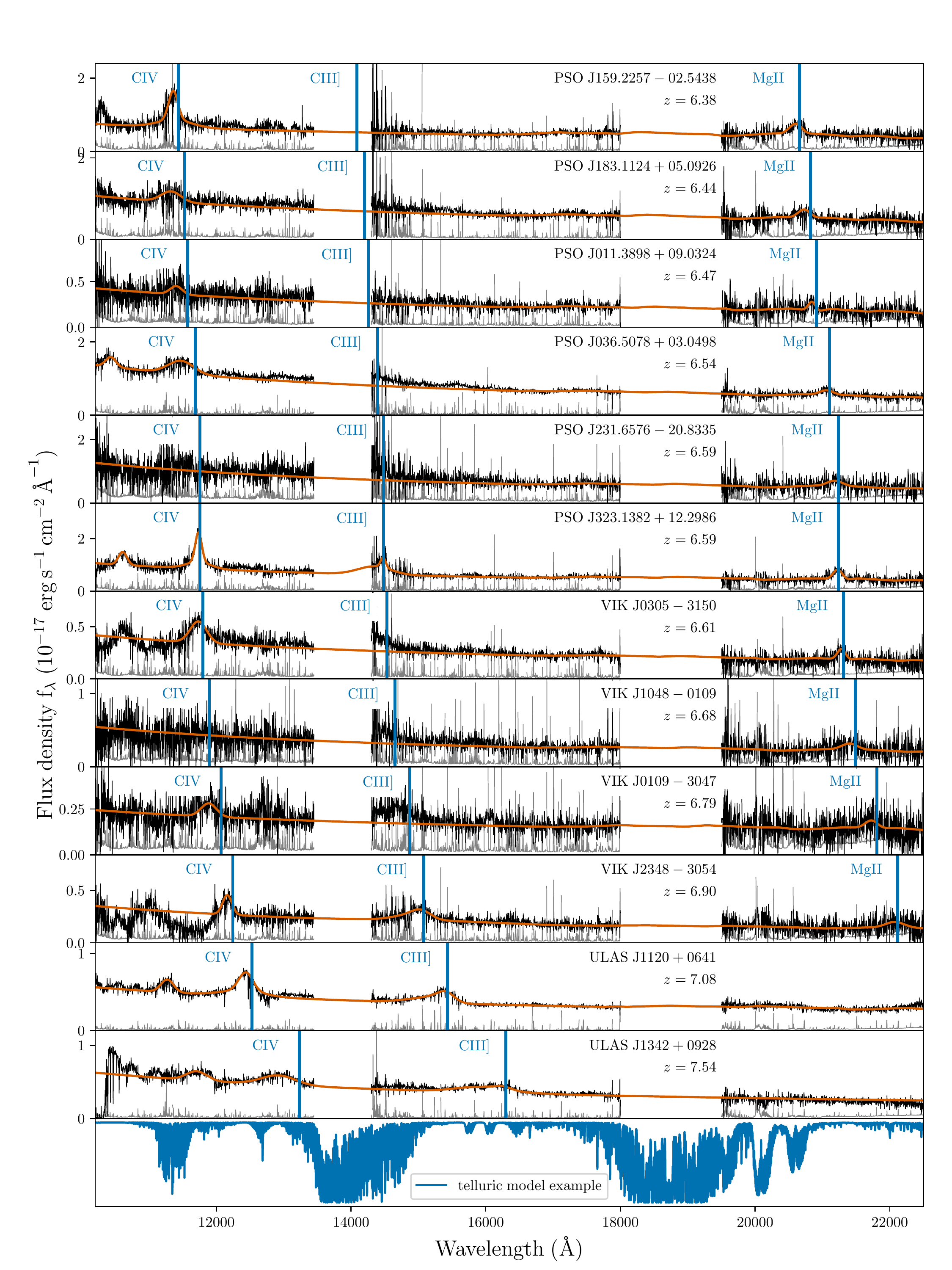}
\caption{Same as Figure\,\ref{fig:sample}\label{fig:sample3}}
\end{figure*}



The sample we present herein consists of \Nsamp{} quasars with redshifts between $z=5.78$ and $z=7.54$ (median $z=6.18$). They were selected to have both near-infrared X-SHOOTER spectroscopy as well as ALMA mm observations of the quasar host. The mm observations are crucial as they allow us to place the quasar (Black Hole mass, Eddington ratio, line redshifts, etc.) in context with the galaxy (systemic redshift, dynamical mass, gas mass, etc.). 
While the mm ALMA results have been previously published \citep{Venemans2017c, Decarli2018, Banados2019, Venemans2019, Eilers2020a, Venemans2020} or are in preparation (Neeleman et al. 2020, in prep.), a large fraction of the X-SHOOTER spectroscopy is presented here for the first time.
An overview of the sample is given in Tables\,\ref{table:sample} and \ref{table:sample_add}.

With the exception of four sources, we adopt systemic redshifts measured from the \ciilong{} emission line.
As shown in Figure\,4 of \citet{Decarli2018} the \ciilong{}-based redshifts provide a substantial improvement over the quasar discovery redshifts. Their comparison includes a large fraction of our sample.

Figure\,\ref{fig:lbol_z} shows the X-SHOOTER/ALMA sample in the plane of bolometric luminosity and redshift, compared to other samples with near-infrared spectroscopy from the literature. The quasars in our sample can be considered luminous with a median absolute magnitude of $M_{1450}=-26.5$ ($-29.0$ to $-24.4$), as determined from their spectral fits. 
With the exception of SDSS~J0100+2802 ($\log(L_{\rm{bol}}/\rm{erg}\,\rm{s}^{-1})=48.19$), all other quasars lie in a narrow range of bolometric luminosities, $\log(L_{\rm{bol}}/\rm{erg}\,\rm{s}^{-1})=46.67$ to $47.67$ (median $47.26$). Details on how the bolometric luminosity was calculated from the spectra is provided in Section\,\ref{sec:fitanalysis}.

The recently published compilation of 50 quasars with GNIRS spectroscopy \citep{Shen2019a} is shown as blue diamonds in Figure\,\ref{fig:lbol_z}. Compared to our work their quasar sample is at slightly lower redshifts (median $z=5.97$) and on average less luminous (median $\log(L_{\rm{bol}}/\rm{erg}\,\rm{s}^{-1})=47.05$).

\subsection{The X-SHOOTER spectroscopy}

The X-SHOOTER spectrograph \citep{Vernet2011} covers the wavelength range from $300\,\rm{nm}$ to $2500\,\rm{nm}$ with three spectral arms (UVB: $300-559.5\,\rm{nm}$, VIS: $559.5-1024\,\rm{nm}$, NIR: $1024-2480\,\rm{nm}$). By design the spectral format for the three arms is fixed, resulting in the same wavelength coverage for all observations.

For the purpose of this work we focus on the near-infrared spectroscopy to study the broad \siiv{}, \civ{}, \ciii{} and \mgii{} quasar emission lines. The X-SHOOTER NIR spectroscopy of our sample was collected from a variety of observing programs listed in Table\,\ref{table:sample_add}.   
Total exposure times of the near-infrared observations vary between 2400\,s and 80400\,s (median 7200\,s). The observations were taken with slit widths of $0\farcs6$, $0\farcs9$, and $1\farcs2$ resulting in resolutions of $R\sim8100$, $\sim5600$, and $\sim4300$ for the NIR arm.

\subsection{Data reduction of the X-SHOOTER near-infrared spectroscopy}

In order to guarantee a homogeneous analysis we reduce the X-SHOOTER NIR spectra using the newly developed open source Python Spectroscopic Data Reduction Pipeline, \texttt{PypeIt}\footnote{https://github.com/pypeit/PypeIt} \citep{PypeitProchaska2019, PypeitProchaska2020}. 
We include six quasar spectra in our sample, which were already reduced with \texttt{PypeIt} and presented in \citet{Eilers2020a}. 
The pipeline uses supplied flat field images to automatically trace the echelle orders and correct for the detector illumination. Difference imaging of dithered AB pairs and a 2D BSpline fitting procedure are used to perform sky subtraction on the 2D images. 
Object traces are automatically identified and extracted to produce 1D spectra using the optimal spectrum extraction technique \citep{Horne1986}. 
We apply a relative flux correction to all 1D spectra using X-SHOOTER flux standards, which were taken at most 6-months apart from the observations. 
All flux-calibrated 1D spectra of each quasar are then co-added and corrected for telluric absorption using \texttt{PypeIt}. A telluric model is fit to correct the absorbed science spectrum up to a best-fit PCA model \citep{Davies2018d} of said spectrum. The telluric model is based on telluric model grids produced from the Line-By-Line Radiative Transfer Model \citep[LBLRTM4][]{Clough2005, Gullikson2014}. 
In the last step we apply an absolute flux calibration to the fully reduced quasar spectra. 
All quasars in our sample have available $J$-band photometry measurements in the literature, while only a sub-set has $K$-band measurements. Therefore, we normalized the spectra using the $J$-band magnitudes (see Table\,\ref{table:sample}). The near-infrared quasar spectra were not corrected for Galactic extinction, which is negligible at the observed wavelengths.

{\movetabledown=1.5in 
\begin{deluxetable*}{lrrcccccc}
\rotate 
\tabletypesize{\footnotesize} 
\tablecaption{The X-SHOOTER/ALMA sample of high-redshift quasars (1) - General quasar properties \label{table:sample}}
\tablehead{\colhead{Quasar Name} &\colhead{R.A. (J2000)} &\colhead{Decl. (J2000)} &\colhead{$z_{\rm{sys}}$} &\colhead{Method} &\colhead{Reference} &\colhead{Cross} &\colhead{Modeled Lines} &J-band \\ 
\nocolhead{} &\colhead{(hh:mm:ss.sss)} &\colhead{(dd:mm:ss.ss)} &\nocolhead{} &\colhead{($z_{\rm{sys}}$)} &\colhead{($z_{\rm{sys}}$)} &\colhead{Reference} & \nocolhead{} &\colhead{(AB mag)} 
} 
\startdata 
PSO~J004.3936+17.0862 & 00:17:34.467& +17:05:10.70& $5.8165\pm0.0023$ & [CII]& \cite{Eilers2020a}& g&\ion{C}{4}(1G), C~III], \ion{Mg}{2} &$20.67\pm0.16$ \\ 
PSO~J007.0273+04.9571 & 00:28:06.560& +04:57:25.68& $6.0015\pm0.0002$ & [CII]& \cite{Venemans2020}& f&\ion{C}{4}, C~III], \ion{Mg}{2} &$19.77\pm0.11$ \\ 
PSO~J009.7355--10.4316 & 00:38:56.522& -10:25:53.90& $6.0040\pm0.0003$ & [CII]& \cite{Venemans2020}& &\ion{C}{4}(1G) &$19.93\pm0.07$ \\ 
PSO~J011.3898+09.0324 & 00:45:33.568& +09:01:56.96& $6.4694\pm0.0025$ & [CII]& \cite{Eilers2020a}& g&\ion{C}{4}(1G), \ion{Mg}{2} &$20.80\pm0.13$ \\ 
VIK~J0046--2837 & 00:46:23.645& -28:37:47.34& $5.9926\pm0.0028$ & MgII& This work& &\ion{Mg}{2} &$20.96\pm0.09$ \\ 
SDSS~J0100+2802 & 01:00:13.027& +28:02:25.84& $6.3269\pm0.0002$ & [CII]& \cite{Venemans2020}& e&\ion{Mg}{2} &$17.64\pm0.02$ \\ 
VIK~J0109--3047 & 01:09:53.131& -30:47:26.31& $6.7904\pm0.0003$ & [CII]& \cite{Venemans2020}& b,c,e&\ion{C}{4}(1G),  \ion{Mg}{2} &$21.27\pm0.16$ \\ 
PSO~J036.5078+03.0498 & 02:26:01.875& +03:02:59.40& $6.5405\pm0.0001$ & [CII]& \cite{Venemans2020}& c,e&\ion{Si}{4}, \ion{C}{4}(1G), \ion{Mg}{2} &$19.51\pm0.03$ \\ 
VIK~J0305--3150 & 03:05:16.916& -31:50:55.90& $6.6139\pm0.0001$ & [CII]& \cite{Venemans2019}& b,c,e&\ion{C}{4}(1G), \ion{Mg}{2} &$20.68\pm0.07$ \\ 
PSO~J056.7168--16.4769 & 03:46:52.044& -16:28:36.88& $5.9670\pm0.0023$ & [CII]& \cite{Eilers2020a}& g&\ion{C}{4}, C~III], \ion{Mg}{2} &$20.25\pm0.10$ \\ 
PSO~J065.4085--26.9543 & 04:21:38.049& -26:57:15.61& $6.1871\pm0.0003$ & [CII]& \cite{Venemans2020}& &\ion{C}{4}(1G), \ion{Mg}{2} &$19.36\pm0.02$ \\ 
PSO~J065.5041--19.4579 & 04:22:00.995& -19:27:28.69& $6.1247\pm0.0006$ & [CII]& \cite{Decarli2018}& &\ion{C}{4}(1G), \ion{Mg}{2} &$19.90\pm0.15$ \\ 
SDSS~J0842+1218 & 08:42:29.430& +12:18:50.50& $6.0754\pm0.0005$ & [CII]& \cite{Venemans2020}& a,f&\ion{C}{4}, \ion{Mg}{2} &$19.78\pm0.03$ \\ 
SDSS~J1030+0524 & 10:30:27.098& +05:24:55.00& $6.3048\pm0.0012$ & LyaH& \cite{Farina2019}& a,e&\ion{C}{4}, \ion{Mg}{2} &$19.79\pm0.08$ \\ 
PSO~J158.69378--14.42107 & 10:34:46.509& -14:25:15.89& $6.0681\pm0.0024$ & [CII]& \cite{Eilers2020a}& g&\ion{C}{4}, \ion{Mg}{2} &$19.19\pm0.06$ \\ 
PSO~J159.2257--02.5438 & 10:36:54.190& -02:32:37.94& $6.3809\pm0.0005$ & [CII]& \cite{Decarli2018}& &\ion{C}{4}, \ion{Mg}{2} &$20.00\pm0.10$ \\ 
SDSS~J1044--0125 & 10:44:33.041& -01:25:02.20& $5.7846\pm0.0005$ & [CII]& \cite{Venemans2020}& e,f&\ion{C}{4}(1G), C~III] &$19.25\pm0.05$ \\ 
VIK~J1048--0109 & 10:48:19.082& -01:09:40.29& $6.6759\pm0.0002$ & [CII]& \cite{Venemans2020}& &\ion{Mg}{2} &$20.65\pm0.17$ \\ 
ULAS~J1120+0641 & 11:20:01.478& +06:41:24.30& $7.0848\pm0.0004$ & [CII]& \cite{Venemans2020}& b,c,e&\ion{Si}{4}, \ion{C}{4}, C~III] &$20.36\pm0.05$ \\ 
ULAS~J1148+0702 & 11:48:03.286& +07:02:08.33& $6.3337\pm0.0028$ & MgII& This work& f&\ion{C}{4}, \ion{Mg}{2} &$20.30\pm0.11$ \\ 
PSO~J183.1124+05.0926 & 12:12:26.984& +05:05:33.49& $6.4386\pm0.0002$ & [CII]& \cite{Venemans2020}& e&\ion{C}{4}(1G), \ion{Mg}{2} &$19.77\pm0.08$ \\ 
SDSS~J1306+0356 & 13:06:08.258& +03:56:26.30& $6.0330\pm0.0002$ & [CII]& \cite{Venemans2020}& a,e&\ion{C}{4}, C~III], \ion{Mg}{2} &$19.71\pm0.10$ \\ 
ULAS~J1319+0950 & 13:19:11.302& +09:50:51.49& $6.1347\pm0.0005$ & [CII]& \cite{Venemans2020}& e&\ion{C}{4}, C~III], \ion{Mg}{2} &$19.70\pm0.03$ \\ 
ULAS~J1342+0928 & 13:42:08.105& +09:28:38.61& $7.5400\pm0.0003$ & [CII]& \cite{Banados2019}& e&\ion{Si}{4}, \ion{C}{4}(1G), C~III] &$20.30\pm0.02$ \\ 
CFHQS~J1509--1749 & 15:09:41.779& -17:49:26.80& $6.1225\pm0.0007$ & [CII]& \cite{Decarli2018}& e&\ion{C}{4}, C~III], \ion{Mg}{2} &$19.80\pm0.08$ \\ 
PSO~J231.6576--20.8335 & 15:26:37.838& -20:50:00.66& $6.5869\pm0.0004$ & [CII]& \cite{Venemans2020}& c,e&\ion{Mg}{2} &$19.66\pm0.05$ \\ 
PSO~J239.7124--07.4026 & 15:58:50.991& -07:24:09.59& $6.1097\pm0.0024$ & [CII]& \cite{Eilers2020a}& g&\ion{C}{4}, \ion{Mg}{2} &$19.35\pm0.08$ \\ 
PSO~J308.0416--21.2339 & 20:32:09.994& -21:14:02.31& $6.2355\pm0.0003$ & [CII]& \cite{Venemans2020}& &\ion{C}{4}, \ion{Mg}{2} &$20.17\pm0.11$ \\ 
SDSS~J2054--0005 & 20:54:06.490& -00:05:14.80& $6.0389\pm0.0001$ & [CII]& \cite{Venemans2020}& &\ion{C}{4}(1G), C~III], \ion{Mg}{2} &$20.12\pm0.06$ \\ 
CFHQS~J2100--1715 & 21:00:54.619& -17:15:22.50& $6.0807\pm0.0004$ & [CII]& \cite{Venemans2020}& g&\ion{C}{4}(1G), \ion{Mg}{2} &$21.42\pm0.10$ \\ 
PSO~J323.1382+12.2986 & 21:32:33.189& +12:17:55.26& $6.5872\pm0.0004$ & [CII]& \cite{Venemans2020}& c,e&\ion{Si}{4}, \ion{C}{4}, C~III], \ion{Mg}{2} &$19.74\pm0.03$ \\ 
VIK~J2211--3206 & 22:11:12.391& -32:06:12.95& $6.3394\pm0.0010$ & [CII]& \cite{Decarli2018}& &\ion{C}{4}(1G), \ion{Mg}{2} &$19.62\pm0.03$ \\ 
CFHQS~J2229+1457 & 22:29:01.649& +14:57:09.00& $6.1517\pm0.0005$ & [CII]& \cite{Willott2015}& g&\ion{C}{4} &$21.95\pm0.07$ \\ 
PSO~J340.2041--18.6621 & 22:40:49.001& -18:39:43.81& $6.0007\pm0.0020$ & LyaH& \cite{Farina2019}& &\ion{C}{4}, C~III],  \ion{Mg}{2} &$20.28\pm0.08$ \\ 
SDSS~J2310+1855 & 23:10:38.880& +18:55:19.70& $6.0031\pm0.0002$ & [CII]& \cite{Wang2013}& f&\ion{C}{4}(1G), \ion{Mg}{2} &$18.88\pm0.05$ \\ 
VIK~J2318--3029 & 23:18:33.103& -30:29:33.36& $6.1456\pm0.0002$ & [CII]& \cite{Venemans2020}& &\ion{C}{4}(1G), \ion{Mg}{2} &$20.20\pm0.06$ \\ 
VIK~J2348--3054 & 23:48:33.336& -30:54:10.24& $6.9007\pm0.0005$ & [CII]& \cite{Venemans2020}& b,c,e&\ion{C}{4}(1G), C~III], \ion{Mg}{2} &$21.14\pm0.08$ \\ 
PSO~J359.1352--06.3831 & 23:56:32.452& -06:22:59.26& $6.1719\pm0.0002$ & [CII]& \cite{Venemans2020}& g&\ion{C}{4}(1G), \ion{Mg}{2} &$19.85\pm0.10$ \\ 
\enddata 
\tablecomments{\textbf{Quasar coordinates are available in decimal degrees in an on-line table summarized in Table\,\ref{table:master_table_overview}}. Redshift method abbreviations: [CII] - peak of the \ensuremath{\text{\textsc{[Cii]}}_{158\,\mu\text{m}}} line, MgII - peak of the \textsc{Mg ii} 2798\,\AA\ line, LyaH - determined from the Lyman-$\alpha$ halo.}
\tablerefs{The cross references in the table denote previous publications analyzing near-infrared spectroscopy of these quasars. The references are: a=\citet{DeRosa2011}, b=\citet{DeRosa2014}, c=\citet{Mazzucchelli2017}, d=\citet{Onoue2019}, e=\citet{Meyer2019c}, f=\citet{Shen2019a}, g=\citet{Eilers2020a}).}
\end{deluxetable*} 
}

{\movetabledown=2.0in 
\begin{deluxetable*}{lcccc}
\rotate 
\tabletypesize{\footnotesize} 
\tablecaption{The X-SHOOTER/ALMA sample of high-redshift quasars (2) - Information on X-SHOOTER spectroscopy and discovery reference\label{table:sample_add}}
\tablecolumns{8}\tablehead{\colhead{Quasar Name} &\colhead{Exp. Time (s)} & \colhead{X-SHOOTER Proposal ID} &\colhead{PI} &\colhead{Discovery Ref.} 
} 
\startdata 
PSO~J004.3936+17.0862 & 3600 & 0101.B-0272(A) & Eilers& \cite{Banados2016}\\ 
PSO~J007.0273+04.9571 & 2400 & 098.B-0537(A) & Farina& \cite{Banados2014, Jiang2015}\\ 
PSO~J009.7355--10.4316 & 4800 & 097.B-1070(A) & Farina& \cite{Banados2016}\\ 
PSO~J011.3898+09.0324 & 3600 & 0101.B-0272(A) & Eilers& \cite{Mazzucchelli2017}\\ 
VIK~J0046--2837 & 12000 & 097.B-1070(A) & Farina& \cite{Decarli2018}\\ 
SDSS~J0100+2802 & 10800 & 096.A-0095(A) & Pettini& \cite{Wu2015}\\ 
VIK~J0109--3047 & 24000 & 087.A-0890(A), 088.A-0897(A) & De Rosa, De Rosa& \cite{Venemans2013}\\ 
PSO~J036.5078+03.0498 & 14400 & 0100.A-0625(A), 0102.A-0154(A) & D'Odorico, D'Odorico& \cite{Venemans2015a}\\ 
VIK~J0305--3150 & 16800 & 098.B-0537(A) & Farina& \cite{Venemans2013}\\ 
PSO~J056.7168--16.4769 & 7200 & 097.B-1070(A) & Farina& \cite{Banados2016}\\ 
PSO~J065.4085--26.9543 & 2400 & 098.B-0537(A) & Farina& \cite{Banados2016}\\ 
PSO~J065.5041--19.4579 & 4800 & 097.B-1070(A) & Farina& \cite{Banados2016}\\ 
SDSS~J0842+1218 & 7200 & 097.B-1070(A) & Farina& \cite{DeRosa2011, Jiang2015}\\ 
SDSS~J1030+0524 & 4800 & 086.A-0162(A) & D'Odorico& \cite{Fan2001c}\\ 
PSO~J158.69378--14.42107 & 4320 & 096.A-0418(B) & Shanks& \cite{Chehade2018}\\ 
PSO~J159.2257--02.5438 & 4800 & 098.B-0537(A) & Farina& \cite{Banados2016}\\ 
SDSS~J1044--0125 & 2400 & 084.A-0360(A) & Hjorth& \cite{Fan2000}\\ 
VIK~J1048--0109 & 4800 & 097.B-1070(A) & Farina& \cite{WangFeige2017}\\ 
ULAS~J1120+0641 & 72000 & 286.A-5025(A), 089.A-0814(A), 093.A-0707(A) & Venemans, Becker, Becker& \cite{Mortlock2011}\\ 
ULAS~J1148+0702 & 9600 & 098.B-0537(A) & Farina& \cite{Jiang2016}\\ 
PSO~J183.1124+05.0926 & 4800 & 098.B-0537(A) & Farina& \cite{Mazzucchelli2017}\\ 
SDSS~J1306+0356 & 41400 & 084.A-0390(A) & Ryan-Weber& \cite{Fan2001c}\\ 
ULAS~J1319+0950 & 36000 & 084.A-0390(A) & Ryan-Weber& \cite{Mortlock2009}\\ 
ULAS~J1342+0928 & 80400 & 098.B-0537(A), 0100.A-0898(A) & Farina, Venemans& \cite{Banados2018}\\ 
CFHQS~J1509--1749 & 24000 & 085.A-0299(A), 091.C-0934(B) & D'Odorico, Kaper& \cite{Willott2007}\\ 
PSO~J231.6576--20.8335 & 2400 & 097.B-1070(A) & Farina& \cite{Mazzucchelli2017}\\ 
PSO~J239.7124--07.4026 & 3600 & 0101.B-0272(A) & Eilers& \cite{Banados2016}\\ 
PSO~J308.0416--21.2339 & 9600 & 098.B-0537(A) & Farina& \cite{Banados2016}\\ 
SDSS~J2054--0005 & 7200 & 60.A-9418(A) & Ryan-Weber& \cite{Jiang2008}\\ 
CFHQS~J2100--1715 & 12000 & 097.B-1070(A) & Farina& \cite{Willott2010a}\\ 
PSO~J323.1382+12.2986 & 7200 & 098.B-0537(A) & Farina& \cite{Mazzucchelli2017}\\ 
VIK~J2211--3206 & 5280 & 096.A-0418(A), 098.B-0537(A) & Shanks, Farina& \cite{Decarli2018}\\ 
CFHQS~J2229+1457 & 6000 & 0101.B-0272(A) & Eilers& \cite{Willott2010a}\\ 
PSO~J340.2041--18.6621 & 9600 & 098.B-0537(A) & Farina& \cite{Banados2014}\\ 
SDSS~J2310+1855 & 2400 & 098.B-0537(A) & Farina& \cite{Wang2013, Jiang2016}\\ 
VIK~J2318--3029 & 9600 & 097.B-1070(A) & Farina& \cite{Decarli2018}\\ 
VIK~J2348--3054 & 9200 & 087.A-0890(A) & De Rosa & \cite{Venemans2013}\\ 
PSO~J359.1352--06.3831 & 4800 & 098.B-0537(A) & Farina& \cite{Banados2016, WangFeige2016}\\ 
\enddata 
\end{deluxetable*} 
}

\subsection{Properties of the X-SHOOTER/ALMA sample}\label{sec:sample_properties}

Figures\,\ref{fig:sample},\,\ref{fig:sample2}, and \,\ref{fig:sample3} show the near-infrared spectra of all quasars in the X-SHOOTER/ALMA sample. We over-plot our model fits (see Section\,\ref{sec:fitting}) in solid orange lines and highlight the positions of the broad \civ{}, \ciii{} and \mgii{} emission lines based on the quasar systemic redshift. The spectra are sorted in redshift beginning with the lowest-redshift spectrum. Wavelength ranges affected by strong telluric absorption, as seen in the telluric model example in each figure, have been removed from the spectra for display purposes. Detailed descriptions of the fits for individual quasars are provided in Appendix\,\ref{sec:app_specfitnotes}.

For five quasar spectra we were not able to fit the continuum with our power law and Balmer continuum model across the full wavelength range. These objects are PSO~J009.7355--10.4316, VIK~J0046--2837, PSO~J065.4085--26.9543, PSO~J065.5041--19.4579, CFHQS~J2100--1715 (see classification "D" in Table\,\ref{table:sample_add} of Appendix\,\ref{sec:app_additional_tables}).
In these cases the quasar continuum flux declines blue-ward of the \ciii{} ($\lesssim1900$\,\AA) complex. 
This behavior could be attributed to extinction by the quasar host galaxy or by obscuring material just outside the broad line region, e.g. associated with broad absorption lines (BAL). 
We provide the properties of the broad \civ{} and \mgii{} lines as well as the fluxes and luminosities at $1450\,\rm{\AA}$ and $3000\,\rm{\AA}$ for these five quasars by fitting the regions around the \civ{} and \mgii{} line separately. Due to their intrinsic attenuation, the observed continuum luminosities should be regarded as lower limits for these quasars. 
Throughout this work we clearly state when these quasars are included in the analysis and we specifically highlight them in figures with open, instead of filled orange circles.
After fully excluding instrumental effects an in-depth study of these five sources, including a model for their extinction, is needed to further understand their nature. This is beyond the scope of this paper.

Three quasars in our sample were previously classified as BAL quasars: VIK~J2348--3054 \citep{DeRosa2014}, SDSS~J1044--0125 \citep{Shen2019a}, PSO~J239.7124--07.4026 \citep{Eilers2020a}. We visually classify PSO~J065.5041--19.4579 and VIK~J2211--3206 as BAL quasars by their strong absorption blue-ward of \civ{}. An additional quasar, VIK~J2318-3029, shows an absorption feature at the very blue edge of the spectrum, which potentially indicates a BAL. We will revisit its classification once the optical X-SHOOTER spectrum has been analyzed.
While our sample is not a complete account of high-redshift quasars in this redshift and luminosity range, the BAL fraction of 5/\Nsamp{}$\approx13\%$ is roughly consistent with lower redshift studies \citep[e.g.][]{Trump2006, Maddox2008}. 

Additionally, three quasars in our sample show features associated with proximate damped Lyman-$\alpha$ absorbers (pDLAs), SDSS~J2310+1855 \citep{DOdorico2018}, PSO~J183.1124+05.0926 \citep{Banados2019b}, and PSO~J056.7168--16.4769 \citep{Davies2020a, Eilers2020a}.



\section{Modeling of the NIR spectra}\label{sec:fitting}

\begin{figure*}
\centering
\includegraphics[width=0.92\textwidth]{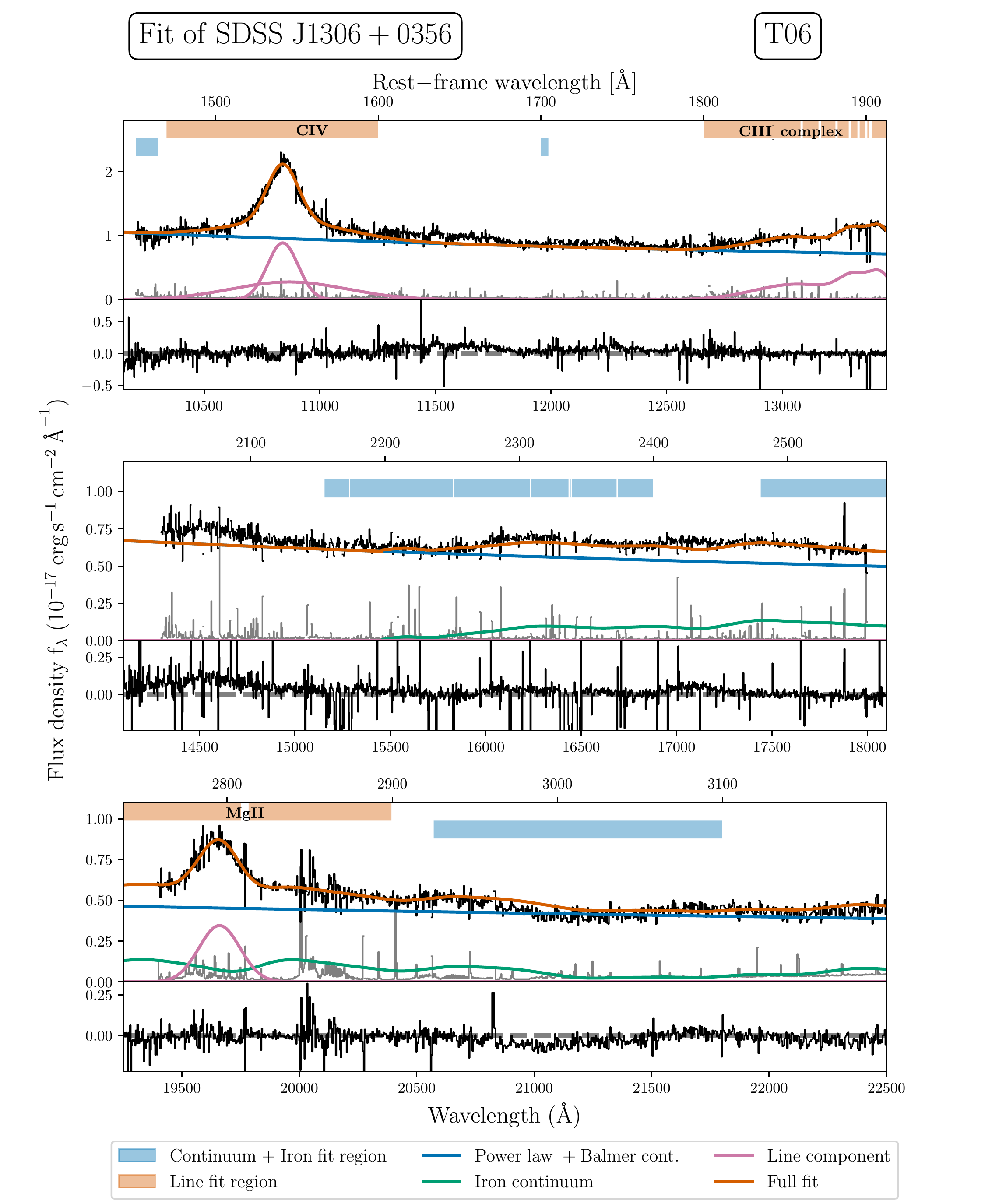}
\caption{Best fit to the near-infrared spectrum of the quasar J1306+0356 at $z=$6.033. The spectrum, binned by 4 pixels, is depicted in black, with flux errors in grey. The combined fit of the continuum and the emission lines is shown as the dark orange line. The combined power law and Balmer continuum model is highlighted in blue, while we show the iron pseudo-continuum in green. Models for the individual emission lines are shown in purple. Light blue and orange bars at the top of each panel show the regions, which constrain the fit for the continuum and emission lines models, respectively. 
As our study focuses on the \civ{} and \mgii{} emission lines, we have not included some other emission  features seen in this spectrum. For example, the emission lines between \civ{} and \ciii{} ](e.g. \heii{} at 1640\,\AA\ or \oiiibr\ at 1663\,\AA; rest-frame) or the broad \feiii{} features at 2000\,\AA-2100\,\AA\ and 2430\,\AA\ (rest-frame) are not included in the fit.
We further note the iron template used in this fit in the top right corner (\citetalias{Tsuzuki2006}). Best-fit figures for all quasars are available as a figure set (83 images) accompanying this paper in the online journal.}
\label{fig:example_fit}
\end{figure*}

Before we start the model fitting we pre-process the spectra. The majority of the X-SHOOTER NIR spectra have a relatively low signal-to-noise ratio (SNR) in the $J$-band (median SNR$=6.2$, 12500-13450\,\AA). Therefore, we bin the spectra by a factor of 4 in wavelength, increasing the median $J$-band SNR to 11.8. 
Additionally, iterative sigma clipping (${>}3\sigma$) masks out the residuals of strong sky lines or intrinsic narrow absorption lines to allow for better fit results.

We then fit the near-infrared spectra using a custom fitting code, which is based on the LMFIT python package \citep{lmfit2014}. The code enables the user to interactively adjust the fit regions and allowed parameter ranges.
A few spectral regions have to be excluded from the fit. We begin by masking out the reddest order of the X-SHOOTER NIR arm ($\lambda_{\rm{obs}} =22500-25000$\,\AA), which is strongly affected by high background noise from scattered light. 
In addition, we mask out regions with generally low signal to noise ratio, which includes wavelength windows with strong telluric absorption as well as the blue edge of the NIR spectrum. These masked regions are $\lambda_{\rm{obs}} = 13450{-}14300\,\text{\AA}, 18000{-}19400\,\text{\AA}$ and $\lambda_{\rm{obs}} \leq 10250\,\text{\AA}$. 
A figure set showing the best-fit models for each quasar, including the regions considered in the continuum and emission line fits accompanies this article on-line. An example of our best-fit model to SDSS~J1306+0356 is shown in Figure\,\ref{fig:example_fit}.

The spectral modeling is a two stage process. 
In a first step we model the continuum components and the broad \siiv{}, \civ{}, \ciii{} and \mgii{} emission lines. The best-fit model is saved.
In the second step we estimate the uncertainties on the fit parameters. We re-sample each spectrum 1000 times and then draw new flux values on a pixel by pixel basis from a Gaussian distribution, where we assumed the original flux value to be the mean and the flux errors as its standard deviation. Each re-sampled spectrum is then automatically fit using our interactively determined best-fit as the initial guess. 


In this section we describe the assumptions and method of the fitting process. In Section\,\ref{sec:fitanalysis} we briefly discuss how we measure the spectral properties and derive related quantities published along this paper from the fits. Additional details on the spectral modeling of individual quasars are given in Appendix\,\ref{sec:app_specfitnotes}.

\subsection{Continuum model}

The rest-frame UV/optical spectrum of quasars is dominated by radiation from the accretion disk, which is well modeled as a single power law component. Additionally blended high order Balmer lines and bound-free Balmer continuum emission give rise to a Balmer pseudo-continuum, which is a non-negligible component in the wavelength range of our near-infrared spectra. Transitions of single and double ionized iron atoms (\feii{} and \feiii{}) produce an additional iron pseudo-continuum, which is especially strong around the broad \mgii{} emission line.
Our model of the quasar continuum includes all three components as discussed below. We do not include emission from the stellar component of the quasar host as this can be regarded as negligible in comparison to the central engine.
In general, intrinsic absorption by the quasar host can attenuate the UV/optical spectrum. However, at this point we do not consider dust attenuation in our model.

The full continuum model is fit to regions that are chosen to be free of narrow and broad quasar emission lines. Discussions of these line free regions are provided in many references in the literature \citep[e.g.][]{Vestergaard2006, Decarli2010, Shen2011, Mazzucchelli2017, Shen2019b}.
As our continuum model includes contributions from the Balmer and iron pseudo-continua, we generally fit our continuum model to the following wavelength windows: $\lambda_{\rm{rest}} = 1445{-}1465\,\text{\AA}$, $1700{-}1705\,\text{\AA}$,  $2155{-}2400\,\text{\AA}$, $2480{-}2675\,\text{\AA}$, and $2900{-}3090\,\text{\AA}$.
These continuum windows are interactively adjusted on a case by case basis to exclude regions with strong sky residuals, unusually large flux errors or broad absorption features. 

\subsubsection{Power law and Balmer continuum}\label{sec:plbc_model}

We model the emission of the accretion disk as a power law normalized at $2500\,\text{\AA}$:
\begin{equation}
    F_{\rm{PL}}(\lambda) = F_{\rm{PL}, 0}\ \left(\frac{\lambda}{2500\,\text{\AA}}\right)^{\alpha_{\lambda}} \ .
\end{equation}
Here $F_{\rm{PL}, 0}$ is the normalization and $\alpha_{\lambda}$ is the slope of the power law.

The X-SHOOTER NIR arm spectral coverage only allows to reach rest-frame wavelengths of $< 3400\,\text{\AA}$ for our quasar sample. As our spectra do not cover the Balmer break at $\lambda_{\rm{BE}} = 3646\,\text{\AA}$, we only model the bound-free emission of the Balmer pseudo-continuum. 
For this we follow the description of \citet{Dietrich2003c}, who assumed the Balmer emission arises from gas clouds of uniform electron temperature that are partially optically thick:
\begin{equation}
\begin{split}
   F_{\rm{BC}}(\lambda) =  F_{\rm{BC}, 0}\ B_{\lambda}(\lambda, T_e) \left(1 - e^{\left(-\tau_{\rm{BE}} (\lambda/\lambda_{\rm{BE}})^3 \right)} \right) \ ,  \\
   \forall\ \lambda \le \lambda_{\rm{BE}} \ ,
 \end{split}
\end{equation}
where $B_{\lambda}(T_e)$ is the Planck function at the electron temperature of $T_e$, $\tau_{\rm{BE}}$ is the optical depth at the Balmer edge and $F_{\rm{BC}, 0}$ is the normalized flux density at the Balmer break \citep{Grandi1982}. 
\citet{Dietrich2003c} discuss that the strength of the Balmer emission ($F_{\rm{BC}, 0}$) can be estimated from the flux density slightly redward of the Balmer edge at $\lambda = 3675\,\text{\AA}$ after subtraction of the power law continuum. However, our wavelength range does not cover this region in the spectra. Therefore, we follow previous studies \citep{DeRosa2011, Mazzucchelli2017, Onoue2020} and fix the Balmer continuum contribution to $30\%$ \citep{Dietrich2003c, Kurk2007, DeRosa2011, Shin2019, Onoue2020} of the power law flux at the Balmer edge by requiring:
\begin{equation}
    F_{\rm{BC}}(3646\,\text{\AA}, F_{\rm{BC}, 0}) = 0.3 \times F_{\rm{PL}}(3646\,\text{\AA})
\end{equation}
This choice does not affect the final results as discussed in \citet{Onoue2020}. We further fix the electron temperature and the optical depth to values of $T_e = 15,000\,\rm{K}$  and $\tau_{\text{BE}}=1$, common values in the literature \citep{Dietrich2003c, Kurk2007, DeRosa2011, Mazzucchelli2017, Shin2019, Onoue2020}.

\subsubsection{Iron pseudo-continuum}

Careful analysis of the quasar continuum as well as the properties of the broad emission lines is complicated by the presence of atomic and ionic iron in the broad-line region. The large number of electron levels in iron atoms lead to a multitude of emission line transitions, especially from \feii{}, throughout the entire spectral region probed in this study. Due to the large velocities of the broad-line region clouds the weak iron emission lines are broadened and blend into a pseudo-continuum, hindering our analysis of the \civ{}, \ciii{}, and \mgii{} emission lines. 
Empirical and semi-empirical iron templates, derived from the narrow-line Seyfert 1 galaxy I Zwicky 1 \citep{Boroson1992a, Vestergaard2001, Tsuzuki2006}, allow to easily incorporate iron emission into spectral fitting routines. For our analysis we use both the \citet[][hereafter T06]{Tsuzuki2006} and the \citet[][hereafter VW01]{Vestergaard2001} template. 

\citetalias{Tsuzuki2006} and \citetalias{Vestergaard2001} discuss that subdividing the iron template into segments may be necessary as the individual emission strengths of the iron multiplets vary across the spectrum. \citetalias{Vestergaard2001} discuss that their undivided iron template over predicts the iron emission in the $\lambda_{\rm{rest}} = 1400-1530\,\text{\AA}$ region.
Based on this insight we divided the iron template into two segments, one covering the \civ{} \citepalias{Vestergaard2001} and one \mgii{} \citepalias{Tsuzuki2006} line and performed test fits on a few spectra. We discovered that we were not able to constrain the weak iron emission around the \civ{} at rest-frame wavelengths of $\lambda_{\rm{rest}} = 1200-2200\,\text{\AA}$. Therefore, we only incorporate an iron template in our continuum model to separate the \mgii{} line from the underlying iron pseudo-continuum at rest-frame wavelengths of $\lambda_{\rm{rest}} = 2200-3500\,\text{\AA}$.

In contrast to the purely empirical \citetalias{Vestergaard2001} template, in which iron emission beneath broad \mgii{} line is not included, \citetalias{Tsuzuki2006} were able to model this iron contribution using a spectral synthesis code and add it to their template. 
The difficulties in fitting the Fe contribution in quasar spectra is discussed in many studies throughout the literature \citep[e.g.][]{Boroson1992b, Vestergaard2001, Tsuzuki2006, Woo2018, Shin2019, Onoue2020}. A detailed analysis on covariance between the iron contribution and the power law fit is given in \citet{DeRosa2011}.
We will expand on this discussion based on the quantitative results of our sample in Section\,\ref{sec:irondiscussion}.

The original iron emission in the I Zwicky 1 templates have an intrinsic width of $ \rm{FWHM} \approx 900\,\text{km}\text{s}^{-1}$. Therefore, to accurately model the iron emission in our spectra, we broaden the iron templates by convolving them with a Gaussian kernel to match the FWHM of the broad \mgii{} line:
\begin{equation}
 \sigma_{\rm{conv}} = \frac{\sqrt{\rm{FWHM}_{\rm{Mg~II}}^2  -\rm{FWHM}_{\rm{I\,Zwicky\,1}}^2}}{2\sqrt{2\ln2}}
\end{equation}
While the broadening of the iron emission is necessary to study quasars \citep[e.g.][]{Boroson1992a}, our approach is most similar to \citetalias{Tsuzuki2006} and \citet{Shin2019}, who also use the FWHM of the \mgii{} line as a proxy for the velocity dispersion of the broad-line region. 
\citet{Shin2019} compare how a similar assumption influences the measurement of the \feii{}/\mgii{} flux ratio. The authors constrain the \feii{} pseudo-continuum velocity dispersion within 10\% of the \mgii{} FWHM and find \feii{}/\mgii{} flux ratios consistent to each other within 7\% compared to leaving the iron FWHM as a free parameter. Hence, we are confident that this assumption only has a minor influence on our best-fit measurements.

In addition to the FWHM, we also set the iron template redshift to the redshift of the broad \mgii{} line. As the iron template and the \mgii{} fits are interdependent, we iteratively fit the full continuum model and the \mgii{} line. In each step we update the iron template parameters after the \mgii{} line fit until the FWHM and the redshift of the \mgii{} line converge.

\subsection{Emission line models}

Our analysis focuses on the broad emission lines \siiv{}, \civ{}, \ciii{} and \mgii{}. All four lines are doublets. 
However, their broad nature along with the modest signal-to-noise of our spectra does not allow us to resolve them. Therefore, these lines are modeled as single broad lines at rest-frame wavelengths of \siiv{} $\lambda1396.76\,\rm{\AA}$ , $\lambda1549.06\,\rm{\AA}$ for \civ{}, $\lambda1908.73\,\rm{\AA}$ for \ciii{}, and $\lambda2798.75\,\rm{\AA}$ for \mgii{} \citep[see][]{VandenBerk2001}.
We provide an overview over the lines modeled in each quasar spectrum in Table\,\ref{table:sample}. 

\subsubsection{\mgii{} emission line}

The majority of the analyzed X-SHOOTER NIR spectra detect the \mgii{} line with a low SN-ratio (median SNR$=8.6$) even in the binned spectra.
Hence, we decided to model the \mgii{} line with a single Gaussian profile only. The line is generally fit over rest-frame wavelengths of $\lambda_{\rm{rest}} = 2700{-}2900\,\rm{\AA}$, similar to \citep{Shen2019b}. We adjust this wavelength range to mask out regions with absorption lines, bad sky subtraction, or noisy telluric correction. We vary the model parameters to find the best fit of the redshift, the FWHM and the amplitude of the Gaussian profile, assuming a rest-frame central wavelength of $\lambda2798.75\,\rm{\AA}$ \citep{VandenBerk2001}.

\subsubsection{\civ{} emission line}

In comparison to the \mgii{} emission line, the \civ{} line is known to often exhibit asymmetric line profiles associated with an outflowing wind component \citep[e.g.][]{Richards2011}. We always start by using two Gaussian profiles to model the \civ{} line, which allows us to account for this asymmetry. However, not all lines are asymmetric and spectra with very low signal-to-noise ratios or strong absorption lines, can often not constrain a two component model. Therefore, we fit the \civ{} line with a single Gaussian component (1G) in nearly half of our sample (see Table\,\ref{table:sample}). 
We fit the \civ{} line in a rest-frame wavelength window of $\lambda_{\rm{rest}} = 1470{-}1600\,\text{\AA}$. Quasars at high redshift are known to exhibit highly blueshifted \civ{} compared to their other emission lines \citep[e.g.][]{Meyer2019c}. Therefore, we have slightly extended the fitting range bluewards compared to \citet{Shen2011, Shen2019b}.  Equivalent to the \mgii{} line fit, the central wavelength (redshift), the FWHM and the amplitude of each Gaussian component are optimized independent of each other to find the best fit. The line properties are then determined from the combined components of the line fit.

\subsubsection{\ciii{} emission line}

The \ciii{} emission line falls into the telluric absorption window between the $J$- and the $H$-band in the redshift range of $z\approx6$ to $6.5$. Thus we were able to determine properties related to the line only for a subset of our X-SHOOTER spectra. 
In addition, the proximity of the \aliii{} $\lambda1857.40\,\rm{\AA}$ and \siiii{} $\lambda1892.02\,\rm{\AA}$ emission lines to the \ciii{} $\lambda1908.73\,\rm{\AA}$ line complicates the modeling. This is especially true in quasar spectra, where these lines are usually blended due to the large velocity dispersion of the broad emission lines.
Each of the three lines is modeled with a single Gaussian profile.
While we allow for independent variations of the amplitude and FWHM of the three Gaussian profiles, they are fit to the same redshift using the rest-frame line centers provided above. The region over which the lines are fit are always adjusted manually due to the proximity to wavelength regions with strong telluric absorption. 

The combination of the three lines provides a reasonable fit in most cases. However, the individual line contributions of the strongly degenerate \siiii{} and \ciii{} lines cannot be separated with certainty. Therefore, we only extract peak redshift measurement from the \ciii{} complex (sum of all three line models) fits and disregard other properties.
In order to properly fit the \ciii{} complex in a few quasar spectra it was necessary to set the contributions of the \aliii{} and \siiii{} lines to zero. These details are given in Appendix\,\ref{sec:app_specfitnotes}.

\subsubsection{\siiv{} emission line}
In quasars at $z\gtrsim6.4$ the broad \siiv{} $\lambda1396.76\,\rm{\AA}$ emission line redshifts into the wavelength range of the X-SHOOTER near-infrared spectra. 
The broad nature of the \siiv{} line results in a line blend with the close-by semi-forbidden O~IV] $\lambda1402.06\,\rm{\AA}$ transition. Because we cannot disentangle the two lines, we decided to model their blend, \siiv{}+O~IV] $\lambda1399.8\,\rm{\AA}$\footnote{http://classic.sdss.org/dr6/algorithms/linestable.html}, using one Gaussian component. 

The throughput and thus the signal-to-noise declines towards the blue edge of the X-SHOOTER near-infrared spectra. In addition, strong broad absorption lines blueward of the \civ{} line complicate the continuum modeling in a few cases. As a result we were only able to successfully fit the \siiv{} line in the spectra of four quasars: PSO~J036.5078{+}03.0498, ULAS~J1120{+}0641, ULAS~J1342{+}0928, and PSO~J323.1382{+}12.2986 (see Tables\,\ref{table:ciiisiiv} and \ref{table:sample_2}).

\subsection{Overview of the fitting process}
We briefly summarize the steps of the fitting process:
\begin{enumerate}
\item We bin every 4 pixels of the fully reduced X-SHOOTER NIR spectra and mask out strong sky line residuals with iterative sigma clipping.
\item We mask out all regions of strong telluric absorption.
\item We add the power law and Balmer continuum model to the fit. The power-law and Balmer continuum redshift is set to the \cii{} or Ly$\alpha$ halo redshift. In the few cases, where no accurate systemic redshift is available, we set the continuum redshift to the best systemic redshift in the literature and re-evaluate this redshift based on our fit to the \mgii{} emission line. 
\item We further add the iron template and set the initial redshift to the systemic redshift from the literature and provide an initial guess for the FWHM.
\item We fit the full continuum model (power law + Balmer continuum + iron template).
\item Then we add the \mgii{} emission line model and fit it to determine its FWHM and redshift.
\item We iteratively re-fit the full continuum model and the \mgii{} emission line (steps 5 \& 6), applying the \mgii{} FWHM and redshift to the iron template until the \mgii{} line fit converges. This takes about four to six iterations.
\item In the next step we include the \civ{} model and if possible the \ciii{} and \siiv{} models in the line fit.
\item The best model fit and its parameters are then saved. A separate routine re-samples the science spectrum 1000 times using its noise properties, bins the spectrum by every 4 pixels and then re-fits it using the saved model with the best-fit parameters as the first guess.
\end{enumerate}

\section{Analysis of the fits} \label{sec:fitanalysis}

For each best-fit in the re-fitting process we not only determine the values of all fit parameters, but we also calculate all derived quantities. This extends, for example, to black hole mass estimates and \feii{}/\mgii{} flux ratios. We re-sample and re-fit each spectrum 1000 times, resulting in distributions for each fit parameter and derived quantity. The results presented in this paper quote the median of this distribution and the associated uncertainties are the $15.9$ and $84.1$ percentile values. 
A machine readable on-line table summarizes all fit results. Table\,\ref{table:master_table_overview} in Appendix\,\ref{sec:master_table} presents the columns of this table for an overview. 

\subsection{Continuum}

We derive the flux densities and luminosities at wavelengths $1350\,\rm{\AA}$, $1400\,\rm{\AA}$, $1450\,\rm{\AA}$, $2100\,\rm{\AA}$, $2500\,\rm{\AA}$, and $3000\,\rm{\AA}$ from the power law continuum model, including the Balmer continuum contribution. 
In the case of the five spectra, for which we could not model the continuum with a power law, we perform local fits to the continuum around the regions of the \civ{} and \mgii{} lines to determine the continuum fluxes at $1400\,\rm{\AA}$, $1450\,\rm{\AA}$, and $3000\,\rm{\AA}$.

Based on the flux density at $1450\,\rm{\AA}$, we also calculate the apparent and absolute magnitudes, $m_{1450}$ and $M_{1450}$. In some cases the flux densities at $1350\,\rm{\AA}$ ($z\lesssim6.59$), $1400\,\rm{\AA}$ ($z\lesssim6.32$), and $1450\,\rm{\AA}$ ($z\lesssim6.08$) were measured by extrapolating the fit blueward, outside of the spectral range of the X-SHOOTER NIR arm.

For our main analysis we estimate the bolometric luminosity following \citet{Shen2011}:
\begin{equation}
    L_{\rm{bol}} = 5.15\cdot \lambda L_{\lambda, 3000}
\end{equation}


We further determine the integrated flux and the luminosity of the \feii{} pseudo-continuum in the wavelength range of $2200-3090\,\rm{\AA}$, to construct \feii{}/\mgii{} flux ratios as discussed in Section\,\ref{sec:results_feIImgIIratios}.


\subsubsection{Emission Lines}

For the \siiv{}, \civ{} and \mgii{} lines we calculate the peak wavelength from the maximum flux value of the full line model (all components). As most of the \civ{} line models consist of two Gaussian components, these are added before the peak of the line model is determined. The line redshift follows from the peak wavelength of the line model and the corresponding rest-frame wavelength of the line.
Velocity shifts of the lines are derived from their line redshift in comparison to the systemic \cii{} redshift using \textit{linetools} \citep{linetools2016} including relativistic corrections.
We also compute the FWHM, equivalent width (EW), integrated flux and integrated luminosity of all lines using the full line model, hence, taking into account all line components. 
In the re-sampling process catastrophic fits of multi-component lines can occur, where the component peaks are too separated to allow a successful determination of the FWHM. These cases are rare and not taken into account for the final FWHM measurements. All FWHM measurements are corrected for instrumental line broadening introduced by the resolution of the X-SHOOTER spectrograph.

We already discussed the complications in inferring line properties of the strongly blended lines in the \ciii{} complex. Therefore, we only extract the \ciii{} complex redshift from the peak flux of the full \ciii{} complex (sum of all three lines). 

\subsubsection{Black hole mass estimates}
We provide estimates of the black hole masses along with this paper. While these results will be discussed in detail in Farina et al. (in preparation), we include a discussion on how they were estimated in Appendix\,\ref{sec:app_bhmasses} for completeness.

\section{Systematic effects on \mgii{} measurements introduced by the choice of the iron template} \label{sec:irondiscussion}

The broad \mgii{} line lies in a spectral region where a plethora of \feii{} emission lines form a strong pseudo-continuum in many quasar spectra. Hence, it is important to take the contribution of \feii{} emission into account when modeling the \mgii{} line to derive unbiased properties. 
This can either be achieved by using scaled and broadened (semi-)empirical iron templates or calculating full model spectra using a spectral synthesis code. In this work we have chosen the former approach, adopting the iron templates of \citetalias{Vestergaard2001} and \citetalias{Tsuzuki2006}.
The empirical iron template of \citetalias{Vestergaard2001} has been widely used in the literature \citep[e.g.][]{DeRosa2011, Mazzucchelli2017}. It is derived from the spectrum of the narrow-line Seyfert 1 galaxy I Zwicky\,1 and covers the entire UV rest-frame range of the AGN. However, at the time the authors were not able to estimate the strength of the \feii{} pseudo-continuum beneath the \mgii{} line.
Therefore, they made the conscious decision to underestimate the iron continuum contribution and set it to zero beneath the \mgii{} line (see their Section 3.4.1 in \citetalias{Vestergaard2001}).
A few years later \citetalias{Tsuzuki2006} used the spectral synthesis code CLOUDY \citep{Ferland1998} to model the iron contribution beneath the \mgii{} line and created a semi-empirical iron template based on these synthetic iron spectra and the observed spectrum of I Zwicky\,1.
Differences in the iron flux contribution of various iron templates in the literature account for one of the major uncertainties in modeling the \mgii{} line as well as the iron flux itself \citep[e.g.][]{Dietrich2003c, Kurk2007, Woo2018, Shin2019}.


\begin{figure}
    \centering
    \includegraphics[width=0.48\textwidth]{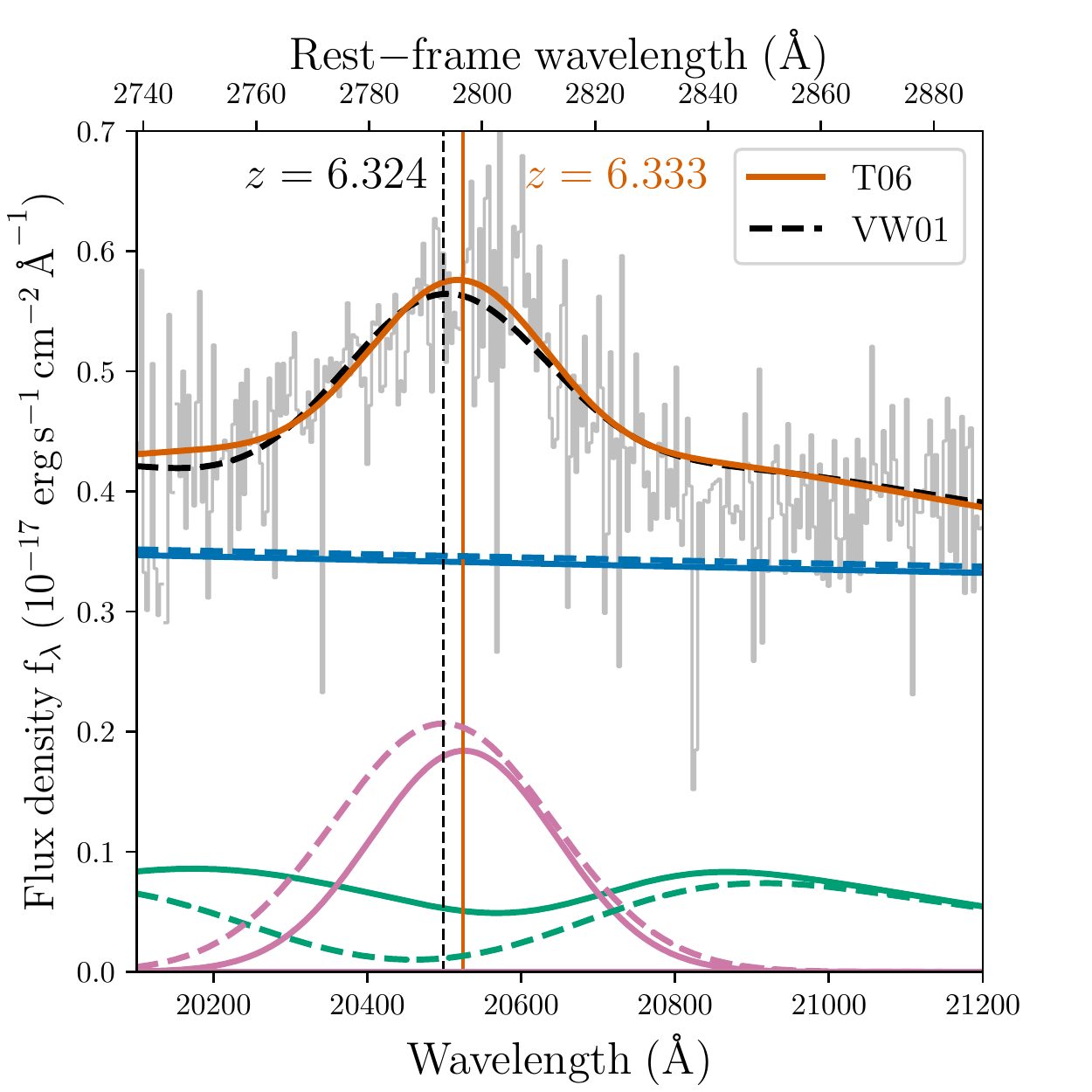}
    \caption{Comparison of the model fits to quasar ULAS~J1148+0702 using the two different iron template of \citetalias{Tsuzuki2006} and \citetalias{Vestergaard2001}. Solid lines denote the \citetalias{Tsuzuki2006} model fit, while dashed line show the \citetalias{Vestergaard2001} model fit. The model fit components are colored as in Figure\,\ref{fig:example_fit}. While the full fit (black/orange) lines are very similar, the iron template (green) and the \mgii{} line components (purple) differ considerably, resulting in different best-fit parameters for the width, amplitude and center of the line.}
    \label{fig:irontemplcomp}
\end{figure}

In fitting each of our spectra with both the \citetalias{Vestergaard2001} and the \citetalias{Tsuzuki2006} iron template, we assess these differences quantitatively to understand possible biases in our measurements.
In Figure\,\ref{fig:irontemplcomp} we show both fits around the \mgii{} line for ULAS~J1148+0702 at $z=6.339$. Solid lines refer to the fit with the \citetalias{Tsuzuki2006} template and dashed lines refer to the model fit using the \citetalias{Vestergaard2001} iron template. This example highlights how the full fit of both models (solid orange and grey dashed lines) is nearly identical. On the other hand the line fit component (purple lines) and the iron template component (green lines) are significantly different. 
All derived fit parameters of the \mgii{} line (FWHM, integrated line flux and central wavelength) as well as the integrated flux of the iron component are affected.

\tabletypesize{\footnotesize} 
\begin{deluxetable}{lrr}
\tablecaption{Comparison of the best-fit properties based on model fits with two different iron templates. We contrast the median values for a subsample of 28 quasars with secure [CII] redshifts.\label{table:fecomp}}
\tablehead{\colhead{Property} &\colhead{Median} &\colhead{Median} \\ 
\nocolhead{} &\colhead{T06 template} &\colhead{VW01 template} 
} 
\decimals 
\startdata 
$\log(L_{3000}/\rm{erg}\,\rm{s}^{-1})$ & 46.57 & 46.58 \\ 
$\Delta v(\rm{MgII}{-}\rm{[CII]})\//(\rm{km}\,\rm{s}^{-1})$ & -390.61 & -612.17 \\ 
$\rm{FWHM}_{\rm{Mg\,II}}\//(\rm{km}\,\rm{s}^{-1})$ & 2955.52 & 3785.64 \\ 
$\log(M_{\rm{BH,\ MgII}}/M_\odot)$ & 9.12 & 9.33 \\ 
$L_{\rm{bol}}(L_{3000})/L_{\rm{Edd}}$ & 1.23 & 0.75 \\ 
$F_{\rm{MgII}}\//(10^{-17}\ \rm{erg}\,\rm{s}^{-1}\,\rm{cm}^{-2}\,\rm{\AA}^{-1})$ & 59.10 & 78.84 \\ 
$F_{\rm{FeII}}\//(10^{-17}\ \rm{erg}\,\rm{s}^{-1}\,\rm{cm}^{-2}\,\rm{\AA}^{-1})$ & 363.25 & 319.06 \\ 
$F_{\rm{FeII}}/F_{\rm{MgII}}$ & 6.70 & 4.30 
\enddata 
\end{deluxetable}

\begin{figure*}
    \centering
    \includegraphics[width=\textwidth]{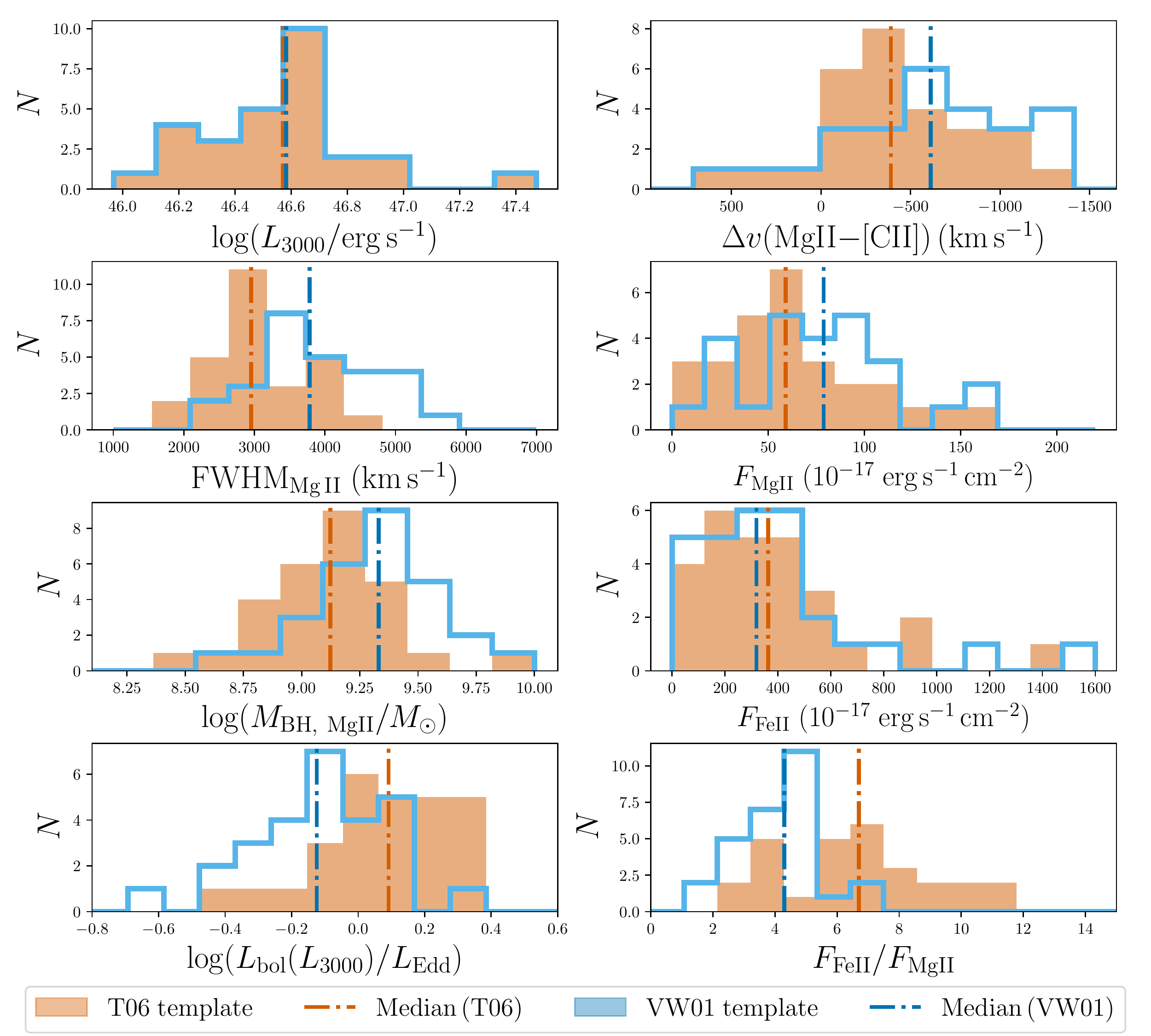}
    \caption{Histograms of the main \mgii{} and \feii{} properties highlighting the differences due to the use of either the \citetalias{Tsuzuki2006} or the \citetalias{Vestergaard2001} iron template in the model fit.
    The FWHM of \mgii{} and its integrated flux, $F_{\rm{MgII}}$, are most affected by the choice of the iron template. By extension, all dependent properties are affected as well.
    A total of 27 quasars with a successful fit of the \mgii{} line and secure \cii{} redshifts contribute to the histograms. The dashed-dotted lines show the median of the distributions. Results based on the \citetalias{Tsuzuki2006} template are colored orange, while results from the \citetalias{Vestergaard2001} template are colored blue. The  panels, from top left to bottom right, show the luminosity at 3000\,\AA ($L_{3000}$), the blueshift of \mgii{} with respect to the \cii{} line, the FWHM of \mgii{}, the \mgii{} integrated flux, the derived black hole mass using the relation of \citet{Vestergaard2009}, the integrated flux of the iron template between $2200{-}3090\,\rm{\AA}$, the Eddington luminosity ratio based on the shown Black Hole (BH) mass, and the flux ratio of the iron and \mgii{} fluxes.}
    \label{fig:feIIcomparison}
\end{figure*}

In Figure\,\ref{fig:feIIcomparison} we compare the model fit results for each template in a sub-sample of 28 quasars, which have both successful \mgii{} line fits and \cii{} redshifts. These 28 quasars are marked in Table\,\ref{table:cIVmgII} for reproducibility. Orange colors in Figure\,\ref{fig:feIIcomparison} refer to the \citetalias{Tsuzuki2006} iron template and blue colors to the \citetalias{Vestergaard2001} template. Dashed-dotted lines in the figure depict the median values of each property, which we also compare in Table\,\ref{table:fecomp}.

As suggested by the example fit in Figure\,\ref{fig:irontemplcomp}, measurements of the joint power law and Balmer continuum model (solid and dashed blue lines) remain largely unaffected by the choice of the iron template. The median values of $L_{3000}$ measured from the two different templates are nearly identical (see top left panel in Figure\,\ref{fig:feIIcomparison} and  Table\,\ref{table:fecomp}). 

Figure\,\ref{fig:feIIcomparison} highlights how all other properties show systematic differences.
We illustrate the influence of the templates on the \mgii{} redshift by analysing the \mgii{} velocity shift with respect to the systemic redshift of the \cii{} line, $\Delta v(\rm{MgII}{-}\rm{[CII]})$. 
The distributions for the velocity shifts appear similar (Figure\,\ref{fig:feIIcomparison}, top right) at first, but the difference in median velocity shift is non-negligible between the templates, $\sim200\,\rm{km}\,\rm{s}^{-1}$ (see Table\,\ref{table:fecomp}), considering the absolute values for the median velocity shifts are around $-400$ to $-600\,\rm{km}\,\rm{s}^{-1}$. 
In consequence, measurements of the \civ{}-\mgii{} velocity shift will be affected. However, the often large \civ{} blueshifts of $\sim1000\,\rm{km}\,\rm{s}^{-1}$ will render this bias less relevant.
The reason for these differences is the asymmetry in the iron contribution underlying the \mgii{} line in the \citetalias{Tsuzuki2006} template, whereas the missing iron emission in the \citetalias{Vestergaard2001} template is largely symmetric around the line.
The comparison of the iron templates (green lines) in Figure\,\ref{fig:irontemplcomp} highlights this difference. Hence, the choice of the iron template has an effect on the best-fit redshift of the \mgii{} line.

As previously discussed in \citet{Woo2018}, the measured FWHM of the \mgii{} is also affected by the iron template used.
The model fits of the \mgii{} (purple lines) in Figure\,\ref{fig:irontemplcomp} emphasize this. The choice of the \citetalias{Vestergaard2001} template results in a broader line fit.
The histogram of the FWHM$_{\rm{MgII}}$ in Figure\,\ref{fig:feIIcomparison} (second panel in the left column) makes this systematic shift towards broader FWHM$_{\rm{MgII}}$ evident. Compared to the \citetalias{Tsuzuki2006} template the median FWHM$_{\rm{MgII}}$ measured with the \citetalias{Vestergaard2001} template is broader by  $\sim800\,\rm{km}\,\rm{s}^{-1}$  (Table\,\ref{table:fecomp}). 
Consequently, the derived black hole masses and Eddington luminosity ratios are shifted as can be seen in the lower two panels in the left column of Figure\,\ref{fig:feIIcomparison}. 
We have derived the black hole mass estimates using the relation of \citet{Vestergaard2009}, which was established using the iron template of \citetalias{Vestergaard2001}. The scaling relation as well as the conversion to Eddington luminosity both use the continuum luminosity at 3000\,\AA, $L_{3000}$, which is largely unaffected by the choice of the iron template. 
Therefore, the shifts in distributions and medians of the black hole masses and Eddington luminosity ratios are a direct consequence of the difference in the measured FWHM$_{\rm{MgII}}$.
The larger FWHM$_{\rm{MgII}}$ values from the use of the \citetalias{Vestergaard2001} template result in larger black hole masses and lower Eddington luminosity ratios. It is worth noting that the use of the \citetalias{Vestergaard2001} template compared to the \citetalias{Tsuzuki2006} template moves the median of the Eddington luminosity ratio from a super-Eddington value to a sub-Eddington value for this sub-set of the X-SHOOTER/ALMA sample. 

\subsection{The effect on the \feii{}/\mgii{} ratio}

\begin{figure*}
    \centering
    \includegraphics[width=0.95\textwidth]{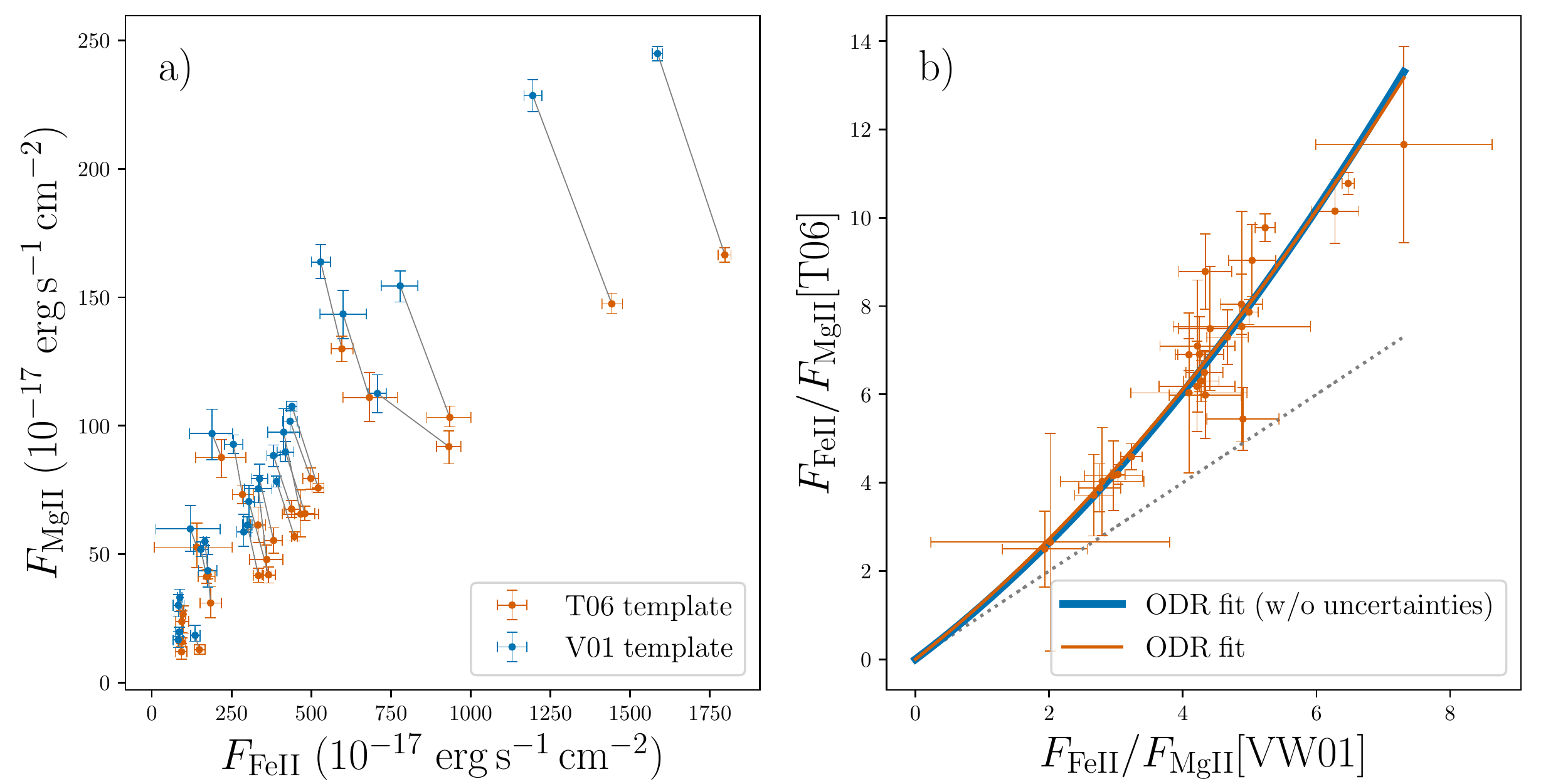}
    \caption{Influence of the iron template on the $F_{\rm{FeII}}/F_{\rm{Mg}II}$ ratios. \textbf{a)}: \mgii{} flux as a function of \feii{} flux. The change from the \citetalias{Tsuzuki2006} (orange) to the \citetalias{Vestergaard2001} (blue) template introduces a diagonal shift (grey line) of the data points toward lower \feii{} and higher \mgii{} flux. \textbf{b)}: A comparison between the flux ratios calculated with both templates. We show our best fit using orthogonal distance regression as the orange solid line. The solid blue line shows the same fit excluding the measurement errors. The $F_{\rm{FeII}}/F_{\rm{Mg}II}$ ratio data based on the \citetalias{Tsuzuki2006} template clearly lie above the 1:1 relation (grey dotted line). 
    }
    \label{fig:feIImgIIcomparison}
\end{figure*}

The choice of the iron template has its most profound effect on the \feii{}/\mgii{} flux ratio, a proxy for the Fe/Mg abundance and therefore for the iron enrichment of BLR gas in high redshift quasars \citep[see e.g.][]{Dietrich2003c}. 
Figure\,\ref{fig:irontemplcomp} shows that the model fits using the two templates result in a nearly equivalent fit of the \mgii{} region. 
The difference between the fits is the flux contribution of the line model and the \feii{} pseudo-continuum to the total fit. 
Using the \citetalias{Vestergaard2001} template results in significantly larger \mgii{} and smaller \feii{} fluxes as can be seen in Figure\,\ref{fig:feIIcomparison} (middle panels in the right column) or in panel a) of Figure\,\ref{fig:feIImgIIcomparison}. The trends have also been discussed in the literature \citep{Dietrich2003c, Shin2019}.
As a consequence, the resulting \feii{}/\mgii{} flux ratios are much smaller compared with the \citetalias{Tsuzuki2006} template (Figure\,\ref{fig:feIIcomparison}, bottom right panel).
However, the effect does not simply shift the $F_{\rm{Mg}II}$ towards larger values when changing from the \citetalias{Tsuzuki2006} to the \citetalias{Vestergaard2001} template. Panel a) in Figure\,\ref{fig:feIImgIIcomparison} clearly shows that quasars with stronger \feii{} emission are effected more significantly. 
As a result the distribution of $F_{\rm{Mg}II}$ in Figure\,\ref{fig:feIIcomparison} (second panel in the right column) also changes its shape and the median value increases strongly from $\sim59$ to $\sim79\times10^{-17}\,\rm{erg}\,\rm{s}^{-1}\,\rm{cm}^{-2}$.
Most model fits to the spectra will result in an equally good fit for both templates. Therefore, a larger $F_{\rm{Mg}II}$ has to result in a reduced iron continuum flux, $F_{\rm{FeII}}$. This is indeed seen in both panel a) of Figure\,\ref{fig:feIImgIIcomparison} and Figure\,\ref{fig:feIIcomparison} (third panel in the right column). 
The effect on the $F_{\rm{FeII}}$ does not appear significant at first. Integrated over rest-frame wavelengths of $2200$ to $3090\,\rm{\AA}$ the iron median flux is much larger than the one of the \mgii{} line. 
However, both effects conspire to result in a severe systematic effect on the \feii{}/\mgii{} flux ratio, $F_{\rm{FeII}}/F_{\rm{Mg}II}$.
Using the \citetalias{Vestergaard2001} template $F_{\rm{FeII}}/F_{\rm{Mg}II}$ reaches a median value of $4.30$, which increases by a factor of $\sim1.5$ to $F_{\rm{FeII}}/F_{\rm{Mg}II}= 6.70$, when the \citetalias{Tsuzuki2006} template is assumed. 
Panel b) in Figure\,\ref{fig:feIImgIIcomparison} shows how $F_{\rm{FeII}}/F_{\rm{Mg}II}$ changes between the two different templates. 
In an attempt to characterize the relationship between the \feii{}/\mgii{} flux ratios resulting from the two different templates, we have fit the data in Figure\,\ref{fig:feIImgIIcomparison} b) with orthogonal distance regression\footnote{We have used the ODR package in the \texttt{scipy}\,\citep{scipy} python library.} modeled by a second order polynomial model without the constant term:
\begin{equation}
    \left(\frac{F_{\rm{FeII}}}{F_{\rm{Mg}II}}\right)_{\rm{T06}} = a \ \left(\frac{F_{\rm{FeII}}}{F_{\rm{Mg}II}}\right)_{\rm{VW01}}^2 + b \ \left(\frac{F_{\rm{FeII}}}{F_{\rm{Mg}II}}\right)_{\rm{VW01}} \label{eq:feIIscaling}
\end{equation}

The model assumes that both flux ratios are equal at the origin. The blue lines in panel b) of Figure\,\ref{fig:feIImgIIcomparison} show our fit results with (orange line, $a=0.083\pm0.021$; $b=1.198\pm0.111$) and without (blue line, $a=0.094\pm0.032$; $b=1.136\pm0.152$) including the uncertainties on the flux ratios.
The non-zero second-order component in both model fits shows that the scaling between the flux ratio values from one template to the other is distinctly non-linear.
The majority of quasars in our sample have flux ratios between 4 to 5 based on the \citetalias{Vestergaard2001} template, resulting in scale factors of 1.53 to 1.61, in good agreement with the median scaling of the flux ratios determined earlier ($\sim1.56$).
These model fits allow us to compare literature values of  $F_{\rm{FeII}}/F_{\rm{Mg}II}$ based on the \citetalias{Vestergaard2001} template to our new values derived using the \citetalias{Tsuzuki2006} template in Section\,\ref{sec:results_feIImgIIratios}.

\subsection{On the future use of different iron templates}
Because the \citetalias{Tsuzuki2006} iron template includes the \feii{} continuum contribution beneath the \mgii{} line we adopt it for our line analysis. 
Future studies focused on the \mgii{} line properties (FWHM, redshift, line flux), the \feii{} continuum and the $F_{\rm{FeII}}/F_{\rm{Mg}II}$ should consider using the \citetalias{Tsuzuki2006} iron template or any equivalent iron template, which includes the \feii{} continuum beneath the \mgii{} line. 

However, one has to be very careful, when it comes to the derivation of BH mass estimates and, subsequently, Eddington luminosity ratios based on the \mgii{} line. The majority of single epoch virial estimators \citep[e.g.][both applied in this work]{Vestergaard2009, McLure2004} use the \citetalias{Vestergaard2001} iron template, when constructing the scaling relations of \mgii{} from the H$\beta$ line. 
Therefore, estimates of black hole masses derived from the \mgii{} line need to be based on the same iron template, with which the scaling relation was originally established. Otherwise, one will risk systematic biases in the BH masses and Eddington luminosity ratios as shown in Figure\,\ref{fig:feIIcomparison}.
For example, the use of the \citetalias{Tsuzuki2006} template in combination with the \citet{Vestergaard2009} scaling relation erroneously gave the impression that our sample has a large fraction of quasars showing super-Eddington accretion.
At last, we should remind ourselves that both the \citetalias{Tsuzuki2006} and \citetalias{Vestergaard2001} templates are based on a single, low-redshift, low-luminosity Seyfert galaxy and thus may have limited applicability for the high-redshift quasar population.

\section{Results}\label{sec:results}

\subsection{Iron enrichment traced by high-redshift quasars}\label{sec:results_feIImgIIratios}
One possible way to trace the build up of metals in the galaxy's interstellar material (ISM) is by measuring the abundance ratio of iron to $\alpha$-process elements.
While $\alpha$-process elements are predominantly produced in core-collapse Type\,II supernovae (SNe\,II), which have massive star progenitors, Fe is released into the ISM mainly from Type\,Ia supernovae, which follow the evolution of intermediate, binary stars.
The difference in evolutionary lifetimes leads to a delay between the enrichment of iron compared to $\alpha$-process elements, which has been estimated to be around 1\,Gyr \citep[e.g.][]{Matteucci1986}. However, it has also been shown that this delay can be as short as $\sim0.2-0.6\,\rm{Gyr}$ \citep{Matteucci1994, Friaca1998, Matteucci2001} in the case of elliptical galaxies. 

In high redshift quasars we can measure the \feii{}/\mgii{} flux ratio, which has been widely used in the literature as a proxy for the Fe/Mg abundance ratio to estimate iron enrichment in the quasar's BLR \citep[e.g.][and references therein]{Dietrich2002a, Iwamuro2002,  Barth2003, Dietrich2003c, Freudling2003, Maiolino2003, Iwamuro2004, Kurk2007, DeRosa2011, DeRosa2014, Mazzucchelli2017, Sameshima2017, Shin2019}.

We construct the \feii{}/\mgii{} flux ratio from the total integrated flux of the \mgii{} line model and the flux of the iron template integrated over the wavelength range of $2200{-}3090$\,\AA. 
The wavelength range has been chosen to be comparable to the literature on this topic, as the choice impacts the \feii{}/\mgii{} flux ratio measurement.
We provide the measured \feii{} and \mgii{} fluxes as well as the \feii{}/\mgii{} flux ratio measured with both the \citetalias{Vestergaard2001} and \citetalias{Tsuzuki2006} template in Table\,\ref{table:feIImgII}. We were able to successfully fit the \mgii{} line and the iron pseudo-continuum in 32 quasars from our sample and calculate $F_{\rm{FeII}}/F_{\rm{MgII}}$ for these objects.

{\movetabledown=2.0in 
\tabletypesize{\footnotesize} 
\begin{deluxetable*}{lcccccc}
\tablecaption{\ion{Fe}{2}/\ion{Mg}{2} flux ratios\label{table:feIImgII}}
\tablecolumns{7}\tablehead{\colhead{Quasar Name} &\colhead{$F_{\rm{FeII}}^{a}$ (T06)} &\colhead{$F_{\rm{MgII}}^{b}$ (T06)} &\colhead{$F_{\rm{FeII}}/F_{\rm{MgII}}^{c}$} &\colhead{$F_{\rm{FeII}}^{a}$ (VW01)} &\colhead{$F_{\rm{MgII}}^{b}$ (VW01)} &\colhead{$F_{\rm{FeII}}/F_{\rm{MgII}}^{c}$} \\ 
\nocolhead{} &\multicolumn{2}{c}{($10^{-17}\,\rm{erg}\,\rm{s}^{-1}\rm{cm}^2$)} &\colhead{(T06)} &\multicolumn{2}{c}{($10^{-17}\,\rm{erg}\,\rm{s}^{-1}\rm{cm}^2$)} &\colhead{(VW01)} \\ } 
\startdata 
PSO~J007.0273+04.9571 & ${141.22}_{-133.59}^{+110.93}$ & ${52.72}_{-7.95}^{+9.31}$ & ${2.66}_{-2.52}^{+2.42}$ & ${120.79}_{-108.11}^{+93.49}$ & ${59.84}_{-8.69}^{+9.04}$ & ${2.02}_{-1.79}^{+1.47}$\\ 
PSO~J011.3898+09.0324 & ${148.96}_{-15.93}^{+18.13}$ & ${12.81}_{-1.84}^{+1.98}$ & ${11.66}_{-2.00}^{+2.45}$ & ${135.87}_{-13.89}^{+15.53}$ & ${18.38}_{-2.32}^{+3.88}$ & ${7.31}_{-1.32}^{+1.29}$\\ 
VIK~J0046--2837 & ${125.01}_{-84.46}^{+89.09}$ & ${30.86}_{-2.55}^{+2.81}$ & ${3.99}_{-2.72}^{+2.90}$ & ${209.38}_{-69.56}^{+64.95}$ & ${40.03}_{-4.24}^{+4.05}$ & ${5.24}_{-1.44}^{+1.27}$\\ 
SDSS~J0100+2802 & ${1797.17}_{-20.89}^{+19.07}$ & ${166.50}_{-2.81}^{+2.72}$ & ${10.78}_{-0.23}^{+0.26}$ & ${1586.25}_{-16.68}^{+15.67}$ & ${244.88}_{-2.77}^{+2.85}$ & ${6.47}_{-0.09}^{+0.09}$\\ 
VIK~J0109--3047 & ${93.10}_{-19.37}^{+17.06}$ & ${11.98}_{-2.91}^{+3.28}$ & ${7.54}_{-1.86}^{+3.36}$ & ${82.92}_{-16.44}^{+15.52}$ & ${16.62}_{-3.01}^{+3.36}$ & ${4.88}_{-1.03}^{+1.45}$\\ 
PSO~J036.5078+03.0498 & ${334.37}_{-16.46}^{+16.31}$ & ${41.67}_{-2.63}^{+2.80}$ & ${8.04}_{-0.69}^{+0.67}$ & ${299.35}_{-14.27}^{+14.90}$ & ${61.32}_{-3.04}^{+3.22}$ & ${4.88}_{-0.32}^{+0.34}$\\ 
VIK~J0305--3150 & ${94.31}_{-12.82}^{+12.11}$ & ${15.72}_{-1.60}^{+1.65}$ & ${5.98}_{-0.94}^{+1.02}$ & ${86.08}_{-11.03}^{+10.50}$ & ${19.67}_{-1.65}^{+1.90}$ & ${4.34}_{-0.54}^{+0.60}$\\ 
PSO~J056.7168--16.4769 & ${447.14}_{-9.30}^{+10.20}$ & ${56.85}_{-1.73}^{+1.79}$ & ${7.87}_{-0.27}^{+0.29}$ & ${390.89}_{-8.54}^{+8.10}$ & ${78.33}_{-2.18}^{+2.16}$ & ${4.99}_{-0.14}^{+0.14}$\\ 
PSO~J065.4085--26.9543 & ${931.77}_{-38.77}^{+37.55}$ & ${91.85}_{-6.67}^{+6.13}$ & ${10.15}_{-0.65}^{+0.80}$ & ${707.43}_{-29.35}^{+27.56}$ & ${112.56}_{-7.53}^{+7.27}$ & ${6.28}_{-0.36}^{+0.42}$\\ 
PSO~J065.5041--19.4579 & ${1442.50}_{-30.98}^{+33.71}$ & ${147.45}_{-3.67}^{+4.12}$ & ${9.78}_{-0.32}^{+0.30}$ & ${1194.71}_{-27.17}^{+28.58}$ & ${228.54}_{-6.17}^{+6.30}$ & ${5.23}_{-0.15}^{+0.13}$\\ 
SDSS~J0842+1218 & ${480.44}_{-32.54}^{+30.44}$ & ${65.79}_{-2.74}^{+2.84}$ & ${7.30}_{-0.61}^{+0.62}$ & ${418.96}_{-26.14}^{+26.40}$ & ${89.75}_{-3.66}^{+4.10}$ & ${4.67}_{-0.31}^{+0.33}$\\ 
SDSS~J1030+0524 & ${411.95}_{-28.55}^{+30.45}$ & ${80.42}_{-5.40}^{+4.91}$ & ${5.14}_{-0.48}^{+0.54}$ & ${360.68}_{-26.43}^{+28.46}$ & ${104.08}_{-5.28}^{+5.45}$ & ${3.47}_{-0.27}^{+0.28}$\\ 
PSO~J158.69378--14.42107 & ${218.33}_{-81.29}^{+76.63}$ & ${87.61}_{-7.83}^{+6.98}$ & ${2.50}_{-0.87}^{+0.85}$ & ${188.87}_{-70.58}^{+64.33}$ & ${96.95}_{-10.15}^{+9.45}$ & ${1.94}_{-0.63}^{+0.58}$\\ 
PSO~J159.2257--02.5438 & ${498.71}_{-25.28}^{+23.75}$ & ${79.47}_{-4.11}^{+4.10}$ & ${6.30}_{-0.48}^{+0.45}$ & ${433.67}_{-22.64}^{+22.10}$ & ${101.75}_{-4.51}^{+4.35}$ & ${4.28}_{-0.26}^{+0.26}$\\ 
VIK~J1048--0109 & ${184.33}_{-33.24}^{+33.85}$ & ${30.98}_{-5.84}^{+6.35}$ & ${6.03}_{-1.59}^{+2.04}$ & ${175.36}_{-29.08}^{+28.78}$ & ${43.57}_{-6.32}^{+6.27}$ & ${4.09}_{-0.87}^{+0.89}$\\ 
ULAS~J1148+0702 & ${353.20}_{-15.70}^{+14.36}$ & ${55.65}_{-2.28}^{+2.41}$ & ${6.33}_{-0.37}^{+0.42}$ & ${312.70}_{-13.61}^{+12.85}$ & ${73.72}_{-2.80}^{+2.63}$ & ${4.24}_{-0.20}^{+0.20}$\\ 
PSO~J183.1124+05.0926 & ${381.98}_{-29.91}^{+27.10}$ & ${55.32}_{-5.03}^{+4.97}$ & ${6.91}_{-0.82}^{+0.88}$ & ${337.82}_{-26.34}^{+25.33}$ & ${79.35}_{-4.91}^{+5.75}$ & ${4.25}_{-0.36}^{+0.40}$\\ 
SDSS~J1306+0356 & ${521.90}_{-19.72}^{+18.09}$ & ${75.78}_{-1.78}^{+1.79}$ & ${6.90}_{-0.39}^{+0.34}$ & ${439.55}_{-17.08}^{+15.88}$ & ${107.51}_{-1.86}^{+1.81}$ & ${4.09}_{-0.17}^{+0.15}$\\ 
ULAS~J1319+0950 & ${176.47}_{-5.96}^{+5.98}$ & ${42.03}_{-1.72}^{+1.74}$ & ${4.19}_{-0.20}^{+0.24}$ & ${166.55}_{-5.07}^{+5.17}$ & ${54.95}_{-1.64}^{+1.60}$ & ${3.03}_{-0.11}^{+0.12}$\\ 
CFHQS~J1509--1749 & ${284.22}_{-31.79}^{+33.95}$ & ${73.20}_{-3.59}^{+3.64}$ & ${3.88}_{-0.53}^{+0.57}$ & ${255.85}_{-28.00}^{+29.54}$ & ${92.76}_{-3.53}^{+3.67}$ & ${2.76}_{-0.31}^{+0.32}$\\ 
PSO~J231.6576--20.8335 & ${467.06}_{-57.82}^{+56.08}$ & ${65.58}_{-8.81}^{+9.42}$ & ${7.09}_{-1.30}^{+1.70}$ & ${413.62}_{-50.14}^{+49.54}$ & ${97.49}_{-8.81}^{+9.15}$ & ${4.22}_{-0.56}^{+0.68}$\\ 
PSO~J239.7124--07.4026 & ${934.55}_{-72.35}^{+66.20}$ & ${103.26}_{-3.66}^{+4.39}$ & ${9.04}_{-0.82}^{+0.81}$ & ${779.01}_{-59.78}^{+55.14}$ & ${154.45}_{-6.25}^{+5.83}$ & ${5.04}_{-0.35}^{+0.35}$\\ 
PSO~J308.0416--21.2339 & ${366.10}_{-19.99}^{+21.38}$ & ${41.90}_{-3.16}^{+3.15}$ & ${8.78}_{-0.83}^{+0.87}$ & ${304.29}_{-16.52}^{+17.68}$ & ${70.48}_{-10.00}^{+6.20}$ & ${4.34}_{-0.40}^{+0.69}$\\ 
SDSS~J2054--0005 & ${360.39}_{-53.72}^{+50.37}$ & ${47.92}_{-4.54}^{+5.67}$ & ${7.49}_{-1.38}^{+1.42}$ & ${333.83}_{-43.41}^{+42.82}$ & ${75.44}_{-5.39}^{+5.21}$ & ${4.41}_{-0.48}^{+0.57}$\\ 
CFHQS~J2100--1715 & ${97.96}_{-8.50}^{+7.84}$ & ${26.61}_{-7.41}^{+3.30}$ & ${3.72}_{-0.52}^{+1.33}$ & ${87.93}_{-7.16}^{+6.81}$ & ${33.12}_{-5.43}^{+3.23}$ & ${2.67}_{-0.29}^{+0.49}$\\ 
PSO~J323.1382+12.2986 & ${438.14}_{-23.21}^{+25.08}$ & ${67.54}_{-3.20}^{+3.33}$ & ${6.50}_{-0.48}^{+0.52}$ & ${381.26}_{-20.68}^{+21.21}$ & ${88.38}_{-4.33}^{+4.13}$ & ${4.32}_{-0.28}^{+0.28}$\\ 
VIK~J2211--3206 & ${596.01}_{-33.29}^{+34.79}$ & ${129.94}_{-4.98}^{+4.94}$ & ${4.59}_{-0.28}^{+0.31}$ & ${529.60}_{-29.27}^{+31.00}$ & ${163.75}_{-6.41}^{+6.72}$ & ${3.23}_{-0.16}^{+0.17}$\\ 
PSO~J340.2041--18.6621 & ${222.90}_{-12.76}^{+12.69}$ & ${51.84}_{-2.60}^{+2.60}$ & ${4.31}_{-0.30}^{+0.29}$ & ${198.87}_{-10.83}^{+10.59}$ & ${63.35}_{-2.94}^{+2.60}$ & ${3.15}_{-0.17}^{+0.18}$\\ 
SDSS~J2310+1855 & ${682.49}_{-83.34}^{+87.51}$ & ${111.01}_{-9.37}^{+9.62}$ & ${6.18}_{-0.97}^{+1.07}$ & ${599.77}_{-72.30}^{+73.34}$ & ${143.50}_{-9.60}^{+9.27}$ & ${4.21}_{-0.57}^{+0.57}$\\ 
VIK~J2318--3029 & ${171.74}_{-26.14}^{+26.26}$ & ${41.24}_{-2.53}^{+2.91}$ & ${4.16}_{-0.74}^{+0.84}$ & ${153.64}_{-22.37}^{+23.43}$ & ${51.86}_{-2.82}^{+2.98}$ & ${2.96}_{-0.44}^{+0.45}$\\ 
VIK~J2348--3054 & ${94.12}_{-18.94}^{+21.18}$ & ${23.74}_{-4.30}^{+4.21}$ & ${4.03}_{-1.09}^{+1.35}$ & ${83.27}_{-16.42}^{+17.84}$ & ${30.08}_{-4.47}^{+4.19}$ & ${2.79}_{-0.62}^{+0.72}$\\ 
PSO~J359.1352--06.3831 & ${333.11}_{-24.54}^{+23.68}$ & ${61.35}_{-6.90}^{+6.96}$ & ${5.44}_{-0.65}^{+0.76}$ & ${288.23}_{-22.01}^{+21.92}$ & ${58.67}_{-5.63}^{+6.81}$ & ${4.90}_{-0.54}^{+0.55}$\\ 
\enddata 
\tablecomments{$^a$ The \ion{Fe}{2} flux is calculated from the iron pseudo-continuum integrated over the wavelength range of $2200-3090\,\rm{\AA}$ \\ $^b$ The \ion{Mg}{2} flux has been integrated over the complete \ion{Mg}{2} line model. \\ $^c$ The \ion{Fe}{2}/\ion{Mg}{2} flux ratio is calculated during the re-sampling process. Therefore, the median value of the flux ratio might deviate from the ratios of the median flux values.}
\end{deluxetable*} 
}

\begin{figure*}
    \centering
    \includegraphics[width=0.95\textwidth]{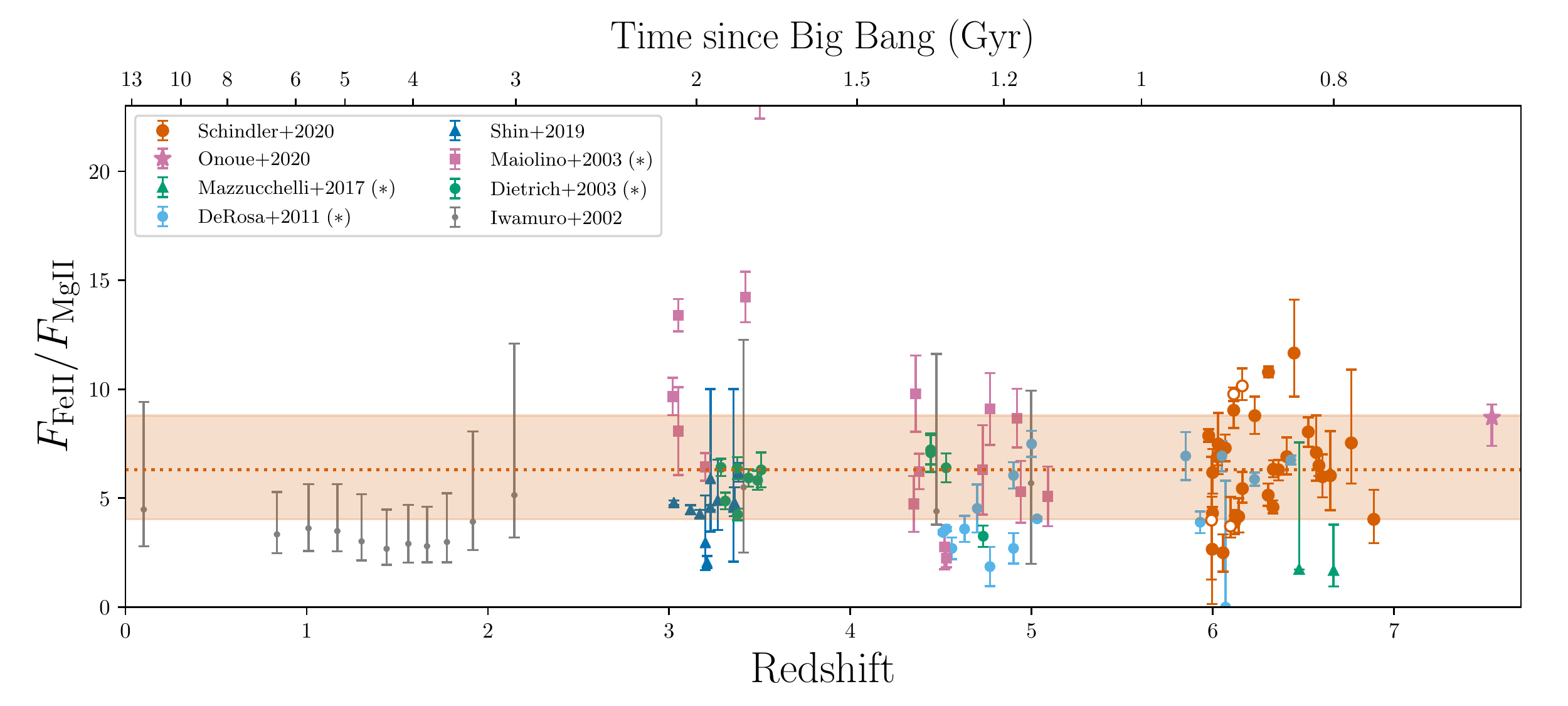}
    \caption{The \feii{}/\mgii{} flux ratio as a function of redshift. The $z>2$ data do not show any significant evolutionary trend with redshift. Our measurements using the \citetalias{Tsuzuki2006} iron template are shown as solid and open orange circles. The open circles refer to spectral fits, in which the continuum was only approximated locally around the \mgii{}, including the iron contribution. We further display the median value of our sample with the dashed orange line and the 16 to 84 percentile region in light orange. Different colored data points show literature values from previous studies \citep{Dietrich2003c, Maiolino2003, DeRosa2011, Mazzucchelli2017, Shin2019, Onoue2020}, which are either using the \citetalias{Tsuzuki2006} template as well or are scaled appropriately (*) from the \citetalias{Vestergaard2001} template using Equation\,\ref{eq:feIIscaling}. At lower redshift we display the median values from the study of \citet{Iwamuro2002}, which are based on their own iron template. We discuss the comparability of the different measurements in Section\,\ref{sec:results_feIImgIIratios} in more detail.}
    \label{fig:feII_ratios}
\end{figure*}
We now turn to compare our results with other measurements in the literature to cover a wide redshift range $0\leq z \leq 7.5$ as shown in Figure\,\ref{fig:feII_ratios}. 
A comparison between \feii{}/\mgii{} flux ratios across many studies is often complicated by differences in their measurements \citep[see e.g.][for discussions]{Kurk2007, DeRosa2011, Shin2019, Onoue2020}. 
Fitting methodology (algorithms, assumed iron template, iron integration wavelength range) varies from study to study, resulting in differences in the measured \feii{}/\mgii{} flux ratio (Section\,\ref{sec:irondiscussion}). 
A good discussion on the impact of the assumed Balmer continuum strength is given in \citet{Onoue2020}. The authors find that reducing the strength of the Balmer continuum model only slightly lowers the measured \feii{}/\mgii{} flux ratio.
We first select studies that provide a \feii{}/\mgii{} flux ratio measurement with the \citetalias{Tsuzuki2006} iron template and the same \feii{} flux integration range used in our work \citep{Shin2019, Onoue2020}. Then we add measurements, for which the \citetalias{Vestergaard2001} template was used in the same integration range \citep{Dietrich2003c, Maiolino2003, Mazzucchelli2017}. To compare this data with our \feii{}/\mgii{} flux ratios, we scale the mean literature values using Equation\,\ref{eq:feIIscaling}.
We further use the results of \citet{DeRosa2011}, of a sample of quasars at $z\approx4.5-5$ and $z\approx5.8-6.5$. The authors use the same rest-frame wavelength range to integrate the \feii{} flux and the \citetalias{Vestergaard2001} template, but add a constant flux density at 2770-2820\,\AA\  equal to 20\% of the mean flux density of the template between 2930-2970\,\AA \citep[see also][]{Kurk2007}. This modification to the \citetalias{Vestergaard2001} template was motivated by the missing Fe flux beneath the \mgii{} line. As discussed in \citet{Shin2019} this modification only slightly increases the average \feii{}/\mgii{} flux ratio by $\sim6\%$. We therefore reduce the \feii{}/\mgii{} accordingly and then apply Equation\,\ref{eq:feIIscaling} to the \citet{DeRosa2011} results.
To populate the redshift range below $z=2$, we add the median values of \citet{Iwamuro2002} to our comparison. However, this comparison is not ideal as the authors extracted their own iron template from the Large Bright Quasar Survey composite spectrum \citep{Francis1991} and integrated the fitted template over a rest-frame wavelength of $2150{-}3300\rm{\AA}$ to calculate the \feii{} flux. 

Figure\,\ref{fig:feII_ratios} shows our results as open and filled orange circles. The open circles refer to quasars, which could not be modeled with a continuous power-law model from \civ{} to \mgii{} (see Section\,\ref{sec:sample_properties}). The uncertainties on the flux ratio measurement reflect the signal-to-noise ratio of the binned spectra.
Different colored symbols show previous results from the literature \citep[][ as discussed above]{Dietrich2003c, Maiolino2003, DeRosa2011, Mazzucchelli2017, Shin2019, Onoue2020}.
Grey data points are the median values from the study of \citet{Iwamuro2002}.

Our results presented in context with the literature data in Figure\,\ref{fig:feII_ratios} do not show any discernible evolutionary trend with redshift. Keeping in mind that the exact measurement (iron template, wavelength integration range, etc) differs from study to study, our result echoes the findings of many previous studies \citep[e.g.][]{Barth2003, Dietrich2003c, Freudling2003, Maiolino2003, Kurk2007, DeRosa2011, DeRosa2014, Mazzucchelli2017, Shin2019, Onoue2020}.
We measure a median value of $F_{\rm{FeII}}/F_{\rm{MgII}} = 6.31_{-2.29}^{+2.49}$ for our sample of 32 quasars. The errors denote the 16 to 84 percentile range on the median measurement. It should be noted that this value is different from the value in Table\,\ref{table:fecomp} ($F_{\rm{FeII}}/F_{\rm{MgII}} = 6.70$) as we now include four more quasars, whose systemic redshifts were determined from the Lyman-$\alpha$ halo or using the \mgii{} line. 
Excluding the four quasars, whose continuum significantly deviates from a power law (open orange circles in Figure\,\ref{fig:feII_ratios}), we calculate a median value of $F_{\rm{FeII}}/F_{\rm{MgII}} = 6.31_{-2.15}^{+1.68}$. The median value is the same as before, but the 16 to 84 percentile range narrows significantly. 

Our \feii{}/\mgii{} flux ratios are similar to values from lower redshift samples even if lower luminosity quasars \citep[][with $L_{\rm{bol}}\approx46.5$]{Shin2019} are considered. 
The only exception are the results of \citet{Iwamuro2002} at $z\approx1-2$. In this redshift range the authors find median values up to $F_{\rm{FeII}}/F_{\rm{MgII}}\sim5$. However, their use of a different iron template might be the cause of the discrepancy in the measurements. 
At $z=7.54$ we have included the FeII/MgII flux ratio of ULAS~1342+0928 from \citet{Onoue2020} measured from a deep GNIRS spectrum. While this quasar is also part of our sample, the \mgii{} line falls into the reddest order, which is dominated by the noise. Therefore, we were not able to constrain the \mgii{} properties. The \feii{}/\mgii{} flux ratio of ULAS~1342+0928 is relatively high compared to our median, but well within the 16-84 percentile range.

\subsubsection{Discussion}

We have so far assumed that the \feii{}/\mgii{} flux ratio in quasars is a good tracer of the Fe/Mg abundance ratio. Therefore, approaching higher and higher redshift, we would expect the \feii{}/\mgii{} flux ratio to first peak and decline significantly following the predictions of Fe/$\alpha$ enrichment \citep[e.g.][their Figure\,17]{Sameshima2017}. However, our results along with the data from the literature (see Figure\,\ref{fig:feII_ratios}) do not show a significant evolution of $F_{\rm{FeII}}/F_{\rm{MgII}}$ at the highest redshifts. As no decrease is evident with redshift, this would indicate, at face value, that the central part of the host galaxy is already sufficiently enriched in iron in all luminous quasars at $z\sim7$, $\sim750\,\rm{Myr}$ after the Big Bang and even in the most distant quasar ULAS~1342+0928 \citep{Onoue2020}, another $\sim70\,\rm{Myr}$ before. Given the delay time of $\sim0.2-0.6\,\rm{Gyr}$ \citep{Matteucci1994, Friaca1998, Matteucci2001}, this would indicate that the first episode of star formation in these quasar hosts would have had to have occurred at $z\gtrsim9$.

However, this result only holds if the physical conditions for the excitation of \feii{} and \mgii{} are the same (or at least similar) in all quasars at all redshifts and if the \feii{}/\mgii{} flux ratio actually traces the Fe/Mg abundance. 
Photo-ionization calculations \citep{Verner2003, Baldwin2004} suggest that the \feii{}/\mgii{} flux ratio does depend on physical parameters of the broad line region, like gas density, micro-turbulence and the properties of the radiation field.
Photo-ionization models of \citet{Sameshima2017} further indicate that the \mgii{} line strength is dependent on the density of the broad-line region gas. In their sample of $\sim17000$ quasars at $z=0.72-1.63$ they identify an observational anti-correlation between the \feii{}/\mgii{} flux ratio and the Eddington luminosity ratio. The authors suspect the accretion rate and the gas density to be inter-dependent, which in turn leads to the anti-correlation with the Eddington luminosity ratio. 

We evaluated the Pearson correlation coefficient and found both properties to be uncorrelated with $\rho=0.02$ and $p=0.93$ in our sample. Yet, we should keep in mind that our quasars only sample a small range of Eddington luminosity ratios.
%
As studies continue to identify non-abundance dependencies of the \feii{}/\mgii{} flux ratio on the physical conditions of the broad line region, there is no doubt that we need to be careful when interpreting it in the context of iron enrichment.
However, ALMA observations of high-redshift quasar host galaxies have detected large amounts of dust \citep[e.g.][]{Venemans2017c}, also suggesting that their ISM is already sufficiently enriched in metals. Future work combining the \feii{}/\mgii{} flux ratio with the ALMA data will shed new light on the chemical enrichment of the highest redshift quasars.



\subsection{Velocity shifts of the broad emission lines}\label{sec:results_BELS}

We focus on the \civ{} and \mgii{} lines, which are available in the majority of the near-infrared spectra. For these lines we measure the peak redshift, the FWHM and the rest-frame equivalent width (EW). These results are summarized in Table\,\ref{table:cIVmgII}. 
In a few spectra we were also able to fit the \ciii{} complex as well as the \siiv{} line. These results are available in Table\,\ref{table:ciiisiiv} (see Appendix\,\ref{sec:app_additional_tables}). For the \siiv{} line we provide the peak redshift, the FWHM and the EW. However, due to the blended nature of the \siiii{} and \ciii{} lines we only report the peak redshift of the entire \ciii{} complex.

\begin{figure*}
    \centering
    \includegraphics[width=0.9\textwidth]{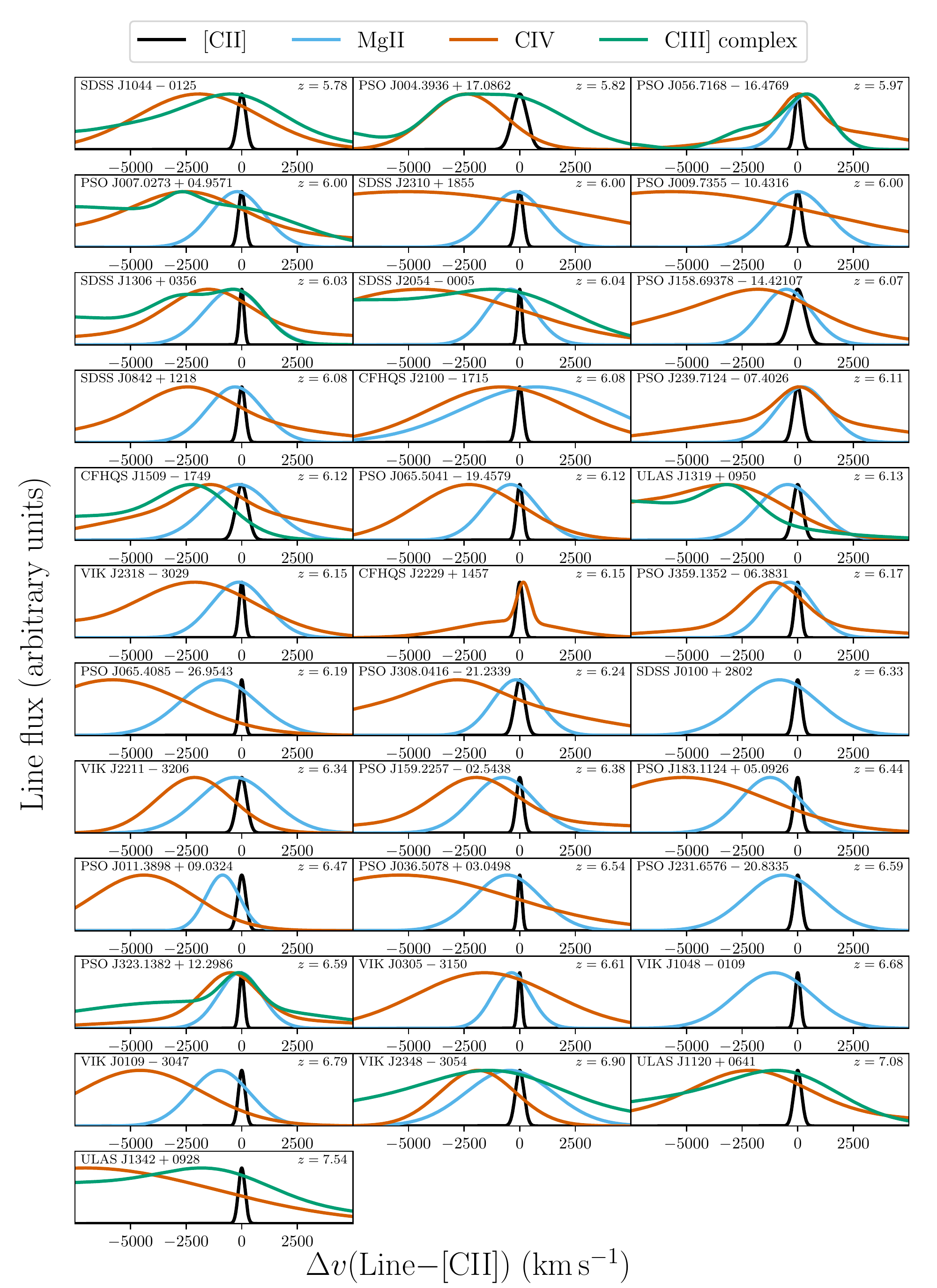}
    \caption{Line models fit to the CIII~], \civ{} and \mgii{} broad emission lines in comparison to the Gaussian line fits of \cii{} line from mm observations for each quasar. We adopted the \cii{} redshift as the systemic redshift and normalized the peak flux of all lines for a better visual comparison between the line widths and velocity shifts. The figure highlights the accuracy of the \cii{} redshift with respect to the broad UV emission lines.}
    \label{fig:linevelocityshifts}
\end{figure*}

We complement our near-infrared measurements of the broad emission lines with the mm results on the \cii{} line from ALMA, where available. The forbidden \cii{} transition traces the cold gas component of the quasar's host galaxy and provides the best estimate of the systemic redshift. 
In Figure\,\ref{fig:linevelocityshifts} we compare the line models of the broad \ciii{}, \civ{} and \mgii{} lines with the host galaxy's \cii{} emission line fit. We have adopted the \cii{} redshift as the systemic redshift and normalized the peak flux of all lines to the same value. It has been shown that the \cii{} redshift is a much more accurate measure \citep[$\sigma_z \sim 10\,\rm{km}\,\rm{s}^{-1}$,][]{Venemans2020} for the systemic redshift than the broad emission lines \citep[$\sigma_z \sim 200\,\rm{km}\,\rm{s}^{-1}$,][]{Shen2016}.
This visual comparison illustrates the narrow nature of the \cii{} line compared to the broad emission lines and highlights its value in determining the quasar's systemic redshift.
In addition, the velocity shifts of the broad lines become strikingly apparent, with the \civ{} line exhibiting extreme blueshifts in nearly all quasars.

\movetabledown=2.0in 
\begin{deluxetable*}{lccccccccc}
\tabletypesize{\footnotesize} 
\rotate 
\tablecaption{Properties of the broad \ion{C}{4} and \ion{Mg}{2} emission line fits using the \citet{Tsuzuki2006} iron template\label{table:cIVmgII}}
\tablehead{\colhead{Quasar Name} &\colhead{$z_{\rm{CIV}}$} &\colhead{FWHM$_{\rm{CIV}}$} &\colhead{EW$_{\rm{CIV}}$} &\colhead{$z_{\rm{MgII}}$} &\colhead{FWHM$_{\rm{MgII}}$} &\colhead{EW$_{\rm{MgII}}$} &\colhead{$\Delta v({\rm{CIV}}{-}\rm{MgII})$} & \colhead{$\Delta v({\rm{CIV}}{-}\rm{[CII]})$} &\colhead{$\Delta v({\rm{MgII}}{-}\rm{[CII]})$} \\ 
\nocolhead{} &\nocolhead{} &\colhead{$(\rm{km}\,\rm{s}^{-1})$} &\colhead{(\AA)} &\nocolhead{} &\colhead{$(\rm{km}\,\rm{s}^{-1})$} &\colhead{(\AA)} & \colhead{$(\rm{km}\,\rm{s}^{-1})$} &\colhead{$(\rm{km}\,\rm{s}^{-1})$} &\colhead{$(\rm{km}\,\rm{s}^{-1})$} \\ } 
\startdata 
PSO~J004.3936+17.0862 & ${5.762}_{-0.004}^{+0.004}$ & ${4071}_{-462}^{+451}$ & ${8.78}_{-1.82}^{+2.21}$ & \dots & \dots & \dots & \dots & ${-2408}_{-192}^{+198}$ & \dots\\ 
PSO~J007.0273+04.9571\tablenotemark{{a}} & ${5.944}_{-0.007}^{+0.007}$ & ${7278}_{-1090}^{+1332}$ & ${26.97}_{-9.29}^{+12.61}$ & ${5.997}_{-0.005}^{+0.005}$ & ${2781}_{-394}^{+1579}$ & ${14.66}_{-2.26}^{+2.72}$ & ${-2250}_{-377}^{+377}$ & ${-2463}_{-320}^{+294}$ & ${-213}_{-211}^{+227}$\\ 
PSO~J009.7355--10.4316 & ${5.872}_{-0.011}^{+0.009}$ & ${15746}_{-2274}^{+2315}$ & ${21.70}_{-4.69}^{+5.31}$ & \dots & \dots & \dots & \dots & ${-5683}_{-483}^{+384}$ & \dots\\ 
PSO~J011.3898+09.0324\tablenotemark{{a}} & ${6.361}_{-0.009}^{+0.009}$ & ${5378}_{-807}^{+994}$ & ${7.39}_{-1.56}^{+1.76}$ & ${6.448}_{-0.003}^{+0.003}$ & ${1780}_{-320}^{+366}$ & ${9.46}_{-1.37}^{+1.45}$ & ${-3523}_{-384}^{+384}$ & ${-4377}_{-362}^{+376}$ & ${-855}_{-105}^{+108}$\\ 
VIK~J0046--2837 & \dots & \dots & \dots & ${5.993}_{-0.002}^{+0.002}$ & ${1737}_{-79}^{+88}$ & ${18.06}_{-1.59}^{+1.74}$ & \dots & \dots & \dots\\ 
SDSS~J0100+2802\tablenotemark{{a}} & \dots & \dots & \dots & ${6.307}_{-0.001}^{+0.001}$ & ${4127}_{-66}^{+67}$ & ${6.53}_{-0.11}^{+0.11}$ & \dots & \dots & ${-825}_{-24}^{+26}$\\ 
VIK~J0109--3047\tablenotemark{{a}} & ${6.672}_{-0.007}^{+0.008}$ & ${6636}_{-798}^{+799}$ & ${12.53}_{-2.24}^{+2.38}$ & ${6.764}_{-0.011}^{+0.010}$ & ${2976}_{-665}^{+577}$ & ${11.05}_{-2.68}^{+3.01}$ & ${-3564}_{-498}^{+498}$ & ${-4573}_{-293}^{+304}$ & ${-1009}_{-426}^{+371}$\\ 
PSO~J036.5078+03.0498\tablenotemark{{a}} & ${6.407}_{-0.003}^{+0.003}$ & ${11640}_{-496}^{+557}$ & ${19.74}_{-0.84}^{+0.90}$ & ${6.526}_{-0.003}^{+0.003}$ & ${3542}_{-279}^{+288}$ & ${10.42}_{-0.66}^{+0.69}$ & ${-4803}_{-145}^{+145}$ & ${-5364}_{-103}^{+104}$ & ${-561}_{-101}^{+102}$\\ 
VIK~J0305--3150\tablenotemark{{a}} & ${6.574}_{-0.002}^{+0.002}$ & ${7277}_{-282}^{+301}$ & ${23.62}_{-1.20}^{+1.32}$ & ${6.605}_{-0.003}^{+0.002}$ & ${1988}_{-290}^{+246}$ & ${11.24}_{-1.17}^{+1.18}$ & ${-1227}_{-130}^{+130}$ & ${-1586}_{-91}^{+88}$ & ${-359}_{-101}^{+87}$\\ 
PSO~J056.7168--16.4769\tablenotemark{{a}} & ${5.968}_{-0.000}^{+0.000}$ & ${2642}_{-50}^{+57}$ & ${63.41}_{-1.99}^{+2.04}$ & ${5.977}_{-0.001}^{+0.001}$ & ${2323}_{-85}^{+89}$ & ${24.49}_{-0.78}^{+0.82}$ & ${-379}_{-36}^{+36}$ & ${59}_{-13}^{+13}$ & ${+438}_{-33}^{+34}$\\ 
PSO~J065.4085--26.9543\tablenotemark{{a}} & ${6.049}_{-0.004}^{+0.004}$ & ${7766}_{-283}^{+268}$ & ${12.84}_{-0.87}^{+0.83}$ & ${6.162}_{-0.002}^{+0.002}$ & ${4032}_{-192}^{+216}$ & ${19.96}_{-1.57}^{+1.45}$ & ${-4758}_{-194}^{+194}$ & ${-5799}_{-162}^{+169}$ & ${-1042}_{-99}^{+102}$\\ 
PSO~J065.5041--19.4579\tablenotemark{{a}} & ${6.071}_{-0.002}^{+0.002}$ & ${5638}_{-215}^{+245}$ & ${77.56}_{-4.44}^{+4.81}$ & ${6.115}_{-0.001}^{+0.001}$ & ${2830}_{-80}^{+87}$ & ${32.20}_{-0.90}^{+1.00}$ & ${-1881}_{-97}^{+97}$ & ${-2282}_{-100}^{+87}$ & ${-401}_{-28}^{+29}$\\ 
SDSS~J0842+1218\tablenotemark{{a}} & ${6.018}_{-0.001}^{+0.001}$ & ${6027}_{-137}^{+135}$ & ${39.97}_{-1.82}^{+2.01}$ & ${6.068}_{-0.001}^{+0.001}$ & ${2935}_{-123}^{+131}$ & ${17.47}_{-0.73}^{+0.79}$ & ${-2122}_{-74}^{+74}$ & ${-2423}_{-48}^{+55}$ & ${-301}_{-52}^{+53}$\\ 
SDSS~J1030+0524 & ${6.285}_{-0.010}^{+0.009}$ & ${4733}_{-679}^{+517}$ & ${32.77}_{-2.19}^{+2.37}$ & ${6.305}_{-0.002}^{+0.002}$ & ${2941}_{-220}^{+203}$ & ${19.69}_{-1.35}^{+1.25}$ & ${-851}_{-402}^{+402}$ & ${-827}_{-417}^{+373}$ & ${+24}_{-72}^{+74}$\\ 
PSO~J158.69378--14.42107\tablenotemark{{a}} & ${6.028}_{-0.010}^{+0.014}$ & ${7703}_{-339}^{+369}$ & ${32.33}_{-3.47}^{+6.60}$ & ${6.056}_{-0.003}^{+0.002}$ & ${2661}_{-172}^{+182}$ & ${11.18}_{-1.06}^{+0.98}$ & ${-1220}_{-521}^{+521}$ & ${-1724}_{-427}^{+595}$ & ${-504}_{-111}^{+91}$\\ 
PSO~J159.2257--02.5438\tablenotemark{{a}} & ${6.333}_{-0.001}^{+0.002}$ & ${4921}_{-183}^{+210}$ & ${54.71}_{-3.51}^{+3.83}$ & ${6.362}_{-0.002}^{+0.002}$ & ${3297}_{-208}^{+235}$ & ${24.66}_{-1.35}^{+1.31}$ & ${-1192}_{-97}^{+97}$ & ${-1958}_{-61}^{+67}$ & ${-766}_{-68}^{+78}$\\ 
SDSS~J1044--0125 & ${5.741}_{-0.020}^{+0.013}$ & ${6478}_{-1090}^{+1363}$ & ${17.20}_{-4.45}^{+6.33}$ & \dots & \dots & \dots & \dots & ${-1912}_{-869}^{+576}$ & \dots\\ 
VIK~J1048--0109\tablenotemark{{a}} & \dots & \dots & \dots & ${6.648}_{-0.008}^{+0.009}$ & ${3955}_{-839}^{+727}$ & ${18.20}_{-3.52}^{+3.86}$ & \dots & \dots & ${-1076}_{-321}^{+340}$\\ 
ULAS~J1120+0641 & ${7.027}_{-0.001}^{+0.001}$ & ${6952}_{-86}^{+91}$ & ${33.10}_{-0.99}^{+1.02}$ & \dots & \dots & \dots & \dots & ${-2136}_{-27}^{+32}$ & \dots\\ 
ULAS~J1148+0702 & ${6.273}_{-0.006}^{+0.006}$ & ${5734}_{-295}^{+295}$ & ${27.27}_{-1.78}^{+2.67}$ & ${6.334}_{-0.002}^{+0.002}$ & ${4151}_{-169}^{+166}$ & ${18.99}_{-0.82}^{+0.87}$ & ${-2476}_{-251}^{+251}$ & \dots & \dots\\ 
PSO~J183.1124+05.0926\tablenotemark{{a}} & ${6.313}_{-0.007}^{+0.007}$ & ${8927}_{-649}^{+768}$ & ${13.37}_{-1.36}^{+1.65}$ & ${6.408}_{-0.004}^{+0.004}$ & ${3132}_{-263}^{+259}$ & ${14.78}_{-1.35}^{+1.32}$ & ${-3873}_{-333}^{+333}$ & ${-5114}_{-277}^{+305}$ & ${-1242}_{-151}^{+172}$\\ 
SDSS~J1306+0356\tablenotemark{{a}} & ${5.998}_{-0.000}^{+0.000}$ & ${5236}_{-99}^{+83}$ & ${47.89}_{-1.77}^{+1.76}$ & ${6.024}_{-0.001}^{+0.001}$ & ${3107}_{-74}^{+73}$ & ${20.24}_{-0.53}^{+0.51}$ & ${-1136}_{-34}^{+34}$ & ${-1499}_{-18}^{+20}$ & ${-363}_{-29}^{+29}$\\ 
ULAS~J1319+0950\tablenotemark{{a}} & ${6.058}_{-0.002}^{+0.002}$ & ${8933}_{-110}^{+118}$ & ${18.67}_{-0.39}^{+0.35}$ & ${6.124}_{-0.001}^{+0.001}$ & ${3155}_{-131}^{+138}$ & ${13.17}_{-0.53}^{+0.55}$ & ${-2807}_{-92}^{+92}$ & ${-3261}_{-77}^{+69}$ & ${-454}_{-56}^{+55}$\\ 
ULAS~J1342+0928 & ${7.341}_{-0.003}^{+0.003}$ & ${13969}_{-334}^{+263}$ & ${21.18}_{-0.70}^{+0.54}$ & \dots & \dots & \dots & \dots & ${-7061}_{-94}^{+103}$ & \dots\\ 
CFHQS~J1509--1749\tablenotemark{{a}} & ${6.089}_{-0.001}^{+0.001}$ & ${5537}_{-175}^{+183}$ & ${31.54}_{-1.58}^{+1.87}$ & ${6.119}_{-0.001}^{+0.001}$ & ${3491}_{-171}^{+191}$ & ${16.81}_{-0.86}^{+0.85}$ & ${-1286}_{-76}^{+76}$ & ${-1421}_{-52}^{+49}$ & ${-135}_{-57}^{+57}$\\ 
PSO~J231.6576--20.8335\tablenotemark{{a}} & \dots & \dots & \dots & ${6.571}_{-0.007}^{+0.007}$ & ${3894}_{-585}^{+569}$ & ${17.52}_{-2.42}^{+2.53}$ & \dots & \dots & ${-645}_{-292}^{+289}$\\ 
PSO~J239.7124--07.4026\tablenotemark{{a}} & ${6.111}_{-0.011}^{+0.004}$ & ${3633}_{-481}^{+827}$ & ${31.24}_{-2.63}^{+4.44}$ & ${6.115}_{-0.001}^{+0.001}$ & ${2723}_{-115}^{+122}$ & ${17.61}_{-0.68}^{+0.82}$ & ${-158}_{-315}^{+315}$ & ${67}_{-453}^{+172}$ & ${+225}_{-38}^{+40}$\\ 
PSO~J308.0416--21.2339\tablenotemark{{a}} & ${6.168}_{-0.004}^{+0.005}$ & ${8035}_{-861}^{+749}$ & ${33.16}_{-1.41}^{+1.61}$ & ${6.231}_{-0.002}^{+0.002}$ & ${2515}_{-151}^{+140}$ & ${11.43}_{-0.85}^{+0.83}$ & ${-2657}_{-218}^{+218}$ & ${-2823}_{-188}^{+200}$ & ${-167}_{-101}^{+98}$\\ 
SDSS~J2054--0005\tablenotemark{{a}} & ${5.936}_{-0.007}^{+0.007}$ & ${10795}_{-1669}^{+2049}$ & ${21.51}_{-4.41}^{+6.17}$ & ${6.029}_{-0.004}^{+0.003}$ & ${2527}_{-279}^{+370}$ & ${19.50}_{-1.86}^{+2.25}$ & ${-4013}_{-337}^{+337}$ & ${-4428}_{-295}^{+316}$ & ${-415}_{-158}^{+128}$\\ 
CFHQS~J2100--1715\tablenotemark{{a}} & ${6.060}_{-0.010}^{+0.008}$ & ${7433}_{-999}^{+2324}$ & ${13.10}_{-2.96}^{+4.97}$ & ${6.097}_{-0.011}^{+0.009}$ & ${7726}_{-2572}^{+1007}$ & ${27.83}_{-7.87}^{+3.74}$ & ${-1562}_{-570}^{+570}$ & ${-866}_{-433}^{+338}$ & ${+697}_{-459}^{+380}$\\ 
PSO~J323.1382+12.2986\tablenotemark{{a}} & ${6.575}_{-0.001}^{+0.001}$ & ${3286}_{-83}^{+93}$ & ${39.27}_{-1.48}^{+1.48}$ & ${6.585}_{-0.001}^{+0.001}$ & ${2291}_{-142}^{+122}$ & ${20.81}_{-0.99}^{+1.03}$ & ${-421}_{-60}^{+60}$ & ${-494}_{-26}^{+22}$ & ${-72}_{-56}^{+54}$\\ 
VIK~J2211--3206\tablenotemark{{a}} & ${6.287}_{-0.002}^{+0.002}$ & ${3996}_{-246}^{+250}$ & ${12.53}_{-1.43}^{+1.54}$ & ${6.332}_{-0.001}^{+0.001}$ & ${3890}_{-166}^{+191}$ & ${24.87}_{-1.09}^{+1.12}$ & ${-1814}_{-108}^{+108}$ & ${-2128}_{-87}^{+100}$ & ${-314}_{-54}^{+53}$\\ 
CFHQS~J2229+1457 & ${6.156}_{-0.000}^{+0.000}$ & ${886}_{-49}^{+51}$ & ${83.95}_{-8.09}^{+10.57}$ & \dots & \dots & \dots & \dots & ${164}_{-16}^{+13}$ & \dots\\ 
PSO~J340.2041--18.6621 & ${5.994}_{-0.000}^{+0.000}$ & ${1767}_{-42}^{+44}$ & ${38.01}_{-1.41}^{+1.53}$ & ${5.998}_{-0.001}^{+0.001}$ & ${2055}_{-97}^{+94}$ & ${20.58}_{-1.11}^{+1.03}$ & ${-153}_{-51}^{+51}$ & ${-275}_{-12}^{+10}$ & ${-122}_{-49}^{+51}$\\ 
SDSS~J2310+1855\tablenotemark{{a}} & ${5.885}_{-0.003}^{+0.003}$ & ${18297}_{-708}^{+651}$ & ${63.97}_{-4.60}^{+5.02}$ & ${5.999}_{-0.003}^{+0.003}$ & ${2870}_{-207}^{+210}$ & ${15.05}_{-1.30}^{+1.37}$ & ${-4937}_{-159}^{+159}$ & ${-5103}_{-110}^{+113}$ & ${-166}_{-109}^{+118}$\\ 
VIK~J2318--3029\tablenotemark{{a}} & ${6.095}_{-0.003}^{+0.002}$ & ${6733}_{-339}^{+397}$ & ${26.78}_{-1.72}^{+1.92}$ & ${6.142}_{-0.002}^{+0.002}$ & ${2913}_{-170}^{+168}$ & ${16.34}_{-1.03}^{+1.23}$ & ${-1995}_{-132}^{+132}$ & ${-2129}_{-108}^{+103}$ & ${-134}_{-80}^{+80}$\\ 
VIK~J2348--3054\tablenotemark{{a}} & ${6.851}_{-0.003}^{+0.003}$ & ${3982}_{-256}^{+272}$ & ${15.09}_{-1.48}^{+1.55}$ & ${6.888}_{-0.010}^{+0.009}$ & ${4495}_{-870}^{+1012}$ & ${22.04}_{-4.01}^{+4.17}$ & ${-1408}_{-377}^{+377}$ & ${-1891}_{-123}^{+108}$ & ${-484}_{-373}^{+344}$\\ 
PSO~J359.1352--06.3831\tablenotemark{{a}} & ${6.146}_{-0.001}^{+0.001}$ & ${3520}_{-117}^{+123}$ & ${49.88}_{-2.59}^{+2.54}$ & ${6.163}_{-0.002}^{+0.002}$ & ${2505}_{-171}^{+240}$ & ${14.02}_{-1.60}^{+1.59}$ & ${-720}_{-93}^{+93}$ & ${-1100}_{-27}^{+37}$ & ${-380}_{-92}^{+82}$\\ 
\enddata 
\tablenotetext{{a}}{{These 27 quasars were used in the analysis of Section\,\ref{sec:irondiscussion}.}}
\end{deluxetable*}

\subsubsection{The \civ{}-\mgii{} velocity shift} \label{sec:velocity_shifts}

Contrary to some studies in the literature \citep[e.g.][]{Richards2011, Mazzucchelli2017}, all broad line velocity shifts discussed in this paper are given in the observer's frame. From this perspective a negative \civ{} velocity shift with respect to \mgii{}, $\Delta v(\rm{CIV{-}MgII}) < 0\,\rm{km}\,\rm{s}^{-1}$, could be attributed to an outflowing component with positive velocity, as seen from the point of view of the quasar's SMBH. 

Systematic velocity shifts between quasar emission lines were discovered many years ago \citep[e.g.][]{Gaskell1982} and are still a widely discussed topic in the literature \citep[e.g.][]{VandenBerk2001, Richards2002, Hewett2010, Richards2011, Meyer2019c, Yong2020}.
Correlations between the magnitude of the emission line velocity shifts and their ionization potential \citep{Tytler1992, McIntosh1999b, VandenBerk2001} point towards a common physical origin. Curiously, these correlations are not only found within lines associated with the quasar's BLR close to the accretion disk, but are also found in forbidden narrow lines like [OIII] \citep[e.g. ][]{Zakamska2016} commonly associated with the narrow line region (NLR) at kiloparsec-scales, centered on the quasar.
The broad high-ionization lines like \civ{} or \siiv{}, are known to exhibit especially large blueshifts compared to the broad lower ionization lines (e.g. \mgii{}) or the narrow lines. These line shifts are thought to originate from an outflowing component \citep{Gaskell1982} driven by X-ray radiation and/or line driven winds \citep[e.g.][]{Krolik1986, Murray1995a}. 

\begin{figure*}
    \centering
    \includegraphics[width=0.95\textwidth]{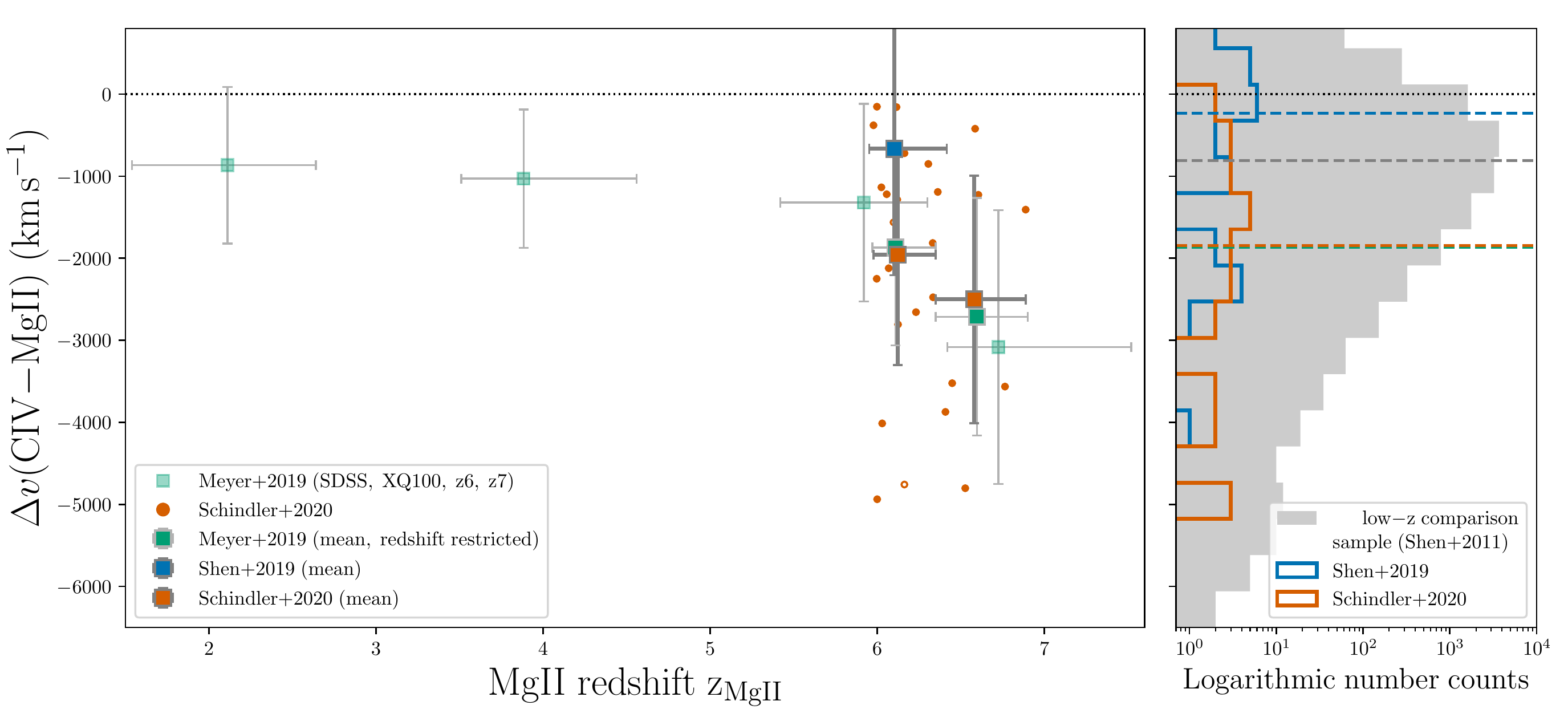}
    \caption{\textbf{Left panel:} \civ{}-\mgii{} velocity shifts of individual high-redshift quasars as a function of the \mgii{} redshift. Our results are shown in orange. We calculate the mean (orange square) and standard deviation (dark grey error bars) \civ{}-\mgii{} for two sub-samples split at $z=6.35$. The error bars in the redshift direction show the redshift bin. We contrast these results with the mean results of the samples of \citet{Meyer2019c} (light green squares). For an appropriate comparison we restrict the z6 and z7 \citet{Meyer2019c} samples to the same redshift ranges as our sample and re-compute the mean (green squares), yielding very good agreement with our data. 
    All velocity shift measurements are based on the peak redshifts of the emission lines. 
    \textbf{Right panel:} Logarithmic \civ{}-\mgii{} velocity shift histograms. We compare our results (orange) to the high-redshift sample of \citet{Shen2019a} (blue). Both measurements are contrasted with low redshift data from \citet{Shen2011} at $1.5 \leq z \leq 2.2$ and restricted to $\log(L_{\rm{bol}}/\rm{erg}\,\rm{s}^{-1})=46.5-47.5$ for a valid comparison to the high redshift quasars. The colored dashed lines show the median of the three distributions as well as the combined z6 and z7 samples of \citet{Meyer2019c} (green). 
    }
    \label{fig:blueshift_redshift_histogramm}
\end{figure*}

The \civ{} high ionization, broad emission line has received special attention in the literature. Not only does the line show the most prominent velocity shifts \citep{Richards2002, Richards2011}, it is also commonly used to estimate the BH masses of high-redshift quasars
\citep{Vestergaard2006}.
Strong \civ{}-\mgii{} blueshifts are ubiquitously found in samples of high-redshift quasars \citep{DeRosa2014, Mazzucchelli2017, Reed2019, Shen2019a, Meyer2019c}. 
In the recent study by \citet{Meyer2019c}, the authors discuss an intriguing redshift evolution in the mean velocity shifts of the \civ{} line compared to lower ionization quasar emission lines (\textsc{C ii}, \textsc{O i} and \mgii{}).
They found the mean \civ{}-\mgii{} blueshift between their $z\sim6$ and $z\sim7$ samples to increase significantly from $-1322\,\rm{km}/\rm{s}$ to $-3082\,\rm{km}/\rm{s}$. 
We display the \civ{}-\mgii{} velocity shifts as a function of \mgii{} redshift in the left panel of Figure\,\ref{fig:blueshift_redshift_histogramm}. We show data on 28 quasars of our sample (open and filled orange circles). The individual values are provided in Table\,\ref{table:cIVmgII}. Where we fit Gaussian profiles to the emission lines and derive the peak redshifts, \citet{Meyer2019c} used a spline fitting algorithm to determine the peak redshifts of the lines.  The mean \civ{}-\mgii{} velocity shifts of \citet{Meyer2019c} are provided in light green squares, emphasizing the evolution at the highest redshifts.
Our sample spans a narrower redshift range and for a valid comparison we cut the \citet{Meyer2019c} "z6" and "z7" samples at the minimum and maximum \mgii{} redshift of our sample. We then divide our sample using the redshift boundary between their "z6" and "z7" samples ($z\approx6.35$).
We show the mean velocity shifts for our (orange) and the redshift restricted sample of \citet[green]{Meyer2019c} including the  sample standard deviation (grey error bars) with squares in the left panel of Figure\,\ref{fig:blueshift_redshift_histogramm}. We further summarize the sub-sample mean properties in Table\,\ref{table:CIVMgIIvshift_z}.
While our $z > 6.35$ redshift sub-sample resembles the "z7" \citet{Meyer2019c} sample (6/9 overlap), our $z \le 6.35$ sub-sample has $4$ times as many quasars, leading to improved sample statistics. 
Our sub-samples not only have virtually the same mean redshift, compared to the \citet{Meyer2019c} redshift restricted samples, they also show very similar $\Delta v(\rm{CIV{-}MgII})$. In the higher redshift bin, where our samples strongly overlap this agreement emphasizes the consistency between their and our measurement methods. 
Based on our two sub-samples, we can confirm a potential evolution of $\Delta v(\rm{CIV{-}MgII})$ at the highest redshifts as the mean velocity shift decreases (the velocity blueshift increases) with redshift by $\sim500\,\rm{km}/\rm{s}$. We should note, however, that there are small differences in the bolometric luminosity and Eddington ratio of the two sub-samples (see Table\,\ref{table:CIVMgIIvshift_z}).

Furthermore, we also display the mean \civ{}-\mgii{} velocity shift calculated from the \citet{Shen2019a} quasar sample.
Equivalent to our approach, the emission line redshifts provided in \citet{Shen2019a} are also measured from the peak of their emission line models.
The 27 quasars from their sample, for which both, the \civ{} and the \mgii{} redshift, were measured, have roughly the same mean redshift ($z=6.10$) as our lower redshift sub-sample. On the other hand, their average \civ{}-\mgii{} blueshift, $\Delta v(\rm{CIV{-}MgII})=-666\,\rm{km}\,\rm{s}^{-1}$, is substantially lower than ours. 
The differences between our and their quasar sample show more clearly in the blue and orange histograms in the right panel of Figure\,\ref{fig:blueshift_redshift_histogramm}. Their sample includes a larger number of quasars that show either no or even positive velocity shifts. This results in a median velocity shift of $-234\,\rm{km}\,\rm{s}^{-1}$ (dashed line), which is strikingly different from the median velocity shift of our or the \citet{Meyer2019c} quasars ($\Delta v(\rm{CIV{-}MgII}) \approx -1800\,\rm{km}/\rm{s}$, orange and green dashed lines). In addition, their sample includes fewer quasars with extreme velocity shifts $\Delta v(\rm{CIV{-}MgII})<-4000\,\rm{km}\,\rm{s}^{-1}$.

As we will discuss below (see Section\,\ref{sec:vshift_discussion}), the \civ{}-\mgii{} blueshift has been shown to correlate with quasar luminosity. The mean bolometric luminosity of the \citet{Shen2019a} sample, $L_{\rm{bol}}=0.9\ 10^{47}\,\rm{erg}\,\rm{s}^{-1}$, is a factor of two lower than our $z\leq6.35$ redshift sub-sample. Therefore, the luminosity difference between the two samples could be a driving factor for the smaller in \civ{}-\mgii{} blueshifts found in their sample.
However, that does not mean that biases due to different modeling strategies can be excluded. For example, \citet{Shen2019a} add a third-order polynomial to model the continuum, while they do not include a Balmer continuum contribution. Furthermore, the broadening of the iron template is a free parameter in their model, while we fix this parameter to the FWHM of the \mgii{} line. This changes the continuum model and thus leads to differences in the continuum-subtracted emission line profiles. As a result different peak redshifts will be measured even when the same iron template is used. Quantifying these differences requires a full fit of their sample with our methodology, which is beyond the scope of this work.



\tabletypesize{\footnotesize} 
\begin{deluxetable}{lrr}
\tablecaption{Comparing the \civ{}-\mgii{} velocity shift in two sub-samples divided at $z=6.35$ \label{table:CIVMgIIvshift_z}}
\tablehead{\colhead{Property} &\colhead{$z < 6.35$} &\colhead{$z \ge 6.35$} 
} 
\decimals 
\startdata 
\multicolumn{3}{c}{This work}\\
\tableline
Number of quasars & 20 & 8 \\
Mean redshift & 6.12 & 6.57 \\
Mean $\Delta v(\rm{CIV{-}MgII})/(\rm{km}\,\rm{s}^{-1})$ &  -1958.74 & -2501.35 \\ 
Mean $M_{\rm{BH}}/(10^9\,M_{\odot})$ & 2.2 & 1.9\\
Mean $L_{\rm{bol}}/L_{\rm{Edd}}$ & 0.89 & 0.83\\
Mean $L_{\rm{bol}}/(10^{47}\,\rm{erg}\,\rm{s}^{-1})$ & 2.2 & 1.5 \\
Overlap with \citet{Meyer2019c}  & 4  & 6  \\
\tableline
\multicolumn{3}{c}{\citet{Meyer2019c} z\_6 and z\_7 redshift restricted samples}\\
\tableline
Number of quasars & 5 & 9 \\
Mean redshift & 6.11 & 6.60 \\
Mean $\Delta v(\rm{CIV{-}MgII})/(\rm{km}\,\rm{s}^{-1})$ & -1869.91 & -2712.78 \\ 
\enddata 
\end{deluxetable}


We also compare our results with a luminosity-matched sample of 12099 low-redshift SDSS quasars at $1.52 \le z \le 2.2$  \citep[][in grey]{Shen2011}. For a detailed description on the construction of the low-redshift comparison sample see Appendix\,\ref{sec:app_lowzsample}. The velocity shifts for the low-redshift sample are also measured from the peak of the multiple-Gaussian model fit to the broad component, equivalent to our measurement method.
A histogram of the low-redshift velocity shifts is shown in grey in Figure\,\ref{fig:blueshift_redshift_histogramm}. 
While the velocity shift distribution of our sample is fairly flat (median $\Delta v(\rm{CIV{-}MgII}) \approx -1800\,\rm{km}/\rm{s}$), the low-redshift quasars show a peaked distribution with a median of $\Delta v(\rm{CIV{-}MgII}) \approx -800\,\rm{km}/\rm{s}$. 
While both quasar samples span a large range of velocity shifts, we do not find notable \civ{}-\mgii{} velocity redshifts ($\Delta v(\rm{CIV{-}MgII}) > 0\,\rm{km}/\rm{s}$ ) for any quasar in our sample.
On the other hand, quasars with extreme blueshifts ($\Delta v(\rm{CIV{-}MgII}) \approx -5500\,\rm{km}/\rm{s}$) are well represented in the low-redshift sample. In other words, it is always possible to identify low-redshift analogues to all of our high-redshift quasars in terms of bolometric luminosity and velocity shift.

\subsubsection{Properties of the \civ{} emission line and the \civ{}-\mgii{} velocity shift}

\begin{figure*}
    \centering
    \includegraphics[width=0.45\textwidth]{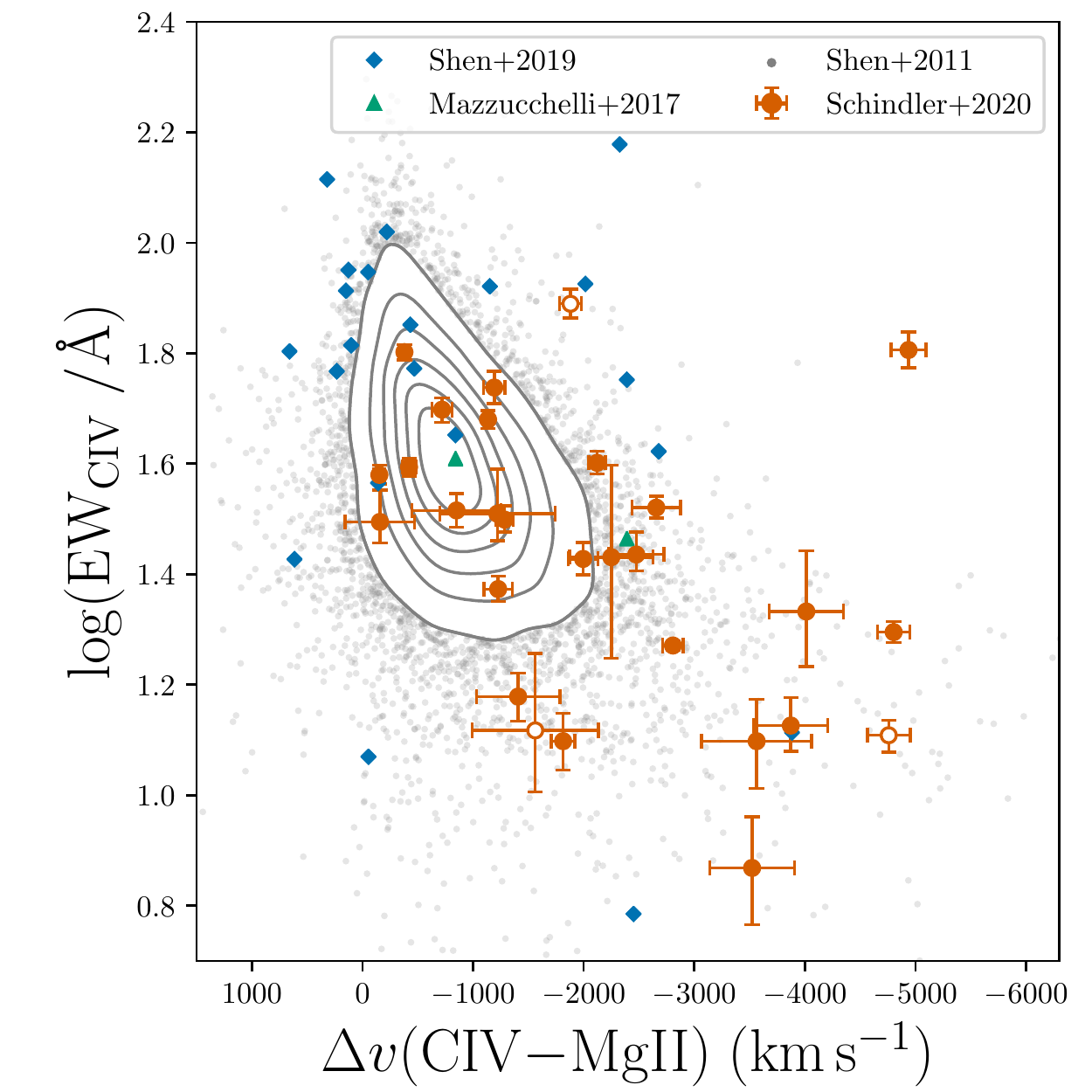}
    \includegraphics[width=0.45\textwidth]{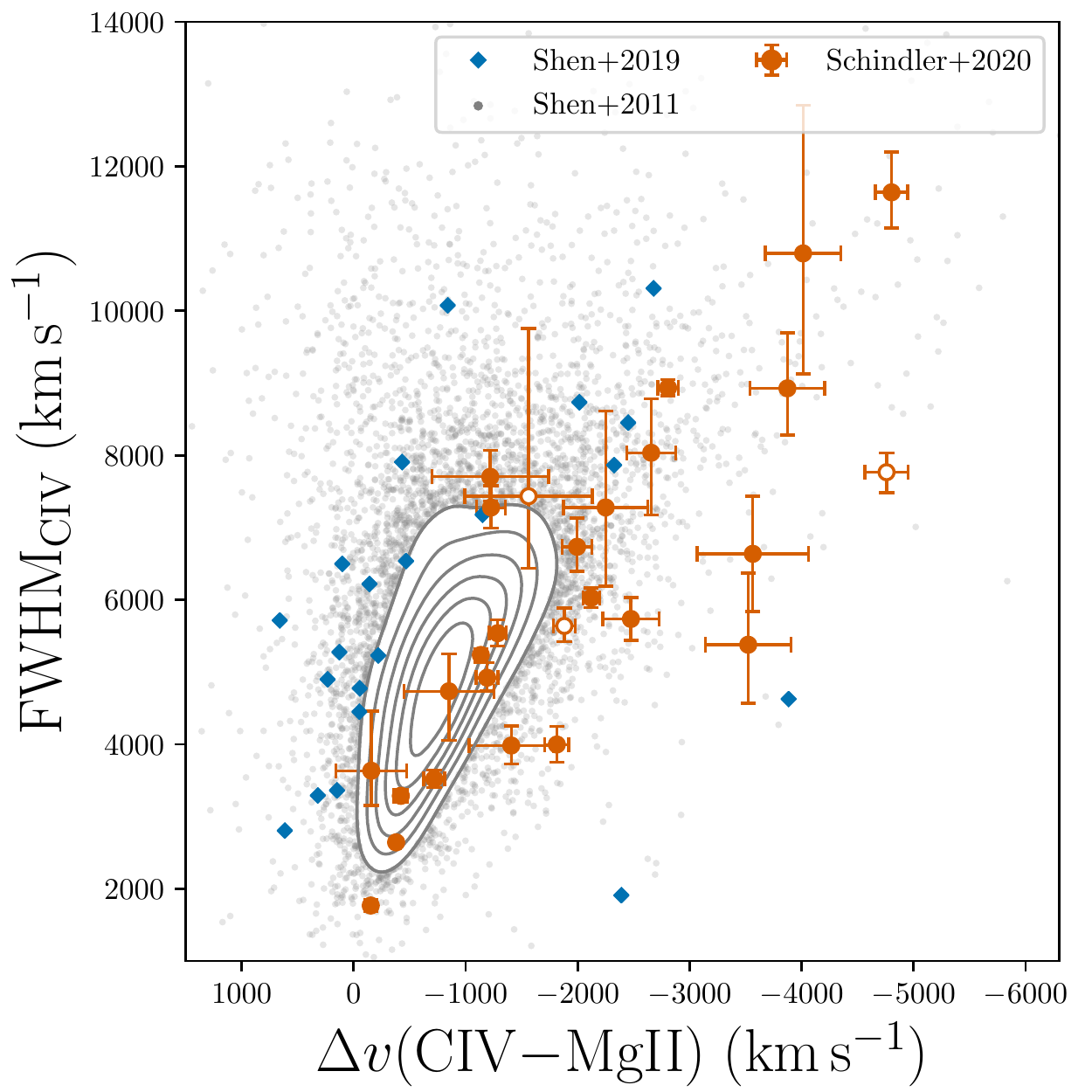}
    \caption{\textbf{Left panel:} \civ{} equivalent width as a function of the \civ{}-\mgii{} velocity shift. We show our data with 68 percentile uncertainties as open and solid orange circles. The open circles refer to spectral fits, in which the continuum was only approximated locally around the \civ{} and \mgii{} lines but not fit across the entire spectrum. Quasars from the recent study of \citet{Shen2019a} are shown as blue diamonds and objects of \citet{Mazzucchelli2017} not covered in our study are shown as green triangles. The grey contours and grey dots show the luminosity-matched low-redshift comparison sample as in Figure\,\ref{fig:blueshift_redshift_histogramm} (also see Appendix\,\ref{sec:app_lowzsample}). Our sample generally follows the low-redshift distribution but preferably occupies a region with low equivalent widths and high \civ{}-\mgii{} blueshifts. \textbf{Right panel:} \civ{} FWHM as a function of the \civ{}-\mgii{} velocity shift. The symbols are analogous to the left panel. Our sample of high-redshift quasars shows a significant correlation between the \civ{} FWHM and the \civ{}-\mgii{} blueshift, possibly indicating a strong non-virial component in the \civ{} emission \citep[e.g.][]{Shen2008, Richards2011, Coatman2016}.}
    \label{fig:civ_context}
\end{figure*}

While the presented X-SHOOTER/ALMA quasar sample is largely homogeneous in terms of its bolometric luminosity and shows mostly high Eddington luminosity ratios ($L_{\rm{bol}}/L_{\rm{Edd}} > 0.1$), it exhibits a large range of \civ{}-\mgii{} velocity shifts.
We show the plane of \civ{} equivalent width and \civ{}-\mgii{} velocity shift, the so called \civ{}-plane, in the left panel of Figure\,\ref{fig:civ_context}. Our quasar sample is shown with filled and open orange circles. We also display all high-redshift quasars of \citet{Shen2019a} and \citet{Mazzucchelli2017}, which are not included in the X-SHOOTER/ALMA sample. For comparison we show the low-redshift SDSS quasar sample (as in Figure\,\ref{fig:blueshift_redshift_histogramm}) with grey dots and contours. 
\citet{Richards2011} discussed that quasars with stronger  \civ{}-\mgii{} velocity blueshifts show weaker \civ{} EWs and quasars with weaker velocity blueshifts show stronger \civ{} EWs. However, there is also a population of quasars in the lower left part of the plane, quasars with both weak \civ{} EWs and weak velocity shifts, while the upper right part of the plane is not populated at all.
Our high-redshift quasars follow the same low-redshift trends. The objects with larger $\Delta v(\rm{CIV{-}MgII})$ have generally lower rest-frame \civ{} equivalent widths. Yet, a large fraction of the high-redshift quasars in our sample occupy a region outside of the low-redshift contours with very large blueshifts and predominantly weak \civ{} EWs.
This behavior might well be related to the quasar's luminosity. According to \citet[Baldwin effect]{Baldwin1977} the EW of high ionization lines, like \civ{}, is inversely correlated with the quasar's UV luminosity. Stronger UV emission has also been linked to larger velocity shifts for resonant lines driven by radiation pressure (see Section\,\ref{sec:vshift_discussion}).

We already discussed that the \citet{Shen2019a} quasar sample shows overall more moderate \civ{}-\mgii{} blueshifts. The left panel of Figure\,\ref{fig:civ_context} reveals that for the same \civ{}-\mgii{} blueshifts their quasars also show a tendency for larger EWs. As their quasar sample includes less luminous quasars than ours this systematic difference could possibly be related to the Baldwin effect. However, as discussed above, systematic effects introduced by different data and model fitting techniques cannot be excluded.

\begin{figure*}[ht]
    \centering
    \includegraphics[width=0.95\textwidth]{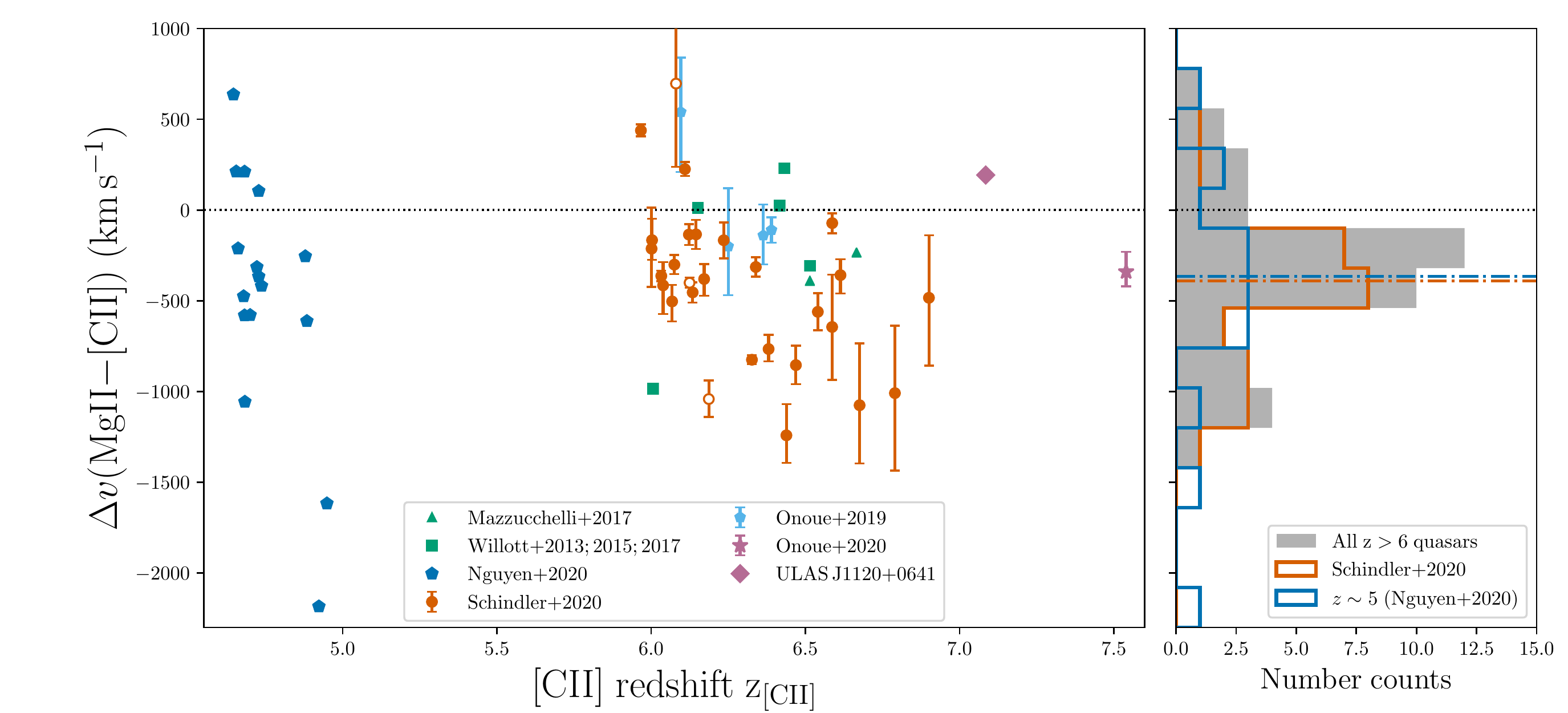}
    \caption{\textbf{Left panel:} \mgii{}-\cii{} velocity shifts of individual high-redshift quasars as a function of the \cii{} redshift. Our results are shown in orange. We include other quasars in the literature with different colored symbols. The velocity shift of ULAS\,J1120+0641 is based on the \mgii{} redshift published by \citet{Meyer2019c} and the \cii{} redshift of \citet{Venemans2020}. We excluded J1208-0200 from the sample of \citet{Onoue2019}. This quasar shows an extreme positive velocity shift, which is likely biased due to the weak \mgii{} emission and contamination by an OH sky line. 
    \textbf{Right panel:} Histograms of the \mgii{}-\cii{} velocity shift. The data from our sample is shown in orange, while all $z>6$ shown in the left panel result in the grey distribution. We contrast the $z>6$ quasars with the $z\sim5$ sample of \citet{Nguyen2020}. On average all quasar distributions show blueshifted \mgii{} emission with respect to the \cii{} redshift. The median \mgii{}-\cii{} velocity shift of our sample (dashed-dotted orange line) is $-415.95\,\rm{km}\,\rm{s}^{-1}$ and shows good agreement with the $z\sim5$ quasars (blue dashed-dotted line).}
    \label{fig:mgiiblueshift_ciiredshift_histogramm}
\end{figure*}

In the right panel of Figure\,\ref{fig:civ_context}, we show the relation between the \civ{} FWHM and the \civ{}-\mgii{} velocity shift. The data is colored analogously to the left panel of the same figure. 
Large samples of SDSS quasars \citep{Shen2008, Shen2011} revealed a significant anti-correlation between the \civ{}-\mgii{} velocity shift and the \civ{} FWHM, indicating a possibly non-virialized component in the \civ{} line \citep{Shen2008, Shen2011, Richards2011}. A non-varying component of \civ{} was later discovered in reverberation mapping data \citep{Denney2012}.
The comparison sample of low-redshift SDSS quasars shows that the stronger the \civ{}-\mgii{} blueshift is, the larger is the measured FWHM of the \civ{} line. While outliers do populate the upper left region of the figure with weak \civ{} velocity shifts and very broad lines, the lower right is largely empty. 
Our sample of high-redshift quasars shows a large range of \civ{}-\mgii{} velocity shifts and displays a prominent anti-correlation with the measured \civ{} FWHM, reminiscent of the luminous $2 \le z \le 2.7$ quasar sample of \citet{Coatman2016}. 
Similar to the \civ{}-plane, half of our sample falls into regions that are sparsely populated by the low-redshift comparison sample of strong \civ{}-\mgii{} blueshifts and large \civ{} FWHM.
As many $z>5$ quasars show considerable \civ{}-\mgii{} blueshifts, which possibly indicate that the \civ{} line is not fully virialized in these objects, BH mass estimates based on this emission line should be considered with great caution.
To mitigate potential biases due to the \civ{} blueshift-FWHM correlation, \citet{Coatman2017} developed a correction, which we apply to \civ{}-based BH masses in our sample. However, we caution against an over-interpretation of the values, as these empirical corrections may have limited applicability \citep{MejiaRestrepo2018a}.

\subsubsection{Broad line velocity shifts relative to the host galaxy \cii{} line}

Measurements of the $158\,\mu\rm{m}$ \cii{} line probe the cold, dense gas of the quasar host galaxy and define a precise systemic redshift independent of the NIR spectral properties.
This redshift measurement allows us to study the velocity shifts of the broad \siiv{},\civ{}, \ciii{} and \mgii{} lines with respect to the galaxy's rest frame. Our measurements are reported in Tables\,\ref{table:cIVmgII} and \ref{table:ciiisiiv}.

Velocity blueshifts of the \mgii{} line with respect to the \cii{} transition have been observed in a number of $z>6$ quasars \citep{Willott2013, Banados2015, Willott2015, Venemans2016, WangRan2016, Willott2017, Venemans2017c, Mazzucchelli2017, Decarli2018}. 
Recently, \citet{Nguyen2020} found similar blueshifts for their sample of $z{\sim}4.8$ quasars.
In Figure\,\ref{fig:mgiiblueshift_ciiredshift_histogramm} we compare the \mgii{}-\cii{} velocity shifts of our sample to other $z\gtrsim6$ quasars in the literature as well as to the $z{\sim}4.8$ sample of \citet{Nguyen2020}. The left panel of this figure shows the \mgii{}-\cii{} velocity shift at $z\gtrsim6$ as a function of \cii{} redshift, while the right panel summarizes the \mgii{}-\cii{} velocity shift distributions in histograms. 
Our sample (orange filled and open circles) consists of 28 quasars and shows a median velocity shift of $\Delta v(\rm{MgII{-}[CII]}) =-390.61_{-455.34}^{256.02}\,\rm{km}\,\rm{s}^{-1}$. Open circles refer to quasars, where we could not fit a power law continuum over the full spectral range and approximated the continuum only closely around the \mgii{} line (see Section\,\ref{sec:sample_properties}).
We compare our sample to other $z>6$ quasars in the literature \citep{Willott2013, Willott2015, Mazzucchelli2017, Willott2017, Onoue2019, Nguyen2020, Onoue2020}.
In the right panel of Figure\,\ref{fig:mgiiblueshift_ciiredshift_histogramm} we compare the velocity shift distributions of our sample (orange) to all $z>6$ quasars (grey, our sample and the literature data) and to the sample of $z\sim4.8$ quasars \citep{Nguyen2020}.
The \mgii{}-\cii{} velocity shift histogram of our sample shows a strong peak (10 quasars) in the $-560$ to $-340\,\rm{km}\,\rm{s}^{-1}$ bin with a broad range of values between $-1250$ to $700\,\rm{km}\,\rm{s}^{-1}$. 
The $z\sim5$ quasar sample of \citet{Nguyen2020} shows a large range of velocity shifts, but the median of their sample ($\Delta v(\rm{MgII{-}[CII]}) =-367\,\rm{km}\,\rm{s}^{-1}$) agrees well with our result. 

\begin{figure}
    \centering
    \includegraphics[width=0.48\textwidth]{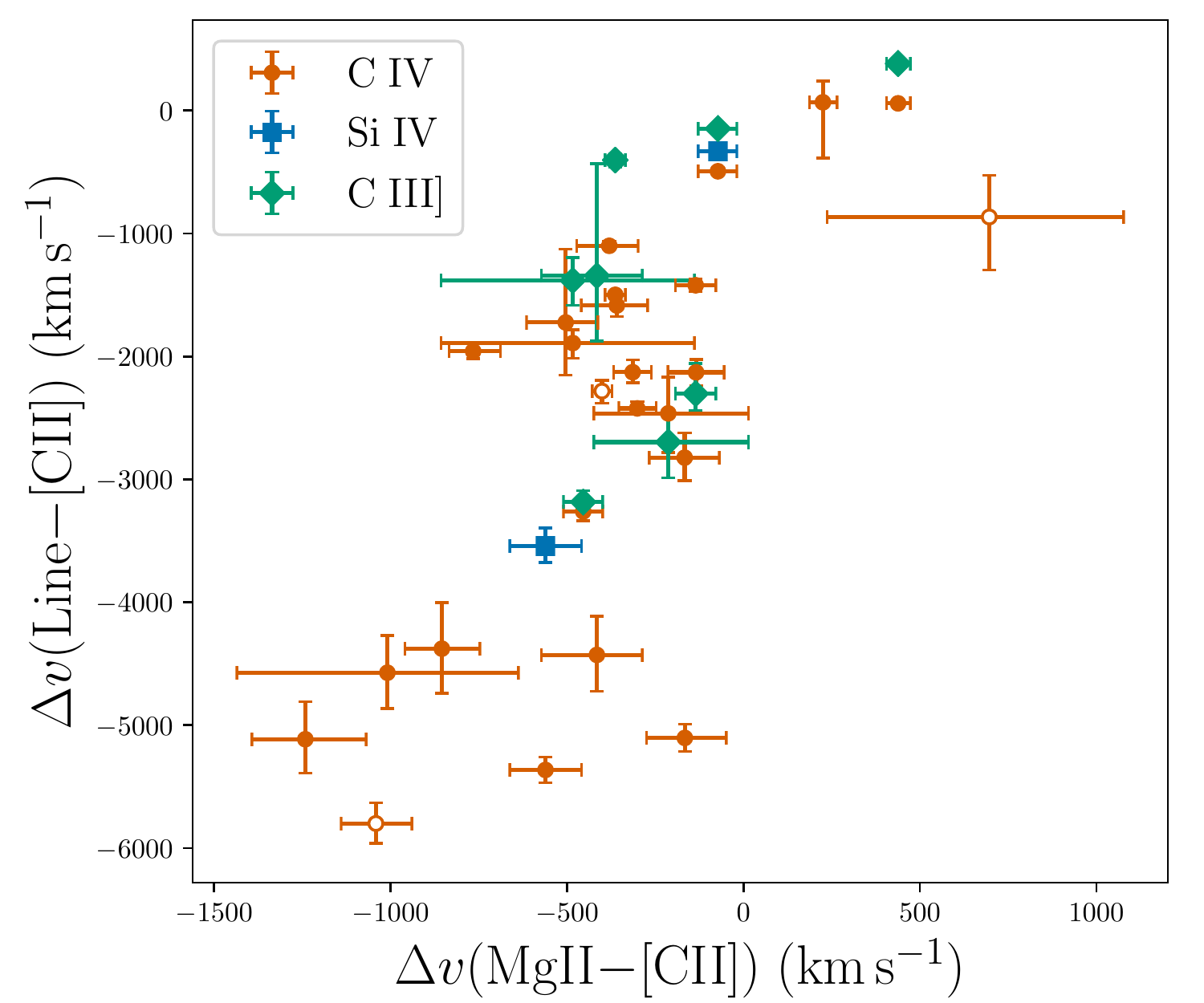}
    \caption{Velocity shifts of the \civ{} (orange circles), \siiv{} (blue squares), and \ciii{} complex peak (green diamonds) with respect to the \mgii{} velocity shift. All velocity shifts are measured with respect to the systemic redshift from the \cii{} line. \siiv{}, \civ{}, and \ciii{} are all high ionization lines, while \mgii{} is regarded as a low ionization line and therefore supposed to originate at a different location in the BLR and under different physical conditions. In addition, the highly blueshifted \civ{} line is suspected to originate in an outflowing wind. The correlation of the \civ{} and \mgii{} velocity shifts with respect to the \cii{} line redshift potentially indicates a common physical origin of the line velocity shifts.}
    \label{fig:mgiiciivshift_vshifts}
\end{figure}

The broad \siiv{} and \civ{} lines as well as the \ciii{} complex also show significant blueshifts with respect to the host galaxy's \cii{} emission (see Tables\,\ref{table:cIVmgII} and \ref{table:ciiisiiv}). 
We display their respective velocity shifts as a function of the \mgii{}-\cii{} velocity shift in Figure\,\ref{fig:mgiiciivshift_vshifts}. Velocity shifts of the \civ{} line are shown with orange circles, while we display velocity shifts of the \siiv{} line and the \ciii{} complex with blue squares and green diamonds, respectively. 
The figure shows a correlation between the \civ{}-\cii{} and \mgii{}-\cii{} velocity shifts. We calculated the Pearson correlation coefficient $\rho$ for the 25 values and found the correlation to be significant with an $\rho=0.71$ and a p-value of $p=7\cdot10^{-5}$.

While the \civ{}-\mgii{} velocity shifts of our high-redshift quasars show a strong anti-correlation with the \civ{} FWHM, we do not find an analogous anti-correlation between the \mgii{}-\cii{} velocity shift and the \mgii{} FWHM ($\rho=0.18$, $p=0.36$). 
As \mgii{}-based black hole masses are based on the assumption that the line traces virialized gas, it is reassuring that the \mgii{} FWHM does not correlate with the \mgii{}-\cii{} velocity shift (see Figure\,\ref{fig:mgiiciivshift_fwhm}).

The \ciii{}-\cii{} velocity shifts also show a positive trend with the \mgii{}-\cii{} velocity shifts and seem to track the \civ{}-\cii{} velocity shifts closely. \citet{Richards2011} also noted that the \ciii{} velocity shifts track the \civ{} velocity shifts if measured in the same reference system. The authors discuss that the \ciii{} complex velocity shift is partly due to a relative flux change of the \siiii{} and \ciii{} lines. In their Figure\,11 they show that the strength of the \siiii{} line increases with \civ{} blueshift, leading to a stronger velocity blueshift of the entire \ciii{} complex \citep[see also][]{Shen2016}.
Unfortunately, our data did not allow to resolve the different contributions of the \siiii{} and \ciii{} lines. Hence, we cannot distinguish between real \ciii{} velocity shifts and the effect of \siiii{} to \ciii{} line ratio changes in our sample. 



\subsubsection{Broad line velocity shifts in relation to other quasar properties}\label{sec:bel_qso_prop}

\begin{figure}
    \centering
    \includegraphics[width=0.50\textwidth]{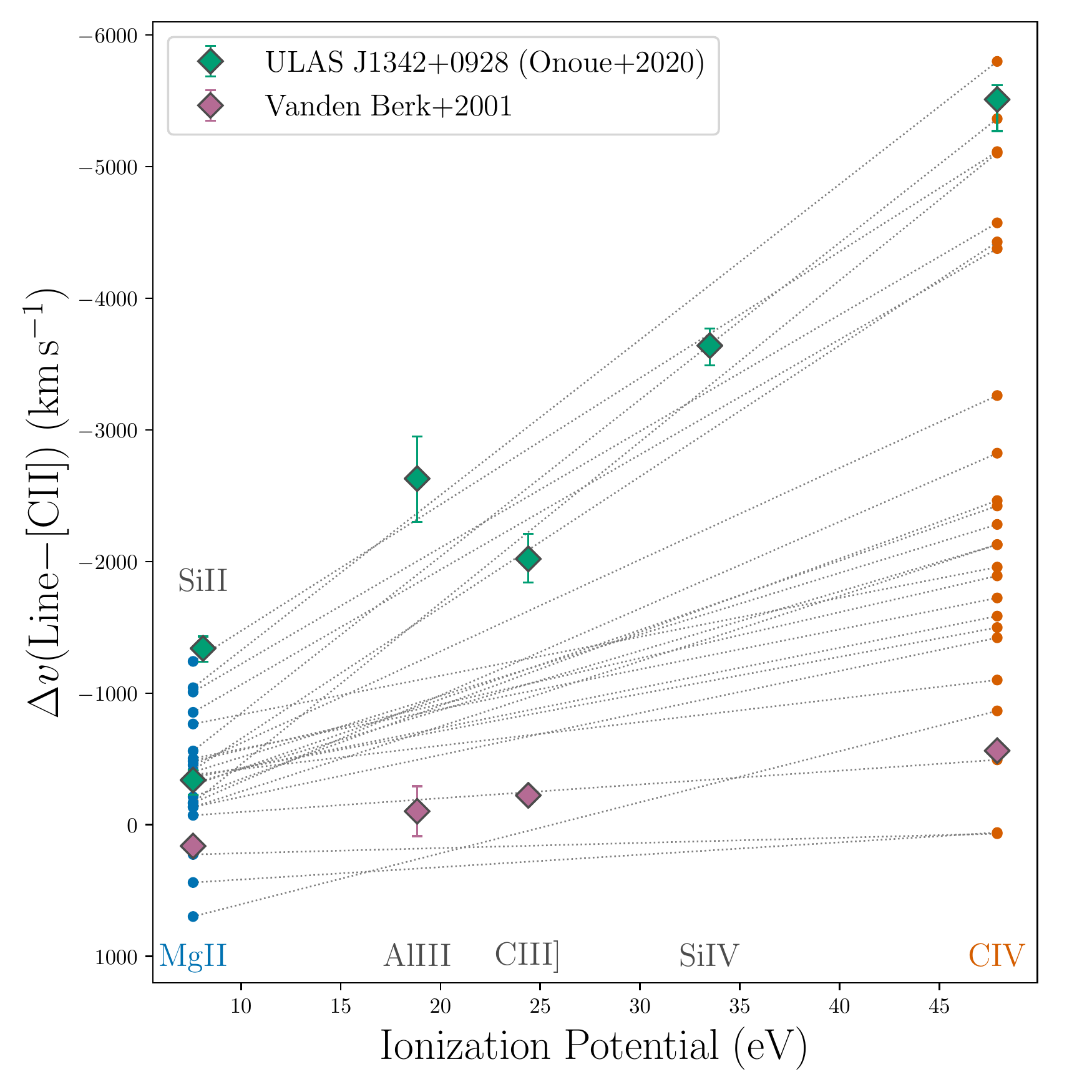}
    \caption{Velocity shifts of emission lines as a function of ionization potential. We only show show \mgii{}-\ciilong{} and \civ{}-\ciilong{} velocity shifts of quasars of our sample with solid blue and orange points. Measurements of the same quasar are connected with a dotted grey line. We compare these measurements with various velocity shifts from ULAS~J1342{+}0928 \citep{Onoue2020}, which were also measured with respect to \ciilong{} in green. The purple data points show velocity shifts with respect to {\sc{[O iii]}} 5007\AA\ from the low-redshift quasar composite of \citet{VandenBerk2001}. Quasars from our sample show a large range of blueshifts.}
    \label{fig:vshifts_ionization}
\end{figure}

The presented X-SHOOTER/ALMA quasar sample spans a rather narrow range of high bolometric luminosity ($46.67\lesssim \log(L_{\rm{bol}}/(\rm{erg}\,\rm{s}^{-1}))\lesssim 47.67$, with the exception of SDSS~J0100{+}2802).
At these luminosities we have measured a large range of \civ{}-\mgii{} velocity shifts ($-5000$ to $0\,\rm{km}\,\rm{s}^{-1}$). We have tested whether the continuum luminosity at $3000$\,\AA\ and the \civ{}-\mgii{} velocity shift is correlated, but did not find any evidence for it ($\rho=0.015$, $p=0.94$, also see Figure\,\ref{fig:civmgii_shift_L3000} in Appendix\,\ref{sec:app_bel_correlations}).
	
Conversely, previous work on large samples of lower redshift quasars \citep{Richards2011, Shen2016} found significant anti-correlations between the velocity shifts of the high-ionization \heii{}, \siiv{}, and \civ{} lines (with respect to \mgii{}) and the quasar luminosity, albeit in lower luminosity samples and over a larger luminosity range \citep[][$44\lesssim \log(L_{\rm{bol}}/(\rm{erg}\,\rm{s}^{-1}))\lesssim 46.5$]{Shen2016}. 
These anti-correlations of the velocity shifts with luminosity have been associated with accretion driven dynamical and/or radiative processes in the BLR that could be responsible for the observed blueshifts \citep{Richards2011}.

The observational bias towards mostly luminous quasars, propagates to our observed distribution of Eddington luminosity ratios as calculated from \mgii{}-based black hole masses. Therefore, our sample shows mostly high Eddington luminosity ratios ($0.1 \lesssim L_{\rm{bol}}/L_{\rm{Edd}} \lesssim $ a few). 
\citet{Coatman2016} suggested that quasars with strong \civ{}-\mgii{} blueshifts are indicative of high Eddington luminosity ratios. 
We do not find any indication for a significant correlation between the \civ{}-\mgii{}, \civ{}-\cii{}, and \mgii{}-\cii{} velocity shifts and the Eddington luminosity ratio in our sample.
Overall we can summarize that broad lines of luminous high-Eddington-luminosity-ratio quasars exhibit a large range of observed velocity (blue)shifts \citep[also see][]{Mazzucchelli2017}.

\citet{VandenBerk2001} reported a correlation between ionization potential and line velocity shifts based on their low-redshift composite spectrum of $z\sim1$ SDSS quasars. Based on a deep GNIRS spectrum of ULAS~J1342{+}0928, \citet{Onoue2020} highlighted that the amounts of line blueshifts are proportional to their respective ionization potentials and their values are much larger compared to the \citet{VandenBerk2001} composite spectrum. In Figure\,\ref{fig:vshifts_ionization} we show  \mgii{}-\cii{} and \civ{}-\cii{} velocity shifts for our quasars in comparison to both, the \citet{VandenBerk2001} composite and ULAS~J1342{+}0928 \citep{Onoue2020}. The velocity shifts from our sample span the entire range between the composite spectrum and ULAS~J1342{+}0928.




There are 12 radio-quiet quasars in our high-redshift quasar sample. So far no quasar is confirmed to be radio-loud, while the radio observations are not yet deep enough to classify the remaining 24 objects. 
Work on low-redshift ($z\le 1$) type-I AGN \citep{Sulentic2007} showed that radio-quiet sources were associated with stronger \civ{} blueshifts, whereas their sample of radio-loud sources showed \civ{} velocity shifts around $0\,\rm{km}/\rm{s}$. Their radio-quiet AGN also showed a first tentative correlation with \civ{} blueshift and \civ{} FWHM not seen in their radio-loud counterparts. We have discussed this correlation in our high-redshift sample (see Figure\,\ref{fig:civ_context}, right panel). Similar trends with radio loudness were seen in the SDSS quasar sample \citep{Richards2002}, in which radio-loud quasars show on average smaller \civ{} blueshifts. 
If the properties of low-redshift AGN and quasars are any guide, the strong \civ{}-\mgii{} blueshifts in high-redshift quasars go hand in hand with the large observed fraction of radio-quiet objects. However, currently no statistically significant sample of radio-loud objects with NIR spectroscopy at $z=6-7$ exists to confirm these correlations with their \civ{} emission line properties. 


\subsubsection{Discussion}\label{sec:vshift_discussion}
The large \civ{} blueshifts seen in some quasars at lower redshifts emerge as a prominent feature in luminous $z\ge6$ quasar samples. The common interpretation explains these blueshifts in the context of an out-flowing, potentially non-virialized component of the \civ{} line \citep[e.g.][]{Richards2002, Richards2011, Mazzucchelli2017}. This is one reason why the validity of \civ{} based BH mass estimates has been scrutinized \citep[e.g.][]{MejiaRestrepo2018a}. 
The X-SHOOTER/ALMA sample not only shows a large range of \civ{} blueshifts, but we also find the \mgii{} line, on average, to be blueshifted with respect to the \cii{} emission of the quasar host galaxy. This has been observed in individual objects or small samples at $z>6$ \citep{WangRan2016, Venemans2016, Mazzucchelli2017, Onoue2020} and was recently reported for quasars at $z\sim4.8$ \citep{Nguyen2020}. However, our larger sample size highlights the high frequency of these blueshifts in luminous, reionization-era quasars. 
Furthermore, we discovered a significant correlation between the \civ{}-\cii{} and \mgii{}-\cii{} velocity shifts, strongly suggesting a common origin likely tied to the physical conditions of the BLR and the accretion process. 
While we could not find correlations of the velocity shifts with either the quasar's luminosity or Eddington luminosity ratio, such correlations have been observed in lower redshift samples \citep{Richards2011}. 
\citet{Shen2016} found the quasar luminosity to be strongly correlated with the \heii{}, \civ{} and \siiv{} blueshifts, whereas they note that the \mgii{} velocity shift is luminosity independent. 
In their sample of $2 < z < 2.7$ quasars \citet{Coatman2016} observed that large \civ{} blueshifts are associated with high Eddington luminosity ratios. 
Indeed, studies of very luminous $z=2-4$ quasars \citep[WISSH quasars sample,][]{Bischetti2017, Vietri2018} find correlations of the \civ{} blueshift with both bolometric luminosity and Eddington luminosity ratio. 
As bolometric luminosity and Eddington luminosity ratio are related quantities, the authors conduct a more detailed analysis and conclude that the fundamental variable is the luminosity rather than the Eddington ratio. In addition, they observe a clear correlation between the \civ{} blueshift and the UV-to-X-ray continuum slope ($\alpha_{\rm{OX}}$), as discussed in \citet{Richards2011}.
Thus quasars with large \civ{} blueshifts show a less ionizing spectral energy distribution dominated by UV rather than by X-ray emission. Such a spectrum would naturally be able to produce winds through radiation line driving \citep{Murray1995a}. 

Let us consider that the broad emission originates from a wind, which emerges from the accretion disc in helical streamlines driven by radiation pressure \citep{Murray1995a}. The wind moving toward the observer is responsible for the blueshifted emission, while the receding side is blocked by the optically-thick accretion disk. According to this model the innermost streamlines with gas in the highest ionization states have the largest rotational and radial velocities, which can naturally explain the larger FWHM of \civ{} with respect to \mgii{} as well as the stratification of the BLR in reverberation mapping observations. The authors also predict that high-ionization lines should be blueshifted relative to low-ionization lines.
While this model might have its shortcomings, it offers a compelling picture to explain the observations of the blueshifted quasar emission lines, especially regarding their correlation with ionization potential \citep{Tytler1992, McIntosh1999b, VandenBerk2001, Onoue2020}. 

Emission line velocity shifts in an axis-symmetric wind model naturally open up discussions on orientation measures based on the \civ{}, \mgii{} and other quasar emission lines \citep[e.g.][]{Richards2002, Meyer2019c, Yong2020}. 
However, the complex nature of the BLR and the limited amount of observational data sets make it hard to disentangle orientation effects, variations in the physical conditions of the BLR, and biases of the quasar samples. 

It is a worthwhile endeavor to expand the presented analysis to quasars at lower redshift and lower luminosities to 
further investigate the kinematic information provided by emission line velocity shifts as they may provide a way to constrain the quasar's orientation.

\section{Summary}\label{sec:conclusion}

We presented new and archival X-SHOOTER near-infrared spectroscopy for \Nsamp{} quasars at $5.78 < z < 7.54$, of which 34 have complimentary \cii{} detection with ALMA. We have discussed the spectral modeling in detail and provide a machine-readable master table on-line, which includes all measured and derived quantities. An overview of that table is given in Appendix\,\ref{sec:master_table}.

\begin{itemize}
\item We have investigated the systematic effects on the \mgii{} line and \feii{} pseudo-continuum properties inferred from different iron templates. We specifically compared the \citetalias{Vestergaard2001} and \citetalias{Tsuzuki2006} iron templates. The \citetalias{Vestergaard2001} template does not include \feii{} emission beneath the \mgii{}, whereas the \citetalias{Tsuzuki2006} template does. As a consequence the \mgii{} flux and FWHM are overestimated using the \citetalias{Vestergaard2001} template and the iron contribution is underestimated. Any inclusion of the \feii{} emission beneath the \mgii{} line leads to a more realistic estimate of the spectral properties, e.g. for the calculation of the \feii{}/\mgii{} flux ratio. 

\item
For estimating SMBH masses care has to be taken to measure the \mgii{} FWHM or $\sigma$ using the same iron template, which was used to establish the single-epoch virial estimators. We provide a relation, which allows to scale the \feii{}/\mgii{} flux ratios as measured with the \citetalias{Vestergaard2001} template up to measurements with the \citetalias{Tsuzuki2006} template.

\item
We analyzed the \feii{}/\mgii{} ratio, a proxy for the BLR iron enrichment for our sample and found a median value of $F_{\rm{FeII}}/F_{\rm{MgII}} = 6.31_{-2.29}^{+2.49}$, where uncertainties give the 16 to 84 percentile region. We conclude that the BLRs of all quasars presented in this study are already enriched in iron.


\item We investigated the properties of the broad emission lines with a focus on velocity shifts and the broad \civ{} and \mgii{} lines. We find that high-redshift quasars show a large range of \civ{}-\mgii{} velocity shifts with an emphasis on large blueshifts, which sets them apart from a luminosity matched sample of $1.52 < z < 2.2$ quasars. We calculate median \civ{}-\mgii{} velocity shift of $\sim-1800\,\rm{km}\,\rm{s}^{-1}$, whereas the low-redshift quasars have a median of  $\sim-800\,\rm{km}\,\rm{s}^{-1}$. We further find the \mgii{} line to be often blueshifted with respect to the \cii{} of the host galaxy measured with ALMA. The velocity shift distribution shows a clear peak around the median,  $\Delta v(\rm{MgII{-}[CII]}) =-390.61_{-455.34}^{256.02}\,\rm{km}\,\rm{s}^{-1}$.

\item We find the velocity shifts of \civ{} and \mgii{}, both with respect to the host galaxy \cii{} line, to be significantly correlated, indicating a common origin likely tied to the physical properties of the BLR and the accretion process. 

\item We did not find evidence for correlations between between the line velocity shifts and the bolometric luminosity or the Eddington ratio, keeping in mind that our sample is dominated by luminous, high Eddington luminosity ratio quasars.
\end{itemize}

\subsection{Do quasar emission line properties evolve with redshift?}
As we discover more and more high-redshift quasars deep within the era of reionization, it would not be surprising, if we saw their emission line properties evolve. Yet, quasar spectra at $z\sim6$ bear surprising resemblance to their low-redshift ($z\approx1-2$) counterparts \citep{Shen2019a}. 
Probing quasars at even higher redshifts than \citet{Shen2019a} our analysis takes a close look at the \civ{} and \mgii{} line as well as the \feii{} contribution. 
As seen from Figure\,\ref{fig:feII_ratios} our median \feii{}/\mgii{} flux ratio agrees well with measurements at lower redshifts ($z=3-5$), showing no significant redshift evolution.
In Figure\,\ref{fig:mgiiblueshift_ciiredshift_histogramm} our data show significant blueshifts between the measurements of the \mgii{}-\cii{} lines. Yet, this is also not unique to $z\gtrsim6$ quasars as similar results are found at $z\sim4.8$ \citep{Nguyen2020}. 
Many quasars in our sample also show large \civ{}-\mgii{} velocity blueshifts ($\Delta v(\rm{CIV}{-}\rm{MgII}) < -2000\,\rm{km}\,\rm{s}^{-1}$) that correlate with smaller \civ{} EW and larger \civ FWHM (see Figure\,\ref{fig:civ_context}). Judging from this figure we can always identify low-redshift ($z=1.52-2.2$) quasars occupying the same region of the \civ{}-\mgii{}/EW or the \civ{}-\mgii{}/FWHM parameter space.
However, the average sample \civ{}-\mgii{} velocity shift does seem to decrease significantly at the highest redshifts (Figure\,\ref{fig:blueshift_redshift_histogramm}). 
This trend, first reported and discussed in \citet{Meyer2019c}, is supported by our analysis on a larger high-redshift sample. Yet, it is unclear whether this apparent redshift evolution presents a physical change in the BLR conditions or a selection bias affecting the highest redshift quasars.
%
%
The advent of the James Webb Space telescope will open up possibilities to probe the rest-frame optical emission of high-redshift quasars, providing access to the hydrogen Balmer lines. These measurements will be instrumental for a comprehensive comparison of high-redshift quasars with the low-redshift quasar population.

\appendix 

\section{Additional tables} \label{sec:app_additional_tables}

We present additional tables detailing further properties of the X-SHOOTER/ALMA sample in this section. Table\,\ref{table:ciiisiiv} includes measurements on the \ciii{} and \siiv{} lines and Table\,\ref{table:sample_2} summarizes additional information on the quasar fits, their continuum measurements and information on classifications.

{\movetabledown=2.0in 
\tabletypesize{\footnotesize} 
\begin{deluxetable*}{lcccccc}
\tablecaption{Properties of the \ciii{} and \siiv{} emission lines \label{table:ciiisiiv}}
\tablecolumns{7}\tablehead{\colhead{Quasar Name} &\colhead{$z_{\rm{CIII]}}$} & \colhead{$\Delta v(\rm{CIII]}{-}\rm{[CII]})$} &\colhead{$z_{\rm{SiIV}}$} & \colhead{FWHM$_{\rm{SiIV}}$} &\colhead{EW$_{\rm{SiIV}}$} &\colhead{$\Delta v({\rm{SiIV}}{-}\rm{[CII]})$} \\ 
\nocolhead{} &\nocolhead{} &\colhead{$(\rm{km}\,\rm{s}^{-1})$} &\nocolhead{} &\colhead{$(\rm{km}\,\rm{s}^{-1})$} &\colhead{(\AA)} &\colhead{$(\rm{km}\,\rm{s}^{-1})$} 
} 
\startdata 
PSO~J004.3936+17.0862 & ${5.805}_{-0.008}^{+0.009}$ & ${-526.67}_{-364.58}^{+364.58}$ & \dots & \dots & \dots & \dots\\ 
PSO~J007.0273+04.9571 & ${5.999}_{-0.006}^{+0.005}$ & ${-117.43}_{-246.70}^{+246.70}$ & \dots & \dots & \dots & \dots\\ 
PSO~J009.7355--10.4316 & \dots & \dots & \dots & \dots & \dots & \dots\\ 
PSO~J011.3898+09.0324 & \dots & \dots & \dots & \dots & \dots & \dots\\ 
VIK~J0046--2837 & \dots & \dots & \dots & \dots & \dots & \dots\\ 
SDSS~J0100+2802 & \dots & \dots & \dots & \dots & \dots & \dots\\ 
VIK~J0109--3047 & \dots & \dots & \dots & \dots & \dots & \dots\\ 
PSO~J036.5078+03.0498 & \dots & \dots & ${6.45}_{-0.00}^{+0.00}$ & ${5137.963}_{-299.849}^{+333.895}$ & ${5.17}_{-0.41}^{+0.44}$ & ${-3542.80}_{-132.91}^{+146.05}$\\ 
VIK~J0305--3150 & \dots & \dots & \dots & \dots & \dots & \dots\\ 
PSO~J056.7168--16.4769 & ${5.978}_{-0.001}^{+0.001}$ & ${456.87}_{-58.51}^{+58.51}$ & \dots & \dots & \dots & \dots\\ 
PSO~J065.4085--26.9543 & \dots & \dots & \dots & \dots & \dots & \dots\\ 
PSO~J065.5041--19.4579 & \dots & \dots & \dots & \dots & \dots & \dots\\ 
SDSS~J0842+1218 & \dots & \dots & \dots & \dots & \dots & \dots\\ 
SDSS~J1030+0524 & \dots & \dots & \dots & \dots & \dots & \dots\\ 
PSO~J158.69378--14.42107 & \dots & \dots & \dots & \dots & \dots & \dots\\ 
PSO~J159.2257--02.5438 & \dots & \dots & \dots & \dots & \dots & \dots\\ 
SDSS~J1044--0125 & ${5.781}_{-0.007}^{+0.009}$ & ${-161.50}_{-362.90}^{+362.90}$ & \dots & \dots & \dots & \dots\\ 
VIK~J1048--0109 & \dots & \dots & \dots & \dots & \dots & \dots\\ 
ULAS~J1120+0641 & ${7.075}_{-0.001}^{+0.002}$ & ${-355.50}_{-55.11}^{+55.11}$ & ${7.05}_{-0.00}^{+0.00}$ & ${5834.922}_{-151.979}^{+158.901}$ & ${9.48}_{-0.31}^{+0.31}$ & ${-1252.58}_{-53.36}^{+49.92}$\\ 
ULAS~J1148+0702 & \dots & \dots & \dots & \dots & \dots & \dots\\ 
PSO~J183.1124+05.0926 & \dots & \dots & \dots & \dots & \dots & \dots\\ 
SDSS~J1306+0356 & ${6.033}_{-0.001}^{+0.001}$ & ${20.94}_{-58.47}^{+58.47}$ & \dots & \dots & \dots & \dots\\ 
ULAS~J1319+0950 & ${6.064}_{-0.002}^{+0.002}$ & ${-2978.86}_{-91.20}^{+91.20}$ & \dots & \dots & \dots & \dots\\ 
ULAS~J1342+0928 & ${7.508}_{-0.004}^{+0.004}$ & ${-1136.24}_{-150.46}^{+150.46}$ & ${7.36}_{-0.01}^{+0.01}$ & ${9094.569}_{-847.096}^{+736.015}$ & ${10.55}_{-0.85}^{+0.73}$ & ${-6309.26}_{-427.67}^{+519.08}$\\ 
CFHQS~J1509--1749 & ${6.074}_{-0.004}^{+0.006}$ & ${-2066.44}_{-214.92}^{+214.92}$ & \dots & \dots & \dots & \dots\\ 
PSO~J231.6576--20.8335 & \dots & \dots & \dots & \dots & \dots & \dots\\ 
PSO~J239.7124--07.4026 & \dots & \dots & \dots & \dots & \dots & \dots\\ 
PSO~J308.0416--21.2339 & \dots & \dots & \dots & \dots & \dots & \dots\\ 
SDSS~J2054--0005 & ${6.029}_{-0.023}^{+0.030}$ & ${-417.21}_{-1123.13}^{+1123.13}$ & \dots & \dots & \dots & \dots\\ 
CFHQS~J2100--1715 & \dots & \dots & \dots & \dots & \dots & \dots\\ 
PSO~J323.1382+12.2986 & ${6.585}_{-0.001}^{+0.001}$ & ${-103.01}_{-41.41}^{+41.41}$ & ${6.58}_{-0.00}^{+0.00}$ & ${4122.430}_{-257.856}^{+255.489}$ & ${10.16}_{-0.61}^{+0.66}$ & ${-331.87}_{-64.41}^{+69.17}$\\ 
VIK~J2211--3206 & \dots & \dots & \dots & \dots & \dots & \dots\\ 
CFHQS~J2229+1457 & \dots & \dots & \dots & \dots & \dots & \dots\\ 
PSO~J340.2041--18.6621 & ${5.998}_{-0.001}^{+0.001}$ & ${-117.07}_{-33.30}^{+32.11}$ & \dots & \dots & \dots & \dots\\ 
SDSS~J2310+1855 & \dots & \dots & \dots & \dots & \dots & \dots\\ 
VIK~J2318--3029 & \dots & \dots & \dots & \dots & \dots & \dots\\ 
VIK~J2348--3054 & ${6.944}_{-0.006}^{+0.006}$ & ${1643.12}_{-214.75}^{+214.75}$ & \dots & \dots & \dots & \dots\\ 
PSO~J359.1352--06.3831 & \dots & \dots & \dots & \dots & \dots & \dots\\ 
\enddata 
\end{deluxetable*} 
}

{\movetabledown=2.0in 
\begin{deluxetable*}{lccccccc}
\rotate 
\tabletypesize{\footnotesize} 
\tablecaption{Additional properties of the X-SHOOTER/ALMA quasar sample. \label{table:sample_2}}
\tablecolumns{8}\tablehead{\colhead{Quasar Name} &\colhead{Classification} & \colhead{Class. Reference} &\colhead{Power law slope} &\colhead{$M_{1450}$} & \colhead{$L_{1450}$} &\colhead{$L_{3000}$} &\colhead{$L_{\rm{bol}}$} \\ 
\nocolhead{} &\nocolhead{} &\nocolhead{} &\nocolhead{} &\colhead{(AB mag)} &\multicolumn{3}{c}{($10^{46}\,\rm{erg}\,\rm{s}^{-1}$)} 
} 
\startdata 
PSO~J004.3936+17.0862 & Y & f, This work & ${-2.03}_{-0.09}^{+0.09}$ & ${-25.95}_{+0.04}^{-0.05}$ & ${2.15}_{-0.09}^{+0.09}$ & ${1.21}_{-0.04}^{+0.04}$ & ${6.21}_{-0.21}^{+0.20}$\\ 
PSO~J007.0273+04.9571 & \dots & \dots & ${-1.17}_{-0.09}^{+0.09}$ & ${-26.51}_{+0.06}^{-0.05}$ & ${3.61}_{-0.19}^{+0.18}$ & ${3.89}_{-0.17}^{+0.14}$ & ${20.05}_{-0.88}^{+0.70}$\\ 
PSO~J009.7355--10.4316 & D & This work & \dots & ${-26.03}_{+0.04}^{-0.04}$ & ${2.32}_{-0.08}^{+0.09}$ & ${3.75}_{-0.04}^{+0.04}$ & ${19.32}_{-0.21}^{+0.21}$\\ 
PSO~J011.3898+09.0324 & \dots & \dots & ${-1.57}_{-0.05}^{+0.06}$ & ${-25.87}_{+0.02}^{-0.02}$ & ${2.00}_{-0.04}^{+0.05}$ & ${1.59}_{-0.03}^{+0.03}$ & ${8.21}_{-0.18}^{+0.16}$\\ 
VIK~J0046--2837 & D & This work & \dots & ${-25.09}_{+0.24}^{-0.19}$ & ${0.97}_{-0.19}^{+0.18}$ & ${1.63}_{-0.09}^{+0.09}$ & ${8.39}_{-0.44}^{+0.44}$\\ 
SDSS~J0100+2802 & Y & b & ${-1.55}_{-0.00}^{+0.00}$ & ${-29.02}_{+0.00}^{-0.00}$ & ${36.51}_{-0.04}^{+0.04}$ & ${29.50}_{-0.03}^{+0.03}$ & ${151.93}_{-0.16}^{+0.16}$\\ 
VIK~J0109--3047 & \dots & \dots & ${-1.10}_{-0.07}^{+0.07}$ & ${-25.41}_{+0.03}^{-0.03}$ & ${1.32}_{-0.04}^{+0.04}$ & ${1.49}_{-0.04}^{+0.04}$ & ${7.66}_{-0.20}^{+0.21}$\\ 
PSO~J036.5078+03.0498 & \dots & \dots & ${-1.66}_{-0.01}^{+0.01}$ & ${-27.15}_{+0.01}^{-0.01}$ & ${6.51}_{-0.03}^{+0.03}$ & ${4.83}_{-0.03}^{+0.03}$ & ${24.89}_{-0.15}^{+0.14}$\\ 
VIK~J0305--3150 & \dots & \dots & ${-1.43}_{-0.03}^{+0.03}$ & ${-25.91}_{+0.01}^{-0.01}$ & ${2.07}_{-0.03}^{+0.03}$ & ${1.83}_{-0.02}^{+0.02}$ & ${9.41}_{-0.12}^{+0.11}$\\ 
PSO~J056.7168--16.4769 & pDLA & f, g & ${-1.71}_{-0.03}^{+0.03}$ & ${-26.26}_{+0.02}^{-0.02}$ & ${2.86}_{-0.04}^{+0.04}$ & ${2.04}_{-0.02}^{+0.02}$ & ${10.52}_{-0.12}^{+0.12}$\\ 
PSO~J065.4085--26.9543 & D & This work & \dots & ${-26.94}_{+0.01}^{-0.01}$ & ${5.38}_{-0.05}^{+0.05}$ & ${3.71}_{-0.11}^{+0.10}$ & ${19.10}_{-0.56}^{+0.50}$\\ 
PSO~J065.5041--19.4579 & BAL, D & This work & \dots & ${-26.11}_{+0.03}^{-0.03}$ & ${2.51}_{-0.07}^{+0.07}$ & ${3.43}_{-0.08}^{+0.08}$ & ${17.67}_{-0.41}^{+0.41}$\\ 
SDSS~J0842+1218 & \dots & \dots & ${-1.44}_{-0.02}^{+0.02}$ & ${-26.69}_{+0.01}^{-0.01}$ & ${4.26}_{-0.05}^{+0.05}$ & ${3.72}_{-0.04}^{+0.04}$ & ${19.17}_{-0.21}^{+0.22}$\\ 
SDSS~J1030+0524 & \dots & \dots & ${-1.25}_{-0.04}^{+0.04}$ & ${-26.76}_{+0.02}^{-0.02}$ & ${4.56}_{-0.07}^{+0.08}$ & ${4.60}_{-0.07}^{+0.07}$ & ${23.70}_{-0.34}^{+0.34}$\\ 
PSO~J158.69378--14.42107 & Y & f & ${-0.73}_{-0.06}^{+0.06}$ & ${-27.07}_{+0.03}^{-0.03}$ & ${6.07}_{-0.15}^{+0.16}$ & ${9.05}_{-0.21}^{+0.22}$ & ${46.61}_{-1.08}^{+1.13}$\\ 
PSO~J159.2257--02.5438 & \dots & \dots & ${-1.27}_{-0.04}^{+0.05}$ & ${-26.47}_{+0.02}^{-0.02}$ & ${3.48}_{-0.07}^{+0.06}$ & ${3.45}_{-0.06}^{+0.06}$ & ${17.78}_{-0.30}^{+0.31}$\\ 
SDSS~J1044--0125 & BAL & d & ${-1.61}_{-0.06}^{+0.06}$ & ${-27.16}_{+0.03}^{-0.03}$ & ${6.58}_{-0.20}^{+0.21}$ & ${5.07}_{-0.08}^{+0.09}$ & ${26.09}_{-0.43}^{+0.45}$\\ 
VIK~J1048--0109 & \dots & \dots & ${-1.62}_{-0.08}^{+0.07}$ & ${-26.20}_{+0.03}^{-0.03}$ & ${2.71}_{-0.08}^{+0.08}$ & ${2.07}_{-0.06}^{+0.06}$ & ${10.66}_{-0.32}^{+0.30}$\\ 
ULAS~J1120+0641 & \dots & \dots & ${-1.25}_{-0.01}^{+0.01}$ & ${-26.40}_{+0.00}^{-0.00}$ & ${3.27}_{-0.01}^{+0.01}$ & ${3.31}_{-0.01}^{+0.01}$ & ${17.05}_{-0.07}^{+0.07}$\\ 
ULAS~J1148+0702 & \dots & \dots & ${-1.17}_{-0.03}^{+0.03}$ & ${-26.31}_{+0.02}^{-0.01}$ & ${3.00}_{-0.04}^{+0.04}$ & ${3.21}_{-0.04}^{+0.04}$ & ${16.51}_{-0.18}^{+0.21}$\\ 
PSO~J183.1124+05.0926 & pDLA & e & ${-1.46}_{-0.03}^{+0.03}$ & ${-26.87}_{+0.02}^{-0.01}$ & ${5.01}_{-0.07}^{+0.06}$ & ${4.33}_{-0.05}^{+0.06}$ & ${22.30}_{-0.28}^{+0.30}$\\ 
SDSS~J1306+0356 & \dots & \dots & ${-1.51}_{-0.02}^{+0.02}$ & ${-26.70}_{+0.01}^{-0.01}$ & ${4.29}_{-0.03}^{+0.04}$ & ${3.56}_{-0.02}^{+0.02}$ & ${18.34}_{-0.13}^{+0.13}$\\ 
ULAS~J1319+0950 & \dots & \dots & ${-1.68}_{-0.01}^{+0.01}$ & ${-26.80}_{+0.00}^{-0.00}$ & ${4.71}_{-0.02}^{+0.02}$ & ${3.45}_{-0.01}^{+0.01}$ & ${17.75}_{-0.06}^{+0.07}$\\ 
ULAS~J1342+0928 & \dots & \dots & ${-1.36}_{-0.00}^{+0.01}$ & ${-26.64}_{+0.00}^{-0.00}$ & ${4.07}_{-0.01}^{+0.02}$ & ${3.77}_{+0.00}^{+0.03}$ & ${19.43}_{+0.00}^{+0.13}$\\ 
CFHQS~J1509--1749 & \dots & \dots & ${-0.93}_{-0.02}^{+0.02}$ & ${-26.56}_{+0.01}^{-0.01}$ & ${3.76}_{-0.05}^{+0.04}$ & ${4.84}_{-0.05}^{+0.04}$ & ${24.95}_{-0.27}^{+0.23}$\\ 
PSO~J231.6576--20.8335 & \dots & \dots & ${-1.72}_{-0.05}^{+0.06}$ & ${-27.07}_{+0.03}^{-0.03}$ & ${6.06}_{-0.14}^{+0.14}$ & ${4.31}_{-0.10}^{+0.10}$ & ${22.21}_{-0.53}^{+0.53}$\\ 
PSO~J239.7124--07.4026 & BAL & f & ${-1.33}_{-0.03}^{+0.03}$ & ${-27.07}_{+0.02}^{-0.01}$ & ${6.04}_{-0.09}^{+0.08}$ & ${5.74}_{-0.09}^{+0.09}$ & ${29.54}_{-0.45}^{+0.45}$\\ 
PSO~J308.0416--21.2339 & \dots & \dots & ${-0.77}_{-0.02}^{+0.02}$ & ${-26.27}_{+0.01}^{-0.01}$ & ${2.89}_{-0.02}^{+0.02}$ & ${4.18}_{-0.04}^{+0.04}$ & ${21.52}_{-0.21}^{+0.21}$\\ 
SDSS~J2054--0005 & \dots & \dots & ${-1.38}_{-0.07}^{+0.07}$ & ${-26.15}_{+0.05}^{-0.04}$ & ${2.60}_{-0.11}^{+0.09}$ & ${2.37}_{-0.07}^{+0.08}$ & ${12.23}_{-0.35}^{+0.43}$\\ 
CFHQS~J2100--1715 & Y, D & f, This work & \dots & ${-24.63}_{+0.05}^{-0.05}$ & ${0.64}_{-0.03}^{+0.03}$ & ${0.93}_{-0.02}^{+0.02}$ & ${4.77}_{-0.11}^{+0.11}$\\ 
PSO~J323.1382+12.2986 & \dots & \dots & ${-1.65}_{-0.02}^{+0.02}$ & ${-26.89}_{+0.01}^{-0.01}$ & ${5.14}_{-0.05}^{+0.05}$ & ${3.85}_{-0.04}^{+0.04}$ & ${19.81}_{-0.21}^{+0.21}$\\ 
VIK~J2211--3206 & BAL & This work & ${-1.36}_{-0.07}^{+0.06}$ & ${-27.09}_{+0.03}^{-0.03}$ & ${6.15}_{-0.16}^{+0.18}$ & ${5.72}_{-0.11}^{+0.11}$ & ${29.45}_{-0.59}^{+0.59}$\\ 
CFHQS~J2229+1457 & \dots & \dots & ${-1.20}_{-0.16}^{+0.16}$ & ${-24.43}_{+0.08}^{-0.07}$ & ${0.53}_{-0.04}^{+0.04}$ & ${0.55}_{-0.03}^{+0.03}$ & ${2.86}_{-0.15}^{+0.17}$\\ 
PSO~J340.2041--18.6621 & BAL & This work & ${-1.36}_{-0.04}^{+0.04}$ & ${-26.23}_{+0.02}^{-0.02}$ & ${2.78}_{-0.05}^{+0.05}$ & ${2.58}_{-0.03}^{+0.03}$ & ${13.30}_{-0.17}^{+0.16}$\\ 
SDSS~J2310+1855 & pDLA & c & ${-1.16}_{-0.03}^{+0.04}$ & ${-27.22}_{+0.02}^{-0.02}$ & ${6.94}_{-0.15}^{+0.14}$ & ${7.53}_{-0.09}^{+0.10}$ & ${38.78}_{-0.48}^{+0.51}$\\ 
VIK~J2318--3029 & \dots & \dots & ${-1.11}_{-0.04}^{+0.03}$ & ${-26.11}_{+0.02}^{-0.02}$ & ${2.49}_{-0.04}^{+0.04}$ & ${2.80}_{-0.04}^{+0.04}$ & ${14.44}_{-0.22}^{+0.21}$\\ 
VIK~J2348--3054 & BAL & a & ${-1.60}_{-0.07}^{+0.07}$ & ${-25.79}_{+0.03}^{-0.03}$ & ${1.86}_{-0.05}^{+0.05}$ & ${1.45}_{-0.04}^{+0.04}$ & ${7.46}_{-0.22}^{+0.20}$\\ 
PSO~J359.1352--06.3831 & \dots & \dots & ${-0.98}_{-0.03}^{+0.03}$ & ${-26.62}_{+0.02}^{-0.02}$ & ${3.99}_{-0.05}^{+0.06}$ & ${4.92}_{-0.06}^{+0.06}$ & ${25.35}_{-0.31}^{+0.28}$\\ 
\enddata 
\tablerefs{Quasar classification: Broad absorption line quasar (BAL), proximate damped Lyman-$\alpha$ absorber (pDLA), young quasars (Y), quasar continuum not described by power law (D). The classification references are: a=\citet{DeRosa2014}, b=\citet{Eilers2017a}, c=\citet{DOdorico2018}, d=\citet{Shen2019a}, e=\citet{Banados2019}, f=\citet{Eilers2020a}, g=\citet{Davies2020a} }
\end{deluxetable*} 
}

\section{Notes on the spectral modeling of individual quasars}\label{sec:app_specfitnotes}

In this section we provide additional information on the model fits of individual quasars. 
As the redshift and the signal-to-noise of the X-SHOOTER spectrum varies from object to object additional assumptions and limitations were necessary to provide an adequate fit. 

For example, in a range of spectra we do not use the fit weights, which are taken to be the squared inverse flux uncertainties, for the continuum model. In these spectra the continuum fit was dominated by higher signal-to-noise in the continuum regions around the \mgii{} line. As a consequence the continuum around the \civ{} line was not properly fit. Disabling the fit weights for the continuum allowed for a proper fit of the continuum model.

An overview over which lines were modeled in each quasar is provided in Table\,\ref{table:sample}. In the table we also indicate in parentheses, in which quasars the \civ{} emission line was fit with only one Gaussian component (1G) instead of two.

\subsubsection*{PSO~J004.3936+17.0862}
In the spectrum of this quasar the \mgii{} line falls into one of the telluric absorption bands. To properly fit the continuum, including the iron contribution, we have assumed a FWHM for the iron template of $\rm{FWHM}_{\rm{FeII}}=2500\,\rm{km}\,\rm{s}^{-1}$ and set the \feii{} redshift to the systemic redshift. 
The low signal-to-noise ratio of this spectrum did not justify to use more than one Gaussian component to model the \civ{} line.

\subsection*{PSO~J007.0273+04.9571}
To properly fit the continuum over the entire observed wavelength range, we disabled the fit weights for the continuum model. While we do fit the \ciii{} line complex, the red-ward part of the \ciii{} line falls into a window of strong telluric absorption. We caution against over-interpreting the resulting \ciii{} properties in this case.

\subsection*{PSO~J009.7355--10.4316}
This quasar has especially weak lines and the continuum does not resemble a power law shape. Hence, we approximated the continuum around the \civ{} line with a simple power-law and fit the line with one Gaussian profile. The \mgii{} line lies very close to one telluric absorption band and model fits were not able to constrain the line properties.

\subsection*{PSO~J011.3898+09.0324}
This quasar has a relatively low signal-to-noise ratio allowing us to fit the \civ{} with one Gaussian component only. 

\subsection*{VIK~J0046--2837}
This quasar has especially weak lines and the continuum does not resemble a power law shape. We approximated the continuum around the \mgii{} line with a simple power-law and fit for the line. The low signal-to-noise ratio in the $J$-band did not allow us to constrain the properties of \civ{} or \ciii{} line. 
 
\subsection*{SDSS~J0100+2802}
This spectrum has a high signal-to-noise ratio. As a consequence the monolithic iron template around the \mgii{} line ($2200-3500\,\rm{\AA}$) was not able to properly model the continuum. Therefore, we divided the iron template into three regions similar to \citet{Tsuzuki2006} ($2200-2660\,\rm{\AA}$, $2660-3000\,\rm{\AA}$, $3000-3500\,\rm{\AA}$) and modeled their amplitudes separately. 
Furthermore, the telluric correction algorithm was not able to fully correct the region around the \civ{} line ($11000-11600\,\rm{\AA}$). This strongly affects any attempts to model the \civ{} line and we decided against including an \civ{} line model.

\subsection*{VIK~J0109--3047}
Even though the redshift of this quasars would allow us to fit the \ciii{} complex, we cannot securely constrain the model due to the low signal-to-noise ratio of the spectrum. Hence, we only fit the \civ{} and \mgii{}, modeling \civ{} with one Gaussian component.

\subsection*{PSO~J036.5078+03.0498}
We have included a \siiv{} line model (single Gaussian component) for this fit and modeled the \civ{} with one Gaussian component only.

\subsection*{VIK~J0305--3150}
While the redshift of this quasar would allow to include the \siiv{} line in the fit, the shape of the spectrum deviates from a power law blue-ward of the \civ{} line. Hence, the \siiv{} line was not included in our fit. The \civ{} line was modeled with a single Gaussian component.

\subsection*{PSO~J056.7168--16.4769}
This spectrum has a relatively high signal-to-noise ratio, which allowed us to fit the \civ{}, \ciii{}, and \mgii{} line. 

\subsection*{PSO~J065.4085--26.9543}
The continuum strongly deviates from a power law shape. We approximated the continuum around the \mgii{} and \civ{} lines and fit them separately with individual continuum models. 
The \civ{} line is rather broad in this spectrum and shows a broad red-ward absorption feature that might well be the result of a poor telluric correction in the $11000-11600\,\rm{\AA}$ region. 
We fit the \civ{} line using a single Gaussian component and caution against over-interpreting the fit results. 

\subsection*{PSO~J065.5041--19.4579}
In the spectrum of this quasar the continuum strongly deviates from a power law shape. We approximated the continuum around the \mgii{} and \civ{} lines and fit them separately with individual continuum models.
The \civ{} line is partially absorbed by a strong blue-ward absorption trough. Thus, we restricted the line fit to the red-ward half of the line and approximated the \civ{} line using only one Gaussian component.  

\subsection*{SDSS~J0842+1218}
In order to properly fit the continuum over the entire wavelength range, it was necessary to disable the fit weights for the continuum model. 
The blue-ward wing of the \ciii{} complex is outside of the telluric absorption band, but its peak is not. Therefore, any line fit would be associated with high uncertainties and we decided against modeling of the \ciii{} in this quasar.

\subsection*{PSO~J158.69378--14.42107}
To properly fit the continuum over the entire wavelength range, it was necessary to disable the fit weights for the continuum model. 

\subsection*{SDSS~J1044--0125}
To properly fit the continuum over the entire wavelength range, it was necessary to disable the fit weights for the continuum model. 
In this spectrum the \mgii{} line falls into one of the telluric absorption bands. To properly fit the continuum, including the iron contribution, we have assumed a FWHM for the iron template of $\rm{FWHM}_{\rm{FeII}}=2500\,\rm{km}\,\rm{s}^{-1}$ and set the \feii{} redshift to the systemic redshift of the quasar. 
The \civ{} line is partially absorbed by a strong blue-ward absorption trough. Thus, we restricted the line fit to the red-ward half of the line and approximated the \civ{} line using only one Gaussian component.  
The \ciii{} complex has a very broad structure in this spectrum. 

\subsection*{VIK~J1048--0109}
The overall low signal-to-noise ratio of this spectrum did not allow to model the \civ{} line.

\subsection*{ULAS~J1120+0641}
Unfortunately, the \mgii{} line of this spectrum falls in the gap between the last two orders of the X-SHOOTER spectrograph. Due to the faint nature of the quasar the extracted traces of the last orders do not overlap and strong artifacts plague the echelle order boundary. Hence, we were unable to fit the \mgii{} line. 
To properly fit the continuum, including the iron contribution, we have assumed a FWHM for the iron template of $\rm{FWHM}_{\rm{FeII}}=2500\,\rm{km}\,\rm{s}^{-1}$ and set the \feii{} redshift to the systemic redshift of the quasar. 
The high redshift of this quasar allows us to successfully model the \siiv{} line (single Gaussian component) as well as the \ciii{} complex. 

\subsection*{PSO~J183.1124+05.0926}
In this spectrum the \civ{} line falls into the wavelength range of $11000-11600\,\rm{\AA}$, where either telluric absorption features could not be fully corrected or intrinsic absorption is present. We further see absorption in the profile of the \mgii{} line. We exclude the worst residuals from the both, \civ{} and \mgii{}, line fits and approximate the \civ{} line using only one Gaussian component. 

\subsection*{SDSS~J1306+0356}
To properly fit the continuum over the entire range, it was necessary to disable the fit weights for the continuum model. 

\subsection*{ULAS~J1319+0950}
While we have included the \ciii{} complex in the fit, we would like to caution against over-interpreting its fit results as it partially falls in a region of strong telluric absorption. In addition, our best fits seems to over-predict the \feii{} pseudo-continuum red-ward of the \mgii{} line. 

\subsection*{ULAS~J1342+0928}
The \mgii{} line is not detected in this spectrum as it falls close to the red edge of the last echelle order, which is dominated by noise. However, we are able to include the \siiv{} line in our model. Due to the extremely broad nature of the \siiv{} and \civ{} lines both were modeled using only a single Gaussian component each. 

\subsection*{CFHQS~J1509--1749}
To properly fit the continuum over the entire range, it was necessary to disable the fit weights for the continuum model. Additionally, the \ciii{} complex falls partly in a region of strong telluric absorption and therefore and we caution against over-interpreting the resulting \ciii{} properties in this case.

\subsection*{PSO~J231.6576--20.8335}
The low signal-to-noise ratio of this spectrum did not allow us to model the \civ{} line successfully.

\subsection*{PSO~J239.7124--07.4026}
To properly fit the continuum over the entire range, it was necessary to disable the fit weights for the continuum model. 

\subsection*{SDSS~J2054--0005}
To properly fit the continuum over the entire range, it was necessary to disable the fit weights for the continuum model. The low signal-to-noise ratio of this spectrum did not justify to use more than one Gaussian component to model the \civ{} line. For the same reason we set the contribution of the \siiii{} line to the \ciii{} complex to zero.

\subsection*{CFHQS~J2100--1715}
The low signal-to-noise ratio of this spectrum did not justify to use more than one Gaussian component to model the \civ{} line.

\subsection*{PSO~J323.1382+12.2986}
The higher redshift of this quasar allows us to successfully model the \siiv{} line with a single Gaussian component. 
The \ciii{} complex falls partially in one of the bands of strong telluric absorption. As a consequence we set the contribution of the \aliii{} line to the \ciii{} complex to zero and caution against over-interpreting the \ciii{} complex properties with the exception of the peak redshift.

\subsection*{VIK~J2211--3206}
In this spectrum the \civ{} line falls into the wavelength range of $11000-11600\,\rm{\AA}$, where telluric absorption features could not be fully corrected. We exclude the worst residuals from the line fit and approximate the \civ{} line using only a single Gaussian component. We also note that a strong absorption feature blue-ward of the \civ{} line complicates the modelling. Hence, we have excluded part of this region from the fit for the line.

\subsection*{CFHQS~J2229+1457}
The low signal-to-noise spectrum did not allow to constrain the \mgii{} with a fit. The strong \civ{} emission was modeled with two Gaussian components.

\subsection*{PSO~J340.2041--18.6621}
The \civ{} and the \ciii{} complex show strong absorption features within their profiles, which have been excluded from the fit. 
While we have included the \ciii{} complex in the fit, we would like to caution against over-interpreting its fit results as the complex partially falls in a region of strong telluric absorption.

\subsection*{SDSS~J2310+1855}
The \civ{} line and the \ciii{} line are unusually broad in this spectrum. In addition, the blue edge of the \civ{} line is either affected by the declining throughput at the blue edge of the spectrum or by absorption. Therefore, we decided to model the \civ{} line with only a single Gaussian component and exclude the \ciii{} complex from the fit. To properly fit the continuum over the entire wavelength range, it was necessary to disable the fit weights for the continuum model.

\subsection*{VIK~J2318--3029}
To properly fit the continuum over the entire range, it was necessary to disable the fit weights for the continuum model. Absorption features within the \civ{} line were masked for the fit.

\subsection*{VIK~J2348--3054}
The \civ{} line is partially absorbed by a strong blue-ward absorption trough. Thus, we restricted the line fit to the red-ward half of the line and approximated the \civ{} line using only a single Gaussian component. To properly fit the \ciii{} complex we mask a strong absorption doublet in its center.

\subsection*{PSO~J359.1352--06.3831}
In the spectrum of this quasar the \mgii{} line falls into a region of strong residuals from telluric absorption features ($19900-20200\,\rm{\AA}$), potentially biasing the derived line properties. We mask out the strongest feature for the fit.

\section{The X-SHOOTER/ALMA master table}\label{sec:master_table}

The X-SHOOTER/ALMA master table of the near-infrared spectral analysis is available as a machine readable table on-line. It has 175 columns, detailed in Table\,\ref{table:master_table_overview} below. For all fit properties we provide the median (\texttt{\_med}) value as well as the differences from the median to the 16th (\texttt{\_low}) and 84th (\texttt{\_upp}) percentile values. Hence, each fit property has three columns in the table. The shorthand \texttt{VW01} refers to fit properties derived from fits with the \citetalias{Vestergaard2001} iron template. All other fit properties were derived using the \citetalias{Tsuzuki2006} iron template.


\startlongtable
\begin{deluxetable*}{cccc}
\tabletypesize{\footnotesize} 
\tablecaption{Description of the on-line only catalog of the X-SHOOTER/ALMA sample of quasars in the epoch of reionization\label{table:master_table_overview}}
\tablehead{\colhead{Column} &\colhead{Name} & \colhead{Unit} &\colhead{Description} 
} 
\startdata 
1 & Name & \dots & Quasar name \\ 
2 & Zsys & \dots & Systemic redshift \\ 
3 & Zsys\_e & \dots & Systemic redshift error \\ 
4 & Z\_method & \dots & Method for systemtic redshift \\ 
5 & z\_ref & \dots & Reference for systemic redshift \\ 
6 & RA & decimal degrees & Right ascension \\ 
7 & Decl & decimal degrees & Declination \\ 
8 & J & AB mag & J-band magnitude \\ 
9 & Je & AB mag & J-band magntiude error \\ 
10 & disc\_ref & \dots & Discovery reference \\ 
11-13 & flux\_1350 & $10^{-17}\,\rm{erg}\,\rm{s}^{-1}\,\rm{cm}^{-2}$\,\AA$^{-1}$ & Continuum model flux at $1350$\,\AA \\ 
14-16 & L\_1350 & $10^{46}\,\rm{erg}\,\rm{s}^{-1}$ & Continuum model luminosity at $1350$\AA \\ 
17-19 & flux\_1450 & $10^{-17}\,\rm{erg}\,\rm{s}^{-1}\,\rm{cm}^{-2}$\,\AA$^{-1}$ & Continuum model flux at $1450$\,\AA \\ 
20-22 & L\_1450 & $10^{46}\,\rm{erg}\,\rm{s}^{-1}$ & Continuum model luminosity at $1450$\AA \\ 
23-25 & flux\_2500 & $10^{-17}\,\rm{erg}\,\rm{s}^{-1}\,\rm{cm}^{-2}$\,\AA$^{-1}$ & Continuum model flux at $2500$\,\AA \\ 
26-28 & L\_2500 & $10^{46}\,\rm{erg}\,\rm{s}^{-1}$ & Continuum model luminosity at $2500$\AA \\ 
29-31 & flux\_3000 & $10^{-17}\,\rm{erg}\,\rm{s}^{-1}\,\rm{cm}^{-2}$\,\AA$^{-1}$ & Continuum model flux at $3000$\,\AA \\ 
32-34 & L\_3000 & $10^{46}\,\rm{erg}\,\rm{s}^{-1}$ & Continuum model luminosity at $3000$\AA \\ 
35-37 & Lbol & $10^{46}\,\rm{erg}\,\rm{s}^{-1}$ & Bolometric luminosity \\ 
38-40 & m1450 & AB mag & Apparent magnitude at 1450\,\AA \\ 
41-43 & M1450 & AB mag & Absolute magnitude at 1450\,\AA \\ 
44-46 & Plslope & \dots & Conntinuum model power law slope \\ 
47-49 & CIV\_wav\_cen & \AA & \civ{} peak wavelength \\ 
50-52 & CIV\_z\_cen & \dots & \civ{} peak redshift \\ 
53-55 & CIV\_vshift & $\rm{km}\,\rm{s}^{-1}$ & \civ{} velocity shift to Zsys \\ 
56-58 & CIV\_FWHM & $\rm{km}\,\rm{s}^{-1}$ & \civ{} FWHM  \\ 
59-61 & CIV\_EW & \AA & \civ{} Rest-frame equivalent width \\ 
62-64 & CIV\_FWHM\_corr & $\rm{km}\,\rm{s}^{-1}$ & \civ{} corrected FWHM \citep{Coatman2017} \\ 
65-67 & CIV\_flux & $\rm{erg}\,\rm{s}^{-1}\,\rm{cm}^{-2}$\,\AA$^{-1}$ & Integrated \civ{} flux \\ 
68-70 & CIV\_L & $\rm{km}\,\rm{s}^{-1}$ & Integrated \civ{} luminosity \\ 
71-73 & CIV\_BHM\_V06 & $10^{9}\,\rm{M}_{\odot}$ & \civ{} Black hole mass \citep{Vestergaard2006} \\ 
74-76 & CIV\_EddR\_V06 & \dots & \civ{} Eddington luminosity ratio \citep{Vestergaard2006} \\ 
77-79 & CIV\_BHM\_Co17 & $10^{9}\,\rm{M}_{\odot}$ & \civ{} Black hole mass \citep{Coatman2017} \\ 
80-82 & CIV\_EddR\_Co17 & \dots & \mgii{} Eddington luminosity ratio \citep{Coatman2017} \\ 
83-85 & MgII\_wav\_cen & \AA & \mgii{} peak wavelength \\ 
86-88 & MgII\_z\_cen & \dots & \mgii{} peak redshift \\ 
89-91 & MgII\_vshift & $\rm{km}\,\rm{s}^{-1}$ & \mgii{} velocity shift to Zsys \\ 
92-94 & MgII\_FWHM & $\rm{km}\,\rm{s}^{-1}$ & \mgii{} FWHM  \\ 
95-97 & MgII\_EW & \AA & \mgii{} Rest-frame equivalent width \\ 
98-100 & MgII\_flux & $\rm{erg}\,\rm{s}^{-1}\,\rm{cm}^{-2}$\,\AA$^{-1}$ & Integrated \mgii{} flux \\ 
101-103 & MgII\_L & $\rm{km}\,\rm{s}^{-1}$ & Integrated \mgii{} luminosity \\ 
104-106 & FeII\_flux & $\rm{erg}\,\rm{s}^{-1}\,\rm{cm}^{-2}$\,\AA$^{-1}$ & Integrated \feii{} flux \\ 
107-109 & FeIIMgII\_ratio & \dots & \feii{}/\mgii{} flux ratio \\ 
110-112 & CIII\_z & \dots & \ciii{} complex model redshift \\ 
113-115 & CIII\_vshift & $\rm{km}\,\rm{s}^{-1}$ & \ciii{} complex peak velocity shift to Zsys \\ 
116-118 & SiIV\_wav\_cen & \AA & \civ{} peak wavelength \\ 
119-121 & SiIV\_z\_cen & \dots & \civ{} peak redshift \\ 
122-124 & SiIV\_vshift & $\rm{km}\,\rm{s}^{-1}$ & \civ{} velocity shift to Zsys \\ 
125-127 & SiIV\_FWHM & $\rm{km}\,\rm{s}^{-1}$ & \civ{} FWHM  \\ 
128-130 & SiIV\_EW & \AA & \civ{} Rest-frame equivalent width \\ 
131-133 & SiIV\_flux & $\rm{erg}\,\rm{s}^{-1}\,\rm{cm}^{-2}$\,\AA$^{-1}$ & Integrated \civ{} flux \\ 
134-136 & SiIV\_L & $\rm{km}\,\rm{s}^{-1}$ & Integrated \civ{} luminosity \\ 
137-139 & VW01\_MgII\_wav\_cen & \AA & \mgii{} peak wavelength \\ 
140-142 & VW01\_MgII\_z\_cen & \dots & \mgii{} peak redshift \\ 
143-145 & VW01\_MgII\_vshift & $\rm{km}\,\rm{s}^{-1}$ & \mgii{} velocity shift to Zsys \\ 
146-148 & VW01\_MgII\_FWHM & $\rm{km}\,\rm{s}^{-1}$ & \mgii{} FWHM  \\ 
149-151 & VW01\_MgII\_EW & \AA & \mgii{} Rest-frame equivalent width \\ 
152-154 & VW01\_MgII\_flux & $\rm{erg}\,\rm{s}^{-1}\,\rm{cm}^{-2}$\,\AA$^{-1}$ & Integrated \mgii{} flux \\ 
155-157 & VW01\_MgII\_L & $\rm{km}\,\rm{s}^{-1}$ & Integrated \mgii{} luminosity \\ 
158-160 & VW01\_MgII\_BHM\_VW09 & $10^{9}\,\rm{M}_{\odot}$ & \mgii{} Black hole mass \citep{Vestergaard2009} \\ 
161-163 & VW01\_MgII\_EddR\_VW09 & \dots & \mgii{} Eddington luminosity ratio \citep{Vestergaard2009} \\ 
164-166 & VW01\_MgII\_BHM\_S11 & $10^{9}\,\rm{M}_{\odot}$ & \mgii{} Black hole mass \citep{Shen2011} \\ 
167-169 & VW01\_MgII\_EddR\_S11 & \dots & \mgii{} Eddington luminosity ratio \citep{Shen2011} \\ 
170 & Resolution & \dots & Lowest resolution of all used observations \\ 
171 & Exptime & \rm{s} & Total exposure time \\ 
172 & ProgramIDs & \dots & ESO proposal program IDs \\ 
173 & PIs & \dots & Principal Investigators \\ 
174 & SNR\_J & \dots & Mean signal-to-noise ratio over 12500-13450\,\AA \\ 
175 & SNR\_J\_binned & \dots & Mean signal-to-noise ratio over 12500-13450\,\AA (binned) 
\enddata 
\end{deluxetable*}

\section{Derivation of the BH masses}\label{sec:app_bhmasses}

In this section we will briefly discuss the calculation of our BH mass estimates. The derived black hole masses are then presented and further discussed in Farina et al. (in prep.).

The properties of the broad emission lines, probes of the BLR gas, allow for first order estimates.
Under the assumption that the line-emitting gas is in virial motion (e.g. a disk with Keplerian rotation) around the SMBH, the line-of-sight velocity dispersion of the gas, measured as the FWHM of the broad emission line ($\rm{FWHM}_{\rm{BLR}}$), traces the gravitational potential of the SMBH mass ($M_{\rm{BH}}$): 
\begin{equation}
    M_{\rm{BH}} = f \cdot \frac{R_{\rm{BLR}}\cdot \rm{FWHM}_{\rm{BLR}}^2}{\rm{G}} \ , 
\end{equation}
where $R_{\rm{BLR}}$ denotes the radius from the SMBH to the line-emitting region for the particular emission line in question. Here, the factor $f$ encapsulates our ignorance on orientation, structure and more complex kinematics of the BLR. While it is generally assumed to be of order unity \citep{Peterson2004, Decarli2010, Mediavilla2020}, it gives rise to significant systematic uncertainties \citep[e.g.][]{Krolik2001a}. 
Reverberation mapping campaigns have found a strong correlation between $R_{\rm{BLR}}$ and the quasar's continuum luminosity \citep[e.g.][]{Kaspi2000, Kaspi2005, Bentz2013} and been successful in measuring BH masses \citep[e.g.][]{Onken2004, Peterson2004}. These results have been recently supported by spatially resolved observations of the BLR in 3C 273 \citep{GravityCollab2018}.
Based on the reverberation mapping results scaling relations have been derived, which allow to estimate a quasar's BH mass solely based on the velocity dispersion of a broad line and the its continuum luminosity. These so-called single-epoch virial mass estimators allow us to estimate the BH mass of a quasar based on a single spectrum and are often written as  
\begin{equation}
M_{\rm{BH}} = 10^{zp(x)} \cdot \left[ \frac{\rm{FWHM}}{1000\,\rm{km}\,\rm{s}^{-1}} \right]^2 \left[ \frac{x L_{\lambda, x}}{10^{44}\,\rm{erg}\,\rm{s}^{-1}} \right]^{b} \,M_\odot
\end{equation}
The zero points $zp$ and the parameter $b$, depend on the broad emission line in question and the monochromatic continuum luminosity $L_{\lambda, x}$ at a given rest-frame wavelength $x$. Single-epoch virial BH mass estimates have a considerable systematic uncertainty due to the unknowns encompassed in the $f$ factor, which surface as scatter in the radius-luminosity relations. These systematic uncertainties can be as large as $\sim0.55\,\rm{dex}$ \citep{Vestergaard2009}.
We derive black hole mass estimates from the properties of the broad \mgii{} and \civ{} emission lines and the adjacent continuum.

\underline{\mgii{}:} For the \mgii{} line we adopt the single-epoch virial mass estimators of \citet[][$zp=6.86$, $b=0.5$, $x=3000\,$\AA]{Vestergaard2009} and \citep[][$zp=6.74$, $b=0.62$, $x=3000\,$\AA]{Shen2011}. 
The scaling relation of \citet{Vestergaard2009} uses single or multiple Gaussian components to model and measure the FWHM of the \mgii{}. In the cases of a multi-component model the FWHM is calculated from the full line model. 
The scaling relation of \citet{Shen2011} uses the radius luminosity relationship of \citet{McLure2004} and re-calibrates the zero point to the H$\beta$ relation of \citet{Vestergaard2006}. 
The FWHM of \mgii{} is always determined with multiple components, with at least a narrow and a broad component both modeled with Gaussian profiles.
While the signal to noise in our spectra does not justify a multi-component fit for \mgii{}, we still argue that both scaling relations are valid in our case as long as the emission line is properly represented by our fit. 

We model the FWHM of the \mgii{} line for BH mass estimates from both relations using the \citetalias{Vestergaard2001} iron template for the \feii{} continuum. As we discuss in Section\,\ref{sec:irondiscussion}, the modeling of the iron pseudo-continuum introduces systematic effects on the measured FWHM of the \mgii{} line. Therefore, our BH mass estimates are based on the FWHM determinations using the \citetalias{Vestergaard2001} iron template analogous to the determinations of the scaling relations.

\underline{\civ{}:} 
Contrary to lower ionization lines, like H$\beta$ or \mgii{} the \civ{} emission line often shows highly asymmetric line profiles correlating with the quasar's luminosity and are commonly associated with an out-flowing wind component \citep[e.g.][]{Richards2011}. 
Outflows that can possibly manifest as a non-reverberating component \citep{Denney2012} can significantly bias BH mass measurements based on \civ{} single-epoch virial estimators. Hence, extensive discussions \citep[e.g.][]{Shen2013review, Coatman2016, MejiaRestrepo2018a} revolve around the reliability of \civ{}-based BH masses and corrections for these biases \citep[e.g.][]{Denney2012, Park2013, Runnoe2013c, MeijaRestrepo2016, Coatman2017, ZuoWenwen2020}.


For a few quasars in our sample the \mgii{} line could not be measured as it falls into a region with extremely low signal-to-noise ratio. In most cases these are telluric absorption regions of the reddest order of the X-SHOOTER NIR spectrum. Thus, we decided to use the \civ{} line to determine BH masses in these cases. We adopt the scaling relation of  \citet[][$zp=6.66$, $b=0.53$, $x=1350\,$\AA]{Vestergaard2006} and correct the BH masses according to Equations\,4 and 6 of \citet{Coatman2017}. 
While our measurement of the \civ{} line properties can be considered equivalent to \citet{Vestergaard2006}, \citet{Coatman2017} modeled the \civ{} line with Gauss-Hermite polynomials. We judge the uncertainties introduced by the different fitting methodology likely to be small compared to the systematic uncertainty on the BH mass estimate itself. 
For their correction \citet{Coatman2017} measured the velocity blueshift from the \civ{} model centroid with respect to the H$\alpha$ Balmer line, which is considered a good proxy of the systemic redshift of the quasar. Instead of the H$\alpha$ line, we have used the systemic redshifts provided in Table\,\ref{table:sample} to derive the \civ{} blueshifts.

\section{Construction of the low-z comparison sample} \label{sec:app_lowzsample}
The low-redshift comparison sample is constructed from the catalog of SDSS DR7 quasars published by \citet{Shen2011} using the updated redshift from \citet{Hewett2010}\footnote{\url{http://www.sdss.org/dr7/products/value_added/index.html}} \citep[also see,][]{Wild2005}. 
We select a sub-sample of quasars broadly following \citet{Richards2011} and \citet{Mazzucchelli2017} to retrieve objects with secure \civ{} and \mgii{} measurements. 
We only consider quasars in the redshift range $1.52 \le z \le 2.2$, where both emission lines are covered by the SDSS spectrograph. We require the \civ{} and \mgii{} line to be well detected: $\rm{FWHM}_{\rm{CIV}} > 1000$ and $\rm{FWHM}_{\rm{CIV}} > 2 \cdot \sigma_{\rm{FWHM}, \rm{CIV}}$ and $\rm{EW}_{\rm{CIV}} > 5\,\rm{\AA}$ and $\rm{EW}_{\rm{CIV}} > 2 \cdot \sigma_{\rm{EW}, \rm{CIV}}$ and $\rm{FWHM}_{\rm{MgII}} > 1000$ and $\rm{FWHM}_{\rm{MgII}} > 2 \cdot \sigma_{\rm{FWHM}, \rm{MgII}}$ and $\rm{EW}_{\rm{MgII}} > 2 \cdot \sigma_{\rm{EW}, \rm{MgII}}$.
Additionally, we only consider quasars without broad absorption lines (\texttt{BAL\_FLAG} == 0) and with valid \civ{} and \mgii{} velocity shifts (\texttt{VOFF\_CIV\_PEAK} $< 20000$ and \texttt{VOFF\_BROAD\_MGII} $< 20000$) measured in relation with their systemic redshifts based on the SDSS pipeline \citep{Stoughton2002}.
We have confirmed that the $\rm{FWHM}_{\rm{CIV}} > 1000$ and  $\rm{FWHM}_{\rm{MgII}} > 1000$ criteria, which are responsible for removing $\sim1\%$ of the spectra, do not remove well measured narrow lines. In the large majority of cases we find $\rm{FWHM}{=}0$ in the catalog. In the few cases, where the catalog reports a non-zero FWHM, our visual inspection identified that the automated model fits had failed.
This sub-set was matched to the updated redshifts of \citet{Hewett2010} to form a sample of 20239 quasars.
Finally, we further limit the low-redshift sample to a similar bolometric luminosity range for a fair comparison with our high-redshift quasars ($46.5 \le \log(L_{\rm{bol}}) \le 47.5$). This reduces the low-redshift sample to to 12099, which we will use for comparison throughout Section\,\ref{sec:results_BELS}.

\section{Additional figures of broad emission line properties}\label{sec:app_bel_correlations}

\begin{figure}
    \centering
    \includegraphics[width=0.48\textwidth]{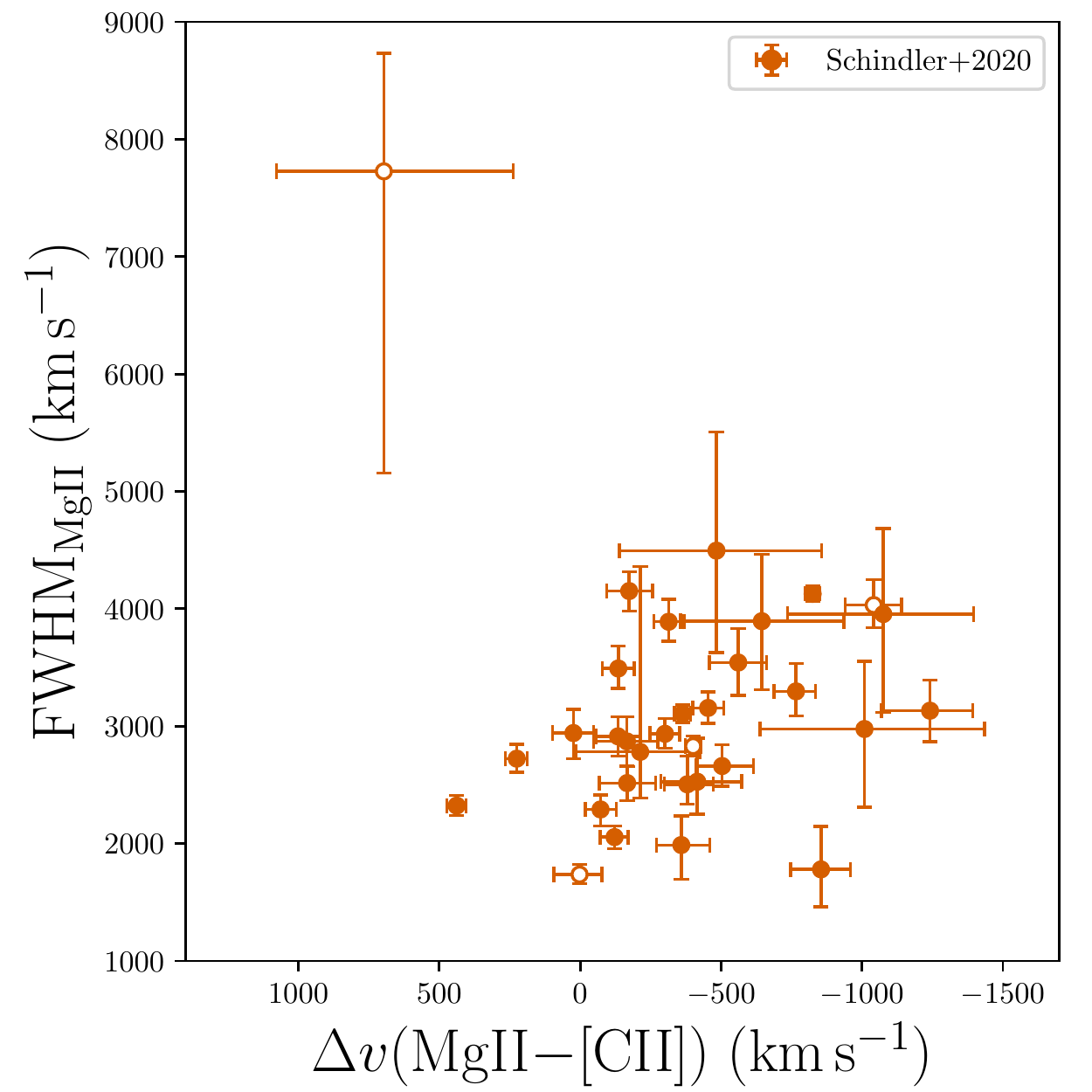}
    \caption{FWHM as a function \mgii{}-\cii{} velocity shift for quasars in our sample. Compared to Figure\,\ref{fig:civ_context} the \mgii{} line does not show a correlation between its velocity shift. Any positive correlation is driven by one outlier, CFHQS~J2100--1715, whose continuum could not be fit with a power law.}
    \label{fig:mgiiciivshift_fwhm}
\end{figure}

\begin{figure}
    \centering
    \includegraphics[width=0.48\textwidth]{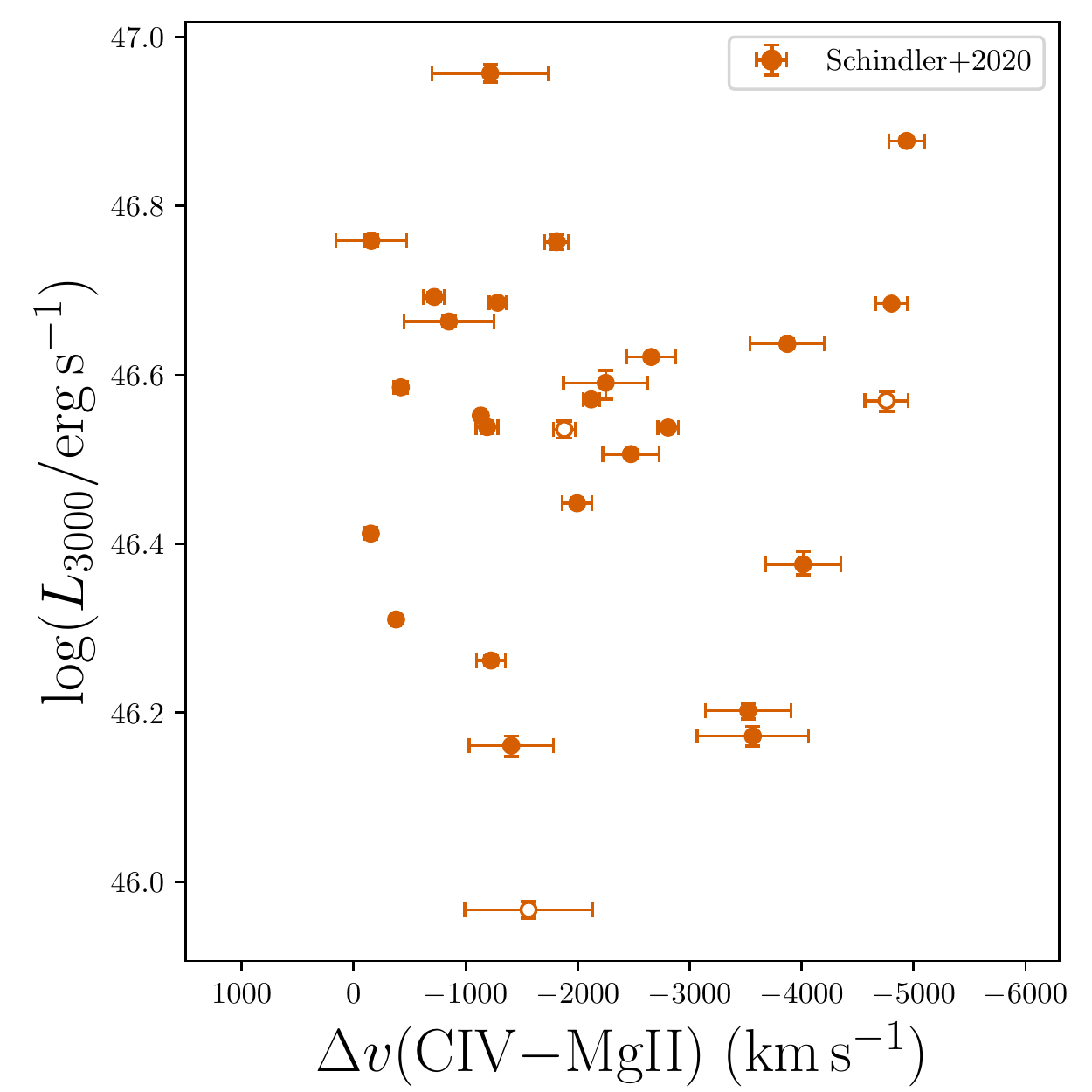}
    \caption{Luminosity at 3000\AA\ as a function of the \civ{}-\mgii{} velocity shift for quasars in our sample. Our sample does not show a correlation between continuum luminosity and \civ{}-\mgii{} blueshift.}
    \label{fig:civmgii_shift_L3000}
\end{figure}

\begin{figure}
    \centering
    \includegraphics[width=0.48\textwidth]{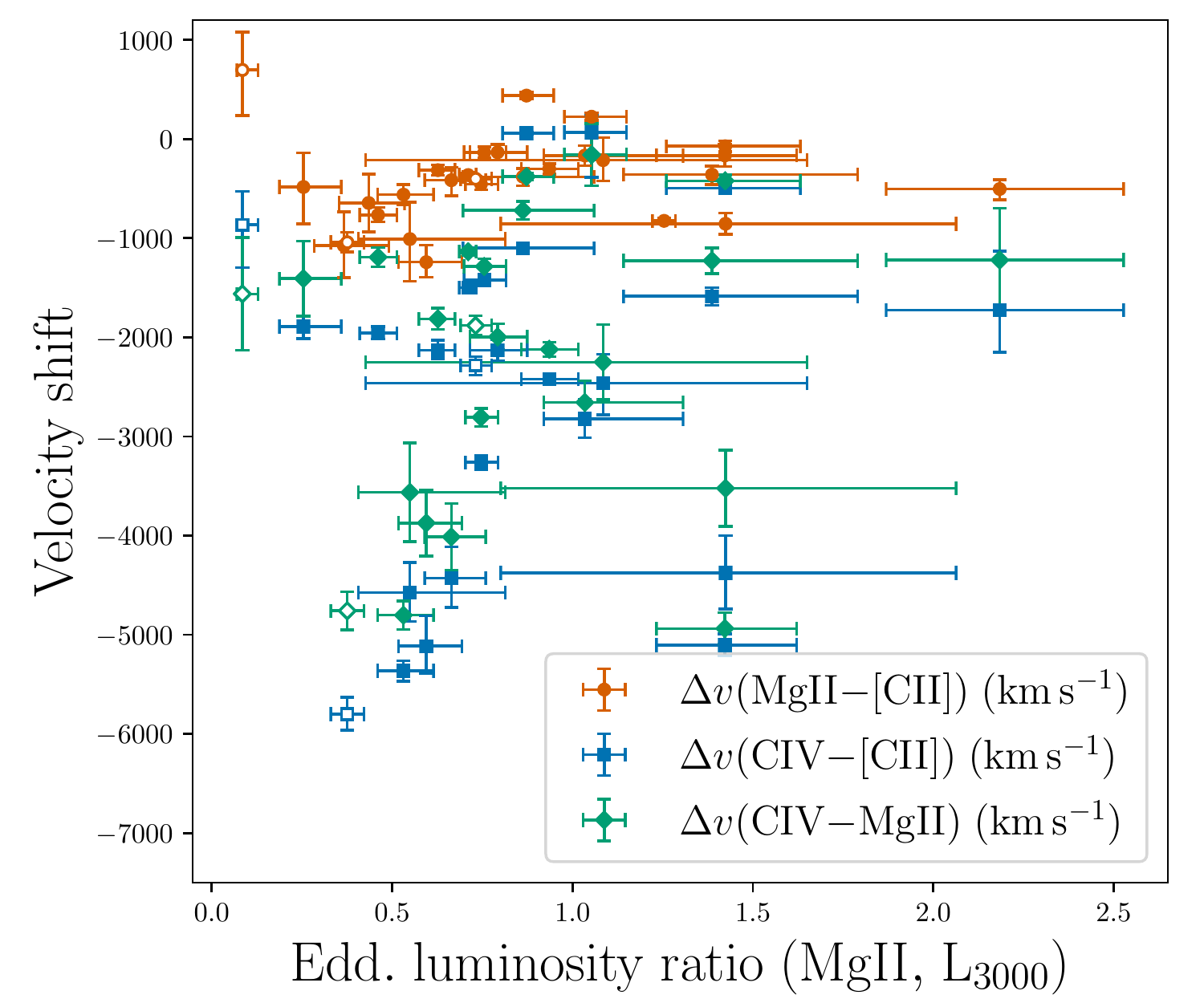}
    \caption{Velocity shifts as a function of the Eddington luminosity ratio calculated from the \mgii{} line using the prescription of \citet{Vestergaard2009} and measured with the \citetalias{Vestergaard2001} iron template. We tested for correlations using the Pearson correlation coefficient and found no significant correlations for $\Delta v(\rm{CIV}{-}\rm{[CII]})$ ($\rho=0.14$, $p=0.49$), $\Delta v(\rm{CIV}{-}\rm{MgII})$ ($\rho=-0.16$, $p=0.45$), and  $\Delta v(\rm{MgII}{-}\rm{[CII]})$ ($\rho=0.09$, $p=0.64$) with the Eddington luminosity ratio. 
    }
    \label{fig:app_vshift_Eddr}
\end{figure}

In this section of the appendix we present additional figures showing broad emission line properties. Figure\,\ref{fig:mgiiciivshift_fwhm} shows the \mgii{} FWHM as a function of the \mgii{}-\cii{} velocity shift. Figure\,\ref{fig:civmgii_shift_L3000} shows the luminosity at 3000\,\AA\ as a function of the \civ{}-\mgii{} velocity shift. There is no indication for any correlations in both of the figures. Lastly, in Figure\,\ref{fig:app_vshift_Eddr} we show the \mgii{}-\cii{}, \mgii{}-\civ{}, and \civ{}-\mgii{} velocity shifts as a function of the Eddington luminosity ratio. The Eddington luminosity ratio was determined from the \mgii{} line using the \citet{Vestergaard2009} relation with the FWHM measured with the \citetalias{Vestergaard2001} iron template. We tested for correlations and found none of the velocity shifts to be correlated with the Eddington luminosity ratio.

\acknowledgements
{
The authors thank R. Meyer for providing line redshifts with improved accuracy for the "z6" and "z7" quasar samples in \citet{Meyer2019c}.
 J. Yang and X.Fan acknowledge the support from the NASA ADAP Grant NNX17AF28G.
A. C. Eilers acknowledges support by NASA through the NASA Hubble Fellowship grant $\#$HF2-51434 awarded by the Space Telescope Science Institute, which is operated by the Association of Universities for Research in Astronomy, Inc., for NASA, under contract NAS5-26555. 
F.Wang acknowledges support provided by NASA through the NASA Hubble Fellowship grant $\#$HST-HF2-51448.001-A awarded by the Space Telescope Science Institute, which is operated by the Association of Universities for Research in Astronomy, Incorporated, under NASA contract NAS5-26555.}
\facilities{VLT:Kueyen (X-SHOOTER)}

\software{Astropy \citep{astropy1, astropy2}, SciPy \citep{scipy}, Numpy \citep{numpy, numpy2020}, Pandas \citep{pandas_software, pandas_paper}, LMFIT \citep{lmfit2014}, Pypeit \citep{PypeitProchaska2019, PypeitProchaska2020}, Extinction \citep{python_extinction}
}

\bibliography{all}{}

\begin{thebibliography}{}
\expandafter\ifx\csname natexlab\endcsname\relax\def\natexlab#1{#1}\fi
\providecommand{\url}[1]{\href{#1}{#1}}
\providecommand{\dodoi}[1]{doi:~\href{http://doi.org/#1}{\nolinkurl{#1}}}
\providecommand{\doeprint}[1]{\href{http://ascl.net/#1}{\nolinkurl{http://ascl.net/#1}}}
\providecommand{\doarXiv}[1]{\href{https://arxiv.org/abs/#1}{\nolinkurl{https://arxiv.org/abs/#1}}}

\bibitem[{{Astropy Collaboration} {et~al.}(2013){Astropy Collaboration},
  {Robitaille}, {Tollerud}, {Greenfield}, {Droettboom}, {Bray}, {Aldcroft},
  {Davis}, {Ginsburg}, {Price-Whelan}, {Kerzendorf}, {Conley}, {Crighton},
  {Barbary}, {Muna}, {Ferguson}, {Grollier}, {Parikh}, {Nair}, {Unther},
  {Deil}, {Woillez}, {Conseil}, {Kramer}, {Turner}, {Singer}, {Fox}, {Weaver},
  {Zabalza}, {Edwards}, {Azalee Bostroem}, {Burke}, {Casey}, {Crawford},
  {Dencheva}, {Ely}, {Jenness}, {Labrie}, {Lim}, {Pierfederici}, {Pontzen},
  {Ptak}, {Refsdal}, {Servillat}, \& {Streicher}}]{astropy1}
{Astropy Collaboration}, {Robitaille}, T.~P., {Tollerud}, E.~J., {et~al.} 2013,
  \aap, 558, A33, \dodoi{10.1051/0004-6361/201322068}

\bibitem[{{Astropy Collaboration} {et~al.}(2018){Astropy Collaboration},
  {Price-Whelan}, {Sip{\H{o}}cz}, {G{\"u}nther}, {Lim}, {Crawford}, {Conseil},
  {Shupe}, {Craig}, {Dencheva}, {Ginsburg}, {Vand erPlas}, {Bradley},
  {P{\'e}rez-Su{\'a}rez}, {de Val-Borro}, {Aldcroft}, {Cruz}, {Robitaille},
  {Tollerud}, {Ardelean}, {Babej}, {Bach}, {Bachetti}, {Bakanov}, {Bamford},
  {Barentsen}, {Barmby}, {Baumbach}, {Berry}, {Biscani}, {Boquien}, {Bostroem},
  {Bouma}, {Brammer}, {Bray}, {Breytenbach}, {Buddelmeijer}, {Burke},
  {Calderone}, {Cano Rodr{\'\i}guez}, {Cara}, {Cardoso}, {Cheedella}, {Copin},
  {Corrales}, {Crichton}, {D'Avella}, {Deil}, {Depagne}, {Dietrich}, {Donath},
  {Droettboom}, {Earl}, {Erben}, {Fabbro}, {Ferreira}, {Finethy}, {Fox},
  {Garrison}, {Gibbons}, {Goldstein}, {Gommers}, {Greco}, {Greenfield},
  {Groener}, {Grollier}, {Hagen}, {Hirst}, {Homeier}, {Horton}, {Hosseinzadeh},
  {Hu}, {Hunkeler}, {Ivezi{\'c}}, {Jain}, {Jenness}, {Kanarek}, {Kendrew},
  {Kern}, {Kerzendorf}, {Khvalko}, {King}, {Kirkby}, {Kulkarni}, {Kumar},
  {Lee}, {Lenz}, {Littlefair}, {Ma}, {Macleod}, {Mastropietro}, {McCully},
  {Montagnac}, {Morris}, {Mueller}, {Mumford}, {Muna}, {Murphy}, {Nelson},
  {Nguyen}, {Ninan}, {N{\"o}the}, {Ogaz}, {Oh}, {Parejko}, {Parley}, {Pascual},
  {Patil}, {Patil}, {Plunkett}, {Prochaska}, {Rastogi}, {Reddy Janga},
  {Sabater}, {Sakurikar}, {Seifert}, {Sherbert}, {Sherwood-Taylor}, {Shih},
  {Sick}, {Silbiger}, {Singanamalla}, {Singer}, {Sladen}, {Sooley},
  {Sornarajah}, {Streicher}, {Teuben}, {Thomas}, {Tremblay}, {Turner},
  {Terr{\'o}n}, {van Kerkwijk}, {de la Vega}, {Watkins}, {Weaver}, {Whitmore},
  {Woillez}, {Zabalza}, \& {Astropy Contributors}}]{astropy2}
{Astropy Collaboration}, {Price-Whelan}, A.~M., {Sip{\H{o}}cz}, B.~M., {et~al.}
  2018, \aj, 156, 123, \dodoi{10.3847/1538-3881/aabc4f}

\bibitem[{{Ba{\~n}ados} {et~al.}(2015){Ba{\~n}ados}, {Decarli}, {Walter},
  {Venemans}, {Farina}, \& {Fan}}]{Banados2015}
{Ba{\~n}ados}, E., {Decarli}, R., {Walter}, F., {et~al.} 2015, \apjl, 805, L8,
  \dodoi{10.1088/2041-8205/805/1/L8}

\bibitem[{{Ba{\~n}ados} {et~al.}(2014){Ba{\~n}ados}, {Venemans}, {Morganson},
  {Decarli}, {Walter}, {Chambers}, {Rix}, {Farina}, {Fan}, {Jiang}, {McGreer},
  {De Rosa}, {Simcoe}, {Wei{\ss}}, {Price}, {Morgan}, {Burgett}, {Greiner},
  {Kaiser}, {Kudritzki}, {Magnier}, {Metcalfe}, {Stubbs}, {Sweeney}, {Tonry},
  {Wainscoat}, \& {Waters}}]{Banados2014}
{Ba{\~n}ados}, E., {Venemans}, B.~P., {Morganson}, E., {et~al.} 2014, \aj, 148,
  14, \dodoi{10.1088/0004-6256/148/1/14}

\bibitem[{{Ba{\~n}ados} {et~al.}(2016){Ba{\~n}ados}, {Venemans}, {Decarli},
  {Farina}, {Mazzucchelli}, {Walter}, {Fan}, {Stern}, {Schlafly}, {Chambers},
  {Rix}, {Jiang}, {McGreer}, {Simcoe}, {Wang}, {Yang}, {Morganson}, {De Rosa},
  {Greiner}, {Balokovi{\'c}}, {Burgett}, {Cooper}, {Draper}, {Flewelling},
  {Hodapp}, {Jun}, {Kaiser}, {Kudritzki}, {Magnier}, {Metcalfe}, {Miller},
  {Schindler}, {Tonry}, {Wainscoat}, {Waters}, \& {Yang}}]{Banados2016}
{Ba{\~n}ados}, E., {Venemans}, B.~P., {Decarli}, R., {et~al.} 2016, \apjs, 227,
  11, \dodoi{10.3847/0067-0049/227/1/11}

\bibitem[{{Ba{\~n}ados} {et~al.}(2018){Ba{\~n}ados}, {Venemans},
  {Mazzucchelli}, {Farina}, {Walter}, {Wang}, {Decarli}, {Stern}, {Fan},
  {Davies}, {Hennawi}, {Simcoe}, {Turner}, {Rix}, {Yang}, {Kelson}, {Rudie}, \&
  {Winters}}]{Banados2018}
{Ba{\~n}ados}, E., {Venemans}, B.~P., {Mazzucchelli}, C., {et~al.} 2018, \nat,
  553, 473, \dodoi{10.1038/nature25180}

\bibitem[{{Ba{\~n}ados} {et~al.}(2019{\natexlab{a}}){Ba{\~n}ados}, {Novak},
  {Neeleman}, {Walter}, {Decarli}, {Venemans}, {Mazzucchelli}, {Carilli},
  {Wang}, {Fan}, {Farina}, \& {Rix}}]{Banados2019}
{Ba{\~n}ados}, E., {Novak}, M., {Neeleman}, M., {et~al.} 2019{\natexlab{a}},
  \apjl, 881, L23, \dodoi{10.3847/2041-8213/ab3659}

\bibitem[{{Ba{\~n}ados} {et~al.}(2019{\natexlab{b}}){Ba{\~n}ados}, {Rauch},
  {Decarli}, {Farina}, {Hennawi}, {Mazzucchelli}, {Venemans}, {Walter},
  {Simcoe}, {Prochaska}, {Cooper}, {Davies}, \& {Chen}}]{Banados2019b}
{Ba{\~n}ados}, E., {Rauch}, M., {Decarli}, R., {et~al.} 2019{\natexlab{b}},
  \apj, 885, 59, \dodoi{10.3847/1538-4357/ab4129}

\bibitem[{{Baldwin}(1977)}]{Baldwin1977}
{Baldwin}, J.~A. 1977, \apj, 214, 679, \dodoi{10.1086/155294}

\bibitem[{{Baldwin} {et~al.}(2004){Baldwin}, {Ferland}, {Korista}, {Hamann}, \&
  {LaCluyz{\'e}}}]{Baldwin2004}
{Baldwin}, J.~A., {Ferland}, G.~J., {Korista}, K.~T., {Hamann}, F., \&
  {LaCluyz{\'e}}, A. 2004, \apj, 615, 610, \dodoi{10.1086/424683}

\bibitem[{Barbary(2016)}]{python_extinction}
Barbary, K. 2016, extinction v0.3.0,  Zenodo, \dodoi{10.5281/zenodo.804967}

\bibitem[{{Barth} {et~al.}(2003){Barth}, {Martini}, {Nelson}, \&
  {Ho}}]{Barth2003}
{Barth}, A.~J., {Martini}, P., {Nelson}, C.~H., \& {Ho}, L.~C. 2003, \apjl,
  594, L95, \dodoi{10.1086/378735}

\bibitem[{{Becker} {et~al.}(2019){Becker}, {Pettini}, {Rafelski}, {D'Odorico},
  {Boera}, {Christensen}, {Cupani}, {Ellison}, {Farina}, {Fumagalli},
  {L{\'o}pez}, {Neeleman}, {Ryan-Weber}, \& {Worseck}}]{Becker2019}
{Becker}, G.~D., {Pettini}, M., {Rafelski}, M., {et~al.} 2019, \apj, 883, 163,
  \dodoi{10.3847/1538-4357/ab3eb5}

\bibitem[{{Bentz} {et~al.}(2013){Bentz}, {Denney}, {Grier}, {Barth},
  {Peterson}, {Vestergaard}, {Bennert}, {Canalizo}, {De Rosa}, {Filippenko},
  {Gates}, {Greene}, {Li}, {Malkan}, {Pogge}, {Stern}, {Treu}, \&
  {Woo}}]{Bentz2013}
{Bentz}, M.~C., {Denney}, K.~D., {Grier}, C.~J., {et~al.} 2013, \apj, 767, 149,
  \dodoi{10.1088/0004-637X/767/2/149}

\bibitem[{{Bischetti} {et~al.}(2017){Bischetti}, {Piconcelli}, {Vietri},
  {Bongiorno}, {Fiore}, {Sani}, {Marconi}, {Duras}, {Zappacosta}, {Brusa},
  {Comastri}, {Cresci}, {Feruglio}, {Giallongo}, {La Franca}, {Mainieri},
  {Mannucci}, {Martocchia}, {Ricci}, {Schneider}, {Testa}, \&
  {Vignali}}]{Bischetti2017}
{Bischetti}, M., {Piconcelli}, E., {Vietri}, G., {et~al.} 2017, \aap, 598,
  A122, \dodoi{10.1051/0004-6361/201629301}

\bibitem[{{Boroson} \& {Green}(1992)}]{Boroson1992a}
{Boroson}, T.~A., \& {Green}, R.~F. 1992, \apjs, 80, 109,
  \dodoi{10.1086/191661}

\bibitem[{{Boroson} \& {Meyers}(1992)}]{Boroson1992b}
{Boroson}, T.~A., \& {Meyers}, K.~A. 1992, \apj, 397, 442,
  \dodoi{10.1086/171800}

\bibitem[{{Chehade} {et~al.}(2018){Chehade}, {Carnall}, {Shanks}, {Diener},
  {Fumagalli}, {Findlay}, {Metcalfe}, {Hennawi}, {Leibler}, {Murphy},
  {Prochaska}, {Irwin}, \& {Gonzalez-Solares}}]{Chehade2018}
{Chehade}, B., {Carnall}, A.~C., {Shanks}, T., {et~al.} 2018, \mnras, 478,
  1649, \dodoi{10.1093/mnras/sty690}

\bibitem[{{Clough} {et~al.}(2005){Clough}, {Shephard}, {Mlawer}, {Delamere},
  {Iacono}, {Cady-Pereira}, {Boukabara}, \& {Brown}}]{Clough2005}
{Clough}, S.~A., {Shephard}, M.~W., {Mlawer}, E.~J., {et~al.} 2005, \jqsrt, 91,
  233, \dodoi{10.1016/j.jqsrt.2004.05.058}

\bibitem[{{Coatman} {et~al.}(2016){Coatman}, {Hewett}, {Banerji}, \&
  {Richards}}]{Coatman2016}
{Coatman}, L., {Hewett}, P.~C., {Banerji}, M., \& {Richards}, G.~T. 2016,
  \mnras, 461, 647, \dodoi{10.1093/mnras/stw1360}

\bibitem[{{Coatman} {et~al.}(2017){Coatman}, {Hewett}, {Banerji}, {Richards},
  {Hennawi}, \& {Prochaska}}]{Coatman2017}
{Coatman}, L., {Hewett}, P.~C., {Banerji}, M., {et~al.} 2017, \mnras, 465,
  2120, \dodoi{10.1093/mnras/stw2797}

\bibitem[{{Davies}(2020)}]{Davies2020a}
{Davies}, F.~B. 2020, \mnras, 494, 2937, \dodoi{10.1093/mnras/staa528}

\bibitem[{{Davies} {et~al.}(2018){Davies}, {Hennawi}, {Ba{\~n}ados}, {Simcoe},
  {Decarli}, {Fan}, {Farina}, {Mazzucchelli}, {Rix}, {Venemans}, {Walter},
  {Wang}, \& {Yang}}]{Davies2018d}
{Davies}, F.~B., {Hennawi}, J.~F., {Ba{\~n}ados}, E., {et~al.} 2018, \apj, 864,
  143, \dodoi{10.3847/1538-4357/aad7f8}

\bibitem[{{De Rosa} {et~al.}(2011){De Rosa}, {Decarli}, {Walter}, {Fan},
  {Jiang}, {Kurk}, {Pasquali}, \& {Rix}}]{DeRosa2011}
{De Rosa}, G., {Decarli}, R., {Walter}, F., {et~al.} 2011, \apj, 739, 56,
  \dodoi{10.1088/0004-637X/739/2/56}

\bibitem[{{De Rosa} {et~al.}(2014){De Rosa}, {Venemans}, {Decarli}, {Gennaro},
  {Simcoe}, {Dietrich}, {Peterson}, {Walter}, {Frank}, {McMahon}, {Hewett},
  {Mortlock}, \& {Simpson}}]{DeRosa2014}
{De Rosa}, G., {Venemans}, B.~P., {Decarli}, R., {et~al.} 2014, \apj, 790, 145,
  \dodoi{10.1088/0004-637X/790/2/145}

\bibitem[{{Decarli} {et~al.}(2010){Decarli}, {Falomo}, {Treves}, {Labita},
  {Kotilainen}, \& {Scarpa}}]{Decarli2010}
{Decarli}, R., {Falomo}, R., {Treves}, A., {et~al.} 2010, \mnras, 402, 2453,
  \dodoi{10.1111/j.1365-2966.2009.16049.x}

\bibitem[{{Decarli} {et~al.}(2018){Decarli}, {Walter}, {Venemans},
  {Ba{\~n}ados}, {Bertoldi}, {Carilli}, {Fan}, {Farina}, {Mazzucchelli},
  {Riechers}, {Rix}, {Strauss}, {Wang}, \& {Yang}}]{Decarli2018}
{Decarli}, R., {Walter}, F., {Venemans}, B.~P., {et~al.} 2018, \apj, 854, 97,
  \dodoi{10.3847/1538-4357/aaa5aa}

\bibitem[{{Denney}(2012)}]{Denney2012}
{Denney}, K.~D. 2012, \apj, 759, 44, \dodoi{10.1088/0004-637X/759/1/44}

\bibitem[{{Dietrich} {et~al.}(2002){Dietrich}, {Appenzeller}, {Vestergaard}, \&
  {Wagner}}]{Dietrich2002a}
{Dietrich}, M., {Appenzeller}, I., {Vestergaard}, M., \& {Wagner}, S.~J. 2002,
  \apj, 564, 581, \dodoi{10.1086/324337}

\bibitem[{{Dietrich} {et~al.}(2003){Dietrich}, {Hamann}, {Appenzeller}, \&
  {Vestergaard}}]{Dietrich2003c}
{Dietrich}, M., {Hamann}, F., {Appenzeller}, I., \& {Vestergaard}, M. 2003,
  \apj, 596, 817, \dodoi{10.1086/378045}

\bibitem[{{D'Odorico} {et~al.}(2018){D'Odorico}, {Feruglio}, {Ferrara},
  {Gallerani}, {Pallottini}, {Carniani}, {Maiolino}, {Cristiani}, {Marconi},
  {Piconcelli}, \& {Fiore}}]{DOdorico2018}
{D'Odorico}, V., {Feruglio}, C., {Ferrara}, A., {et~al.} 2018, \apjl, 863, L29,
  \dodoi{10.3847/2041-8213/aad7b7}

\bibitem[{{Drake} {et~al.}(2019){Drake}, {Farina}, {Neeleman}, {Walter},
  {Venemans}, {Banados}, {Mazzucchelli}, \& {Decarli}}]{Drake2019}
{Drake}, A.~B., {Farina}, E.~P., {Neeleman}, M., {et~al.} 2019, \apj, 881, 131,
  \dodoi{10.3847/1538-4357/ab2984}

\bibitem[{{Eilers} {et~al.}(2017){Eilers}, {Davies}, {Hennawi}, {Prochaska},
  {Luki{\'c}}, \& {Mazzucchelli}}]{Eilers2017a}
{Eilers}, A.-C., {Davies}, F.~B., {Hennawi}, J.~F., {et~al.} 2017, \apj, 840,
  24, \dodoi{10.3847/1538-4357/aa6c60}

\bibitem[{{Eilers} {et~al.}(2020){Eilers}, {Hennawi}, {Decarli}, {Davies},
  {Venemans}, {Walter}, {Ba{\~n}ados}, {Fan}, {Farina}, {Mazzucchelli},
  {Novak}, {Schindler}, {Simcoe}, {Wang}, \& {Yang}}]{Eilers2020a}
{Eilers}, A.-C., {Hennawi}, J.~F., {Decarli}, R., {et~al.} 2020, \apj, 900, 37,
  \dodoi{10.3847/1538-4357/aba52e}

\bibitem[{{Fan} {et~al.}(2000){Fan}, {White}, {Davis}, {Becker}, {Strauss},
  {Haiman}, {Schneider}, {Gregg}, {Gunn}, {Knapp}, {Lupton}, {Anderson},
  {Anderson}, {Annis}, {Bahcall}, {Boroski}, {Brunner}, {Chen}, {Connolly},
  {Csabai}, {Doi}, {Fukugita}, {Hennessy}, {Hindsley}, {Ichikawa},
  {Ivezi{\'c}}, {Loveday}, {Meiksin}, {McKay}, {Munn}, {Newberg}, {Nichol},
  {Okamura}, {Pier}, {Sekiguchi}, {Shimasaku}, {Stoughton}, {Szalay},
  {Szokoly}, {Thakar}, {Vogeley}, \& {York}}]{Fan2000}
{Fan}, X., {White}, R.~L., {Davis}, M., {et~al.} 2000, \aj, 120, 1167,
  \dodoi{10.1086/301534}

\bibitem[{{Fan} {et~al.}(2001){Fan}, {Narayanan}, {Lupton}, {Strauss}, {Knapp},
  {Becker}, {White}, {Pentericci}, {Leggett}, {Haiman}, {Gunn}, {Ivezi{\'c}},
  {Schneider}, {Anderson}, {Brinkmann}, {Bahcall}, {Connolly}, {Csabai}, {Doi},
  {Fukugita}, {Geballe}, {Grebel}, {Harbeck}, {Hennessy}, {Lamb}, {Miknaitis},
  {Munn}, {Nichol}, {Okamura}, {Pier}, {Prada}, {Richards}, {Szalay}, \&
  {York}}]{Fan2001c}
{Fan}, X., {Narayanan}, V.~K., {Lupton}, R.~H., {et~al.} 2001, \aj, 122, 2833,
  \dodoi{10.1086/324111}

\bibitem[{{Fan} {et~al.}(2006){Fan}, {Strauss}, {Richards}, {Hennawi},
  {Becker}, {White}, {Diamond-Stanic}, {Donley}, {Jiang}, {Kim}, {Vestergaard},
  {Young}, {Gunn}, {Lupton}, {Knapp}, {Schneider}, {Brandt}, {Bahcall},
  {Barentine}, {Brinkmann}, {Brewington}, {Fukugita}, {Harvanek}, {Kleinman},
  {Krzesinski}, {Long}, {Neilsen}, {Nitta}, {Snedden}, \& {Voges}}]{Fan2006}
{Fan}, X., {Strauss}, M.~A., {Richards}, G.~T., {et~al.} 2006, \aj, 131, 1203,
  \dodoi{10.1086/500296}

\bibitem[{{Farina} {et~al.}(2019){Farina}, {Arrigoni-Battaia}, {Costa},
  {Walter}, {Hennawi}, {Drake}, {Decarli}, {Gutcke}, {Mazzucchelli},
  {Neeleman}, {Georgiev}, {Eilers}, {Davies}, {Ba{\~n}ados}, {Fan}, {Onoue},
  {Schindler}, {Venemans}, {Wang}, {Yang}, {Rabien}, \& {Busoni}}]{Farina2019}
{Farina}, E.~P., {Arrigoni-Battaia}, F., {Costa}, T., {et~al.} 2019, \apj, 887,
  196, \dodoi{10.3847/1538-4357/ab5847}

\bibitem[{{Ferland} {et~al.}(1998){Ferland}, {Korista}, {Verner}, {Ferguson},
  {Kingdon}, \& {Verner}}]{Ferland1998}
{Ferland}, G.~J., {Korista}, K.~T., {Verner}, D.~A., {et~al.} 1998, \pasp, 110,
  761, \dodoi{10.1086/316190}

\bibitem[{{Francis} {et~al.}(1991){Francis}, {Hewett}, {Foltz}, {Chaffee},
  {Weymann}, \& {Morris}}]{Francis1991}
{Francis}, P.~J., {Hewett}, P.~C., {Foltz}, C.~B., {et~al.} 1991, \apj, 373,
  465, \dodoi{10.1086/170066}

\bibitem[{{Freudling} {et~al.}(2003){Freudling}, {Corbin}, \&
  {Korista}}]{Freudling2003}
{Freudling}, W., {Corbin}, M.~R., \& {Korista}, K.~T. 2003, \apjl, 587, L67,
  \dodoi{10.1086/375338}

\bibitem[{{Friaca} \& {Terlevich}(1998)}]{Friaca1998}
{Friaca}, A. C.~S., \& {Terlevich}, R.~J. 1998, \mnras, 298, 399,
  \dodoi{10.1046/j.1365-8711.1998.01626.x}

\bibitem[{{Gaskell}(1982)}]{Gaskell1982}
{Gaskell}, C.~M. 1982, \apj, 263, 79, \dodoi{10.1086/160481}

\bibitem[{{Grandi}(1982)}]{Grandi1982}
{Grandi}, S.~A. 1982, \apj, 255, 25, \dodoi{10.1086/159799}

\bibitem[{{Gravity Collaboration} {et~al.}(2018){Gravity Collaboration},
  {Sturm}, {Dexter}, {Pfuhl}, {Stock}, {Davies}, {Lutz}, {Cl{\'e}net},
  {Eckart}, {Eisenhauer}, {Genzel}, {Gratadour}, {H{\"o}nig}, {Kishimoto},
  {Lacour}, {Millour}, {Netzer}, {Perrin}, {Peterson}, {Petrucci}, {Rouan},
  {Waisberg}, {Woillez}, {Amorim}, {Brandner}, {F{\"o}rster Schreiber},
  {Garcia}, {Gillessen}, {Ott}, {Paumard}, {Perraut}, {Scheithauer},
  {Straubmeier}, {Tacconi}, \& {Widmann}}]{GravityCollab2018}
{Gravity Collaboration}, {Sturm}, E., {Dexter}, J., {et~al.} 2018, \nat, 563,
  657, \dodoi{10.1038/s41586-018-0731-9}

\bibitem[{{Gullikson} {et~al.}(2014){Gullikson}, {Dodson-Robinson}, \&
  {Kraus}}]{Gullikson2014}
{Gullikson}, K., {Dodson-Robinson}, S., \& {Kraus}, A. 2014, \aj, 148, 53,
  \dodoi{10.1088/0004-6256/148/3/53}

\bibitem[{{Harris} {et~al.}(2020){Harris}, {Jarrod Millman}, {van der Walt},
  {Gommers}, {Virtanen}, {Cournapeau}, {Wieser}, {Taylor}, {Berg}, {Smith},
  {Kern}, {Picus}, {Hoyer}, {van Kerkwijk}, {Brett}, {Haldane}, {Fern{\'a}ndez
  del R{\'\i}o}, {Wiebe}, {Peterson}, {G{\'e}rard-Marchant}, {Sheppard},
  {Reddy}, {Weckesser}, {Abbasi}, {Gohlke}, \& {Oliphant}}]{numpy2020}
{Harris}, C.~R., {Jarrod Millman}, K., {van der Walt}, S.~J., {et~al.} 2020,
  arXiv e-prints, arXiv:2006.10256.
\newblock \doarXiv{2006.10256}

\bibitem[{{Hewett} \& {Wild}(2010)}]{Hewett2010}
{Hewett}, P.~C., \& {Wild}, V. 2010, \mnras, 405, 2302,
  \dodoi{10.1111/j.1365-2966.2010.16648.x}

\bibitem[{{Horne}(1986)}]{Horne1986}
{Horne}, K. 1986, \pasp, 98, 609, \dodoi{10.1086/131801}

\bibitem[{{Iwamuro} {et~al.}(2004){Iwamuro}, {Kimura}, {Eto}, {Maihara},
  {Motohara}, {Yoshii}, \& {Doi}}]{Iwamuro2004}
{Iwamuro}, F., {Kimura}, M., {Eto}, S., {et~al.} 2004, \apj, 614, 69,
  \dodoi{10.1086/423610}

\bibitem[{{Iwamuro} {et~al.}(2002){Iwamuro}, {Motohara}, {Maihara}, {Kimura},
  {Yoshii}, \& {Doi}}]{Iwamuro2002}
{Iwamuro}, F., {Motohara}, K., {Maihara}, T., {et~al.} 2002, \apj, 565, 63,
  \dodoi{10.1086/324540}

\bibitem[{{Izumi} {et~al.}(2018){Izumi}, {Onoue}, {Shirakata}, {Nagao},
  {Kohno}, {Matsuoka}, {Imanishi}, {Strauss}, {Kashikawa}, {Schulze},
  {Silverman}, {Fujimoto}, {Harikane}, {Toba}, {Umehata}, {Nakanishi},
  {Greene}, {Tamura}, {Taniguchi}, {Yamaguchi}, {Goto}, {Hashimoto},
  {Ikarashi}, {Iono}, {Iwasawa}, {Lee}, {Makiya}, {Minezaki}, \&
  {Tang}}]{Izumi2018}
{Izumi}, T., {Onoue}, M., {Shirakata}, H., {et~al.} 2018, \pasj, 70, 36,
  \dodoi{10.1093/pasj/psy026}

\bibitem[{{Izumi} {et~al.}(2019){Izumi}, {Onoue}, {Matsuoka}, {Nagao},
  {Strauss}, {Imanishi}, {Kashikawa}, {Fujimoto}, {Kohno}, {Toba}, {Umehata},
  {Goto}, {Ueda}, {Shirakata}, {Silverman}, {Greene}, {Harikane}, {Hashimoto},
  {Ikarashi}, {Iono}, {Iwasawa}, {Lee}, {Minezaki}, {Nakanishi}, {Tamura},
  {Tang}, \& {Taniguchi}}]{Izumi2019}
{Izumi}, T., {Onoue}, M., {Matsuoka}, Y., {et~al.} 2019, \pasj, 71, 111,
  \dodoi{10.1093/pasj/psz096}

\bibitem[{{Jiang} {et~al.}(2007){Jiang}, {Fan}, {Ivezi{\'c}}, {Richards},
  {Schneider}, {Strauss}, \& {Kelly}}]{Jiang2007}
{Jiang}, L., {Fan}, X., {Ivezi{\'c}}, {\v Z}., {et~al.} 2007, \apj, 656, 680,
  \dodoi{10.1086/510831}

\bibitem[{{Jiang} {et~al.}(2015){Jiang}, {McGreer}, {Fan}, {Bian}, {Cai},
  {Cl{\'e}ment}, {Wang}, \& {Fan}}]{Jiang2015}
{Jiang}, L., {McGreer}, I.~D., {Fan}, X., {et~al.} 2015, \aj, 149, 188,
  \dodoi{10.1088/0004-6256/149/6/188}

\bibitem[{{Jiang} {et~al.}(2008){Jiang}, {Fan}, {Annis}, {Becker}, {White},
  {Chiu}, {Lin}, {Lupton}, {Richards}, {Strauss}, {Jester}, \&
  {Schneider}}]{Jiang2008}
{Jiang}, L., {Fan}, X., {Annis}, J., {et~al.} 2008, \aj, 135, 1057,
  \dodoi{10.1088/0004-6256/135/3/1057}

\bibitem[{{Jiang} {et~al.}(2016){Jiang}, {McGreer}, {Fan}, {Strauss},
  {Ba{\~n}ados}, {Becker}, {Bian}, {Farnsworth}, {Shen}, {Wang}, {Wang},
  {Wang}, {White}, {Wu}, {Wu}, {Yang}, \& {Yang}}]{Jiang2016}
{Jiang}, L., {McGreer}, I.~D., {Fan}, X., {et~al.} 2016, \apj, 833, 222,
  \dodoi{10.3847/1538-4357/833/2/222}

\bibitem[{{Kaspi} {et~al.}(2005){Kaspi}, {Maoz}, {Netzer}, {Peterson},
  {Vestergaard}, \& {Jannuzi}}]{Kaspi2005}
{Kaspi}, S., {Maoz}, D., {Netzer}, H., {et~al.} 2005, \apj, 629, 61,
  \dodoi{10.1086/431275}

\bibitem[{{Kaspi} {et~al.}(2000){Kaspi}, {Smith}, {Netzer}, {Maoz}, {Jannuzi},
  \& {Giveon}}]{Kaspi2000}
{Kaspi}, S., {Smith}, P.~S., {Netzer}, H., {et~al.} 2000, \apj, 533, 631,
  \dodoi{10.1086/308704}

\bibitem[{{Krolik}(2001)}]{Krolik2001a}
{Krolik}, J.~H. 2001, \apj, 551, 72, \dodoi{10.1086/320091}

\bibitem[{{Krolik} \& {Begelman}(1986)}]{Krolik1986}
{Krolik}, J.~H., \& {Begelman}, M.~C. 1986, \apjl, 308, L55,
  \dodoi{10.1086/184743}

\bibitem[{{Kurk} {et~al.}(2007){Kurk}, {Walter}, {Fan}, {Jiang}, {Riechers},
  {Rix}, {Pentericci}, {Strauss}, {Carilli}, \& {Wagner}}]{Kurk2007}
{Kurk}, J.~D., {Walter}, F., {Fan}, X., {et~al.} 2007, \apj, 669, 32,
  \dodoi{10.1086/521596}

\bibitem[{{Lynden-Bell}(1969)}]{LyndenBell1969}
{Lynden-Bell}, D. 1969, \nat, 223, 690, \dodoi{10.1038/223690a0}

\bibitem[{{Maddox} {et~al.}(2008){Maddox}, {Hewett}, {Warren}, \&
  {Croom}}]{Maddox2008}
{Maddox}, N., {Hewett}, P.~C., {Warren}, S.~J., \& {Croom}, S.~M. 2008, \mnras,
  386, 1605, \dodoi{10.1111/j.1365-2966.2008.13138.x}

\bibitem[{{Maiolino} {et~al.}(2003){Maiolino}, {Juarez}, {Mujica}, {Nagar}, \&
  {Oliva}}]{Maiolino2003}
{Maiolino}, R., {Juarez}, Y., {Mujica}, R., {Nagar}, N.~M., \& {Oliva}, E.
  2003, \apjl, 596, L155, \dodoi{10.1086/379600}

\bibitem[{{Maiolino} {et~al.}(2005){Maiolino}, {Cox}, {Caselli}, {Beelen},
  {Bertoldi}, {Carilli}, {Kaufman}, {Menten}, {Nagao}, {Omont}, {Wei{\ss}},
  {Walmsley}, \& {Walter}}]{Maiolino2005}
{Maiolino}, R., {Cox}, P., {Caselli}, P., {et~al.} 2005, \aap, 440, L51,
  \dodoi{10.1051/0004-6361:200500165}

\bibitem[{{Matsuoka} {et~al.}(2016){Matsuoka}, {Onoue}, {Kashikawa}, {Iwasawa},
  {Strauss}, {Nagao}, {Imanishi}, {Niida}, {Toba}, {Akiyama}, {Asami}, {Bosch},
  {Foucaud}, {Furusawa}, {Goto}, {Gunn}, {Harikane}, {Ikeda}, {Kawaguchi},
  {Kikuta}, {Komiyama}, {Lupton}, {Minezaki}, {Miyazaki}, {Morokuma},
  {Murayama}, {Nishizawa}, {Ono}, {Ouchi}, {Price}, {Sameshima}, {Silverman},
  {Sugiyama}, {Tait}, {Takada}, {Takata}, {Tanaka}, {Tang}, \&
  {Utsumi}}]{Matsuoka2016}
{Matsuoka}, Y., {Onoue}, M., {Kashikawa}, N., {et~al.} 2016, \apj, 828, 26,
  \dodoi{10.3847/0004-637X/828/1/26}

\bibitem[{{Matsuoka} {et~al.}(2018){Matsuoka}, {Iwasawa}, {Onoue}, {Kashikawa},
  {Strauss}, {Lee}, {Imanishi}, {Nagao}, {Akiyama}, {Asami}, {Bosch},
  {Furusawa}, {Goto}, {Gunn}, {Harikane}, {Ikeda}, {Izumi}, {Kawaguchi},
  {Kato}, {Kikuta}, {Kohno}, {Komiyama}, {Lupton}, {Minezaki}, {Miyazaki},
  {Morokuma}, {Murayama}, {Niida}, {Nishizawa}, {Oguri}, {Ono}, {Ouchi},
  {Price}, {Sameshima}, {Schulze}, {Shirakata}, {Silverman}, {Sugiyama},
  {Tait}, {Takada}, {Takata}, {Tanaka}, {Tang}, {Toba}, {Utsumi}, {Wang}, \&
  {Yamashita}}]{Matsuoka2018b}
{Matsuoka}, Y., {Iwasawa}, K., {Onoue}, M., {et~al.} 2018, \apjs, 237, 5,
  \dodoi{10.3847/1538-4365/aac724}

\bibitem[{{Matsuoka} {et~al.}(2019{\natexlab{a}}){Matsuoka}, {Iwasawa},
  {Onoue}, {Kashikawa}, {Strauss}, {Lee}, {Imanishi}, {Nagao}, {Akiyama},
  {Asami}, {Bosch}, {Furusawa}, {Goto}, {Gunn}, {Harikane}, {Ikeda}, {Izumi},
  {Kawaguchi}, {Kato}, {Kikuta}, {Kohno}, {Komiyama}, {Koyama}, {Lupton},
  {Minezaki}, {Miyazaki}, {Murayama}, {Niida}, {Nishizawa}, {Noboriguchi},
  {Oguri}, {Ono}, {Ouchi}, {Price}, {Sameshima}, {Schulze}, {Silverman},
  {Sugiyama}, {Tait}, {Takada}, {Takata}, {Tanaka}, {Tang}, {Toba}, {Utsumi},
  {Wang}, \& {Yamashita}}]{Matsuoka2019b}
---. 2019{\natexlab{a}}, \apj, 883, 183, \dodoi{10.3847/1538-4357/ab3c60}

\bibitem[{{Matsuoka} {et~al.}(2019{\natexlab{b}}){Matsuoka}, {Onoue},
  {Kashikawa}, {Strauss}, {Iwasawa}, {Lee}, {Imanishi}, {Nagao}, {Akiyama},
  {Asami}, {Bosch}, {Furusawa}, {Goto}, {Gunn}, {Harikane}, {Ikeda}, {Izumi},
  {Kawaguchi}, {Kato}, {Kikuta}, {Kohno}, {Komiyama}, {Koyama}, {Lupton},
  {Minezaki}, {Miyazaki}, {Murayama}, {Niida}, {Nishizawa}, {Noboriguchi},
  {Oguri}, {Ono}, {Ouchi}, {Price}, {Sameshima}, {Schulze}, {Shirakata},
  {Silverman}, {Sugiyama}, {Tait}, {Takada}, {Takata}, {Tanaka}, {Tang},
  {Toba}, {Utsumi}, {Wang}, \& {Yamashita}}]{Matsuoka2019a}
{Matsuoka}, Y., {Onoue}, M., {Kashikawa}, N., {et~al.} 2019{\natexlab{b}},
  \apj, 872, L2, \dodoi{10.3847/2041-8213/ab0216}

\bibitem[{{Matteucci}(1994)}]{Matteucci1994}
{Matteucci}, F. 1994, \aap, 288, 57

\bibitem[{{Matteucci} \& {Greggio}(1986)}]{Matteucci1986}
{Matteucci}, F., \& {Greggio}, L. 1986, \aap, 154, 279

\bibitem[{{Matteucci} \& {Recchi}(2001)}]{Matteucci2001}
{Matteucci}, F., \& {Recchi}, S. 2001, \apj, 558, 351, \dodoi{10.1086/322472}

\bibitem[{{Mazzucchelli} {et~al.}(2017){Mazzucchelli}, {Ba{\~n}ados},
  {Venemans}, {Decarli}, {Farina}, {Walter}, {Eilers}, {Rix}, {Simcoe},
  {Stern}, {Fan}, {Schlafly}, {De Rosa}, {Hennawi}, {Chambers}, {Greiner},
  {Burgett}, {Draper}, {Kaiser}, {Kudritzki}, {Magnier}, {Metcalfe}, {Waters},
  \& {Wainscoat}}]{Mazzucchelli2017}
{Mazzucchelli}, C., {Ba{\~n}ados}, E., {Venemans}, B.~P., {et~al.} 2017, \apj,
  849, 91, \dodoi{10.3847/1538-4357/aa9185}

\bibitem[{{McIntosh} {et~al.}(1999){McIntosh}, {Rix}, {Rieke}, \&
  {Foltz}}]{McIntosh1999b}
{McIntosh}, D.~H., {Rix}, H.~W., {Rieke}, M.~J., \& {Foltz}, C.~B. 1999, \apjl,
  517, L73, \dodoi{10.1086/312033}

\bibitem[{{McLure} \& {Dunlop}(2004)}]{McLure2004}
{McLure}, R.~J., \& {Dunlop}, J.~S. 2004, \mnras, 352, 1390,
  \dodoi{10.1111/j.1365-2966.2004.08034.x}

\bibitem[{{Mediavilla} {et~al.}(2020){Mediavilla}, {Jim{\'e}nez-vicente},
  {Mej{\'\i}a-restrepo}, {Motta}, {Falco}, {Mu{\~n}oz}, {Fian}, \&
  {Guerras}}]{Mediavilla2020}
{Mediavilla}, E., {Jim{\'e}nez-vicente}, J., {Mej{\'\i}a-restrepo}, J.,
  {et~al.} 2020, \apj, 895, 111, \dodoi{10.3847/1538-4357/ab8ae0}

\bibitem[{{Mej{\'\i}a-Restrepo} {et~al.}(2018){Mej{\'\i}a-Restrepo},
  {Trakhtenbrot}, {Lira}, \& {Netzer}}]{MejiaRestrepo2018a}
{Mej{\'\i}a-Restrepo}, J.~E., {Trakhtenbrot}, B., {Lira}, P., \& {Netzer}, H.
  2018, \mnras, 478, 1929, \dodoi{10.1093/mnras/sty1086}

\bibitem[{{Mej{\'\i}a-Restrepo} {et~al.}(2016){Mej{\'\i}a-Restrepo},
  {Trakhtenbrot}, {Lira}, {Netzer}, \& {Capellupo}}]{MeijaRestrepo2016}
{Mej{\'\i}a-Restrepo}, J.~E., {Trakhtenbrot}, B., {Lira}, P., {Netzer}, H., \&
  {Capellupo}, D.~M. 2016, \mnras, 460, 187, \dodoi{10.1093/mnras/stw568}

\bibitem[{{Meyer} {et~al.}(2019){Meyer}, {Bosman}, \& {Ellis}}]{Meyer2019c}
{Meyer}, R.~A., {Bosman}, S. E.~I., \& {Ellis}, R.~S. 2019, \mnras, 487, 3305,
  \dodoi{10.1093/mnras/stz1504}

\bibitem[{{Michel-Dansac} {et~al.}(2020){Michel-Dansac}, {Blaizot}, {Garel},
  {Verhamme}, {Kimm}, \& {Trebitsch}}]{Michel-Dansac2020}
{Michel-Dansac}, L., {Blaizot}, J., {Garel}, T., {et~al.} 2020, \aap, 635,
  A154, \dodoi{10.1051/0004-6361/201834961}

\bibitem[{{Mortlock} {et~al.}(2009){Mortlock}, {Patel}, {Warren}, {Venemans},
  {McMahon}, {Hewett}, {Simpson}, {Sharp}, {Burningham}, {Dye}, {Ellis},
  {Gonzales-Solares}, \& {Hu{\'e}lamo}}]{Mortlock2009}
{Mortlock}, D.~J., {Patel}, M., {Warren}, S.~J., {et~al.} 2009, \aap, 505, 97,
  \dodoi{10.1051/0004-6361/200811161}

\bibitem[{{Mortlock} {et~al.}(2011){Mortlock}, {Warren}, {Venemans}, {Patel},
  {Hewett}, {McMahon}, {Simpson}, {Theuns}, {Gonz{\'a}les-Solares}, {Adamson},
  {Dye}, {Hambly}, {Hirst}, {Irwin}, {Kuiper}, {Lawrence}, \&
  {R{\"o}ttgering}}]{Mortlock2011}
{Mortlock}, D.~J., {Warren}, S.~J., {Venemans}, B.~P., {et~al.} 2011, \nat,
  474, 616, \dodoi{10.1038/nature10159}

\bibitem[{{Murray} {et~al.}(1995){Murray}, {Chiang}, {Grossman}, \&
  {Voit}}]{Murray1995a}
{Murray}, N., {Chiang}, J., {Grossman}, S.~A., \& {Voit}, G.~M. 1995, \apj,
  451, 498, \dodoi{10.1086/176238}

\bibitem[{Newville {et~al.}(2014)Newville, Stensitzki, Allen, \&
  Ingargiola}]{lmfit2014}
Newville, M., Stensitzki, T., Allen, D.~B., \& Ingargiola, A. 2014, {LMFIT:
  Non-Linear Least-Square Minimization and Curve-Fitting for Python}, 0.8.0,
  Zenodo, \dodoi{10.5281/zenodo.11813}

\bibitem[{{Nguyen} {et~al.}(2020){Nguyen}, {Lira}, {Trakhtenbrot}, {Netzer},
  {Cicone}, {Maiolino}, \& {Shemmer}}]{Nguyen2020}
{Nguyen}, N.~H., {Lira}, P., {Trakhtenbrot}, B., {et~al.} 2020, \apj, 895, 74,
  \dodoi{10.3847/1538-4357/ab8bd3}

\bibitem[{{Onken} {et~al.}(2004){Onken}, {Ferrarese}, {Merritt}, {Peterson},
  {Pogge}, {Vestergaard}, \& {Wandel}}]{Onken2004}
{Onken}, C.~A., {Ferrarese}, L., {Merritt}, D., {et~al.} 2004, \apj, 615, 645,
  \dodoi{10.1086/424655}

\bibitem[{{Onoue} {et~al.}(2019){Onoue}, {Kashikawa}, {Matsuoka}, {Kato},
  {Izumi}, {Nagao}, {Strauss}, {Harikane}, {Imanishi}, {Ito}, {Iwasawa},
  {Kawaguchi}, {Lee}, {Noboriguchi}, {Suh}, {Tanaka}, \& {Toba}}]{Onoue2019}
{Onoue}, M., {Kashikawa}, N., {Matsuoka}, Y., {et~al.} 2019, \apj, 880, 77,
  \dodoi{10.3847/1538-4357/ab29e9}

\bibitem[{{Onoue} {et~al.}(2020){Onoue}, {Ba{\~n}ados}, {Mazzucchelli},
  {Venemans}, {Schindler}, {Walter}, {Hennawi}, {Andika}, {Davies}, {Decarli},
  {Farina}, {Jahnke}, {Nagao}, {Tominaga}, \& {Wang}}]{Onoue2020}
{Onoue}, M., {Ba{\~n}ados}, E., {Mazzucchelli}, C., {et~al.} 2020, \apj, 898,
  105, \dodoi{10.3847/1538-4357/aba193}

\bibitem[{pandas~development team(2020)}]{pandas_software}
pandas~development team, T. 2020, pandas-dev/pandas: Pandas, latest,  Zenodo,
  \dodoi{10.5281/zenodo.3509134}

\bibitem[{{Park} {et~al.}(2013){Park}, {Woo}, {Denney}, \& {Shin}}]{Park2013}
{Park}, D., {Woo}, J.-H., {Denney}, K.~D., \& {Shin}, J. 2013, \apj, 770, 87,
  \dodoi{10.1088/0004-637X/770/2/87}

\bibitem[{{Peterson}(1993)}]{Peterson1993}
{Peterson}, B.~M. 1993, \pasp, 105, 247, \dodoi{10.1086/133140}

\bibitem[{{Peterson} {et~al.}(2004){Peterson}, {Ferrarese}, {Gilbert}, {Kaspi},
  {Malkan}, {Maoz}, {Merritt}, {Netzer}, {Onken}, {Pogge}, {Vestergaard}, \&
  {Wandel}}]{Peterson2004}
{Peterson}, B.~M., {Ferrarese}, L., {Gilbert}, K.~M., {et~al.} 2004, \apj, 613,
  682, \dodoi{10.1086/423269}

\bibitem[{{Planck Collaboration} {et~al.}(2016){Planck Collaboration}, {Ade},
  {Aghanim}, {Arnaud}, {Ashdown}, {Aumont}, {Baccigalupi}, {Banday},
  {Barreiro}, {Bartlett}, \& et~al.}]{PlanckCollaboration2016}
{Planck Collaboration}, {Ade}, P.~A.~R., {Aghanim}, N., {et~al.} 2016, \aap,
  594, A13, \dodoi{10.1051/0004-6361/201525830}

\bibitem[{{Prochaska} {et~al.}(2020){Prochaska}, {Hennawi}, {Westfall},
  {Cooke}, {Wang}, {Hsyu}, {Davies}, \& {Farina}}]{PypeitProchaska2020}
{Prochaska}, J.~X., {Hennawi}, J.~F., {Westfall}, K.~B., {et~al.} 2020, arXiv
  e-prints, arXiv:2005.06505.
\newblock \doarXiv{2005.06505}

\bibitem[{Prochaska {et~al.}(2016)Prochaska, Tejos, Crighton, jnburchett,
  Tuo-Ji, tiffanyhsyu, ktirimba, jhennawi, O'Meara, \& Werk}]{linetools2016}
Prochaska, J.~X., Tejos, N., Crighton, N., {et~al.} 2016, linetools/linetools:
  Second major release, v0.2,  Zenodo, \dodoi{10.5281/zenodo.168270}

\bibitem[{Prochaska {et~al.}(2019)Prochaska, Hennawi, Cooke, Westfall, Wang,
  EmAstro, tiffanyhsyu, Wasserman, Villaume, marijana777, Young, Simha, Wilde,
  Tejos, Isbell, Betts, \& Holden}]{PypeitProchaska2019}
Prochaska, J.~X., Hennawi, J., Cooke, R., {et~al.} 2019, pypeit/PypeIt:
  Releasing for DOI, 0.11.0.1,  Zenodo, \dodoi{10.5281/zenodo.3506873}

\bibitem[{{Reed} {et~al.}(2019){Reed}, {Banerji}, {Becker}, {Hewett},
  {Martini}, {McMahon}, {Pons}, {Rauch}, {Abbott}, {Allam}, {Annis}, {Avila},
  {Bertin}, {Brooks}, {Buckley-Geer}, {Carnero Rosell}, {Carrasco Kind},
  {Carretero}, {Castander}, {Cunha}, {D'Andrea}, {da Costa}, {De Vicente},
  {Desai}, {Diehl}, {Doel}, {Evrard}, {Flaugher}, {Frieman},
  {Garc{\'\i}a-Bellido}, {Gaztanaga}, {Gruen}, {Gschwend}, {Gutierrez},
  {Hollowood}, {Honscheid}, {Hoyle}, {James}, {Kuehn}, {Lahav}, {Lima}, {Maia},
  {Marshall}, {Miquel}, {Ogand o}, {Plazas}, {Roodman}, {Sanchez}, {Scarpine},
  {Schubnell}, {Serrano}, {Sevilla-Noarbe}, {Smith}, {Smith}, {Sobreira},
  {Suchyta}, {Swanson}, {Tarle}, {Thomas}, {Tucker}, \& {Vikram}}]{Reed2019}
{Reed}, S.~L., {Banerji}, M., {Becker}, G.~D., {et~al.} 2019, \mnras, 487,
  1874, \dodoi{10.1093/mnras/stz1341}

\bibitem[{{Richards} {et~al.}(2002){Richards}, {Fan}, {Newberg}, {Strauss},
  {Vanden Berk}, {Schneider}, {Yanny}, {Boucher}, {Burles}, {Frieman}, {Gunn},
  {Hall}, {Ivezi{\'c}}, {Kent}, {Loveday}, {Lupton}, {Rockosi}, {Schlegel},
  {Stoughton}, {SubbaRao}, \& {York}}]{Richards2002}
{Richards}, G.~T., {Fan}, X., {Newberg}, H.~J., {et~al.} 2002, \aj, 123, 2945,
  \dodoi{10.1086/340187}

\bibitem[{{Richards} {et~al.}(2011){Richards}, {Kruczek}, {Gallagher}, {Hall},
  {Hewett}, {Leighly}, {Deo}, {Kratzer}, \& {Shen}}]{Richards2011}
{Richards}, G.~T., {Kruczek}, N.~E., {Gallagher}, S.~C., {et~al.} 2011, \aj,
  141, 167, \dodoi{10.1088/0004-6256/141/5/167}

\bibitem[{{Runnoe} {et~al.}(2013){Runnoe}, {Brotherton}, {Shang}, \&
  {DiPompeo}}]{Runnoe2013c}
{Runnoe}, J.~C., {Brotherton}, M.~S., {Shang}, Z., \& {DiPompeo}, M.~A. 2013,
  \mnras, 434, 848, \dodoi{10.1093/mnras/stt1077}

\bibitem[{{Sameshima} {et~al.}(2017){Sameshima}, {Yoshii}, \&
  {Kawara}}]{Sameshima2017}
{Sameshima}, H., {Yoshii}, Y., \& {Kawara}, K. 2017, \apj, 834, 203,
  \dodoi{10.3847/1538-4357/834/2/203}

\bibitem[{{Shen}(2013)}]{Shen2013review}
{Shen}, Y. 2013, Bulletin of the Astronomical Society of India, 41, 61.
\newblock \doarXiv{1302.2643}

\bibitem[{{Shen} {et~al.}(2008){Shen}, {Greene}, {Strauss}, {Richards}, \&
  {Schneider}}]{Shen2008}
{Shen}, Y., {Greene}, J.~E., {Strauss}, M.~A., {Richards}, G.~T., \&
  {Schneider}, D.~P. 2008, \apj, 680, 169, \dodoi{10.1086/587475}

\bibitem[{{Shen} {et~al.}(2011){Shen}, {Richards}, {Strauss}, {Hall},
  {Schneider}, {Snedden}, {Bizyaev}, {Brewington}, {Malanushenko},
  {Malanushenko}, {Oravetz}, {Pan}, \& {Simmons}}]{Shen2011}
{Shen}, Y., {Richards}, G.~T., {Strauss}, M.~A., {et~al.} 2011, \apjs, 194, 45,
  \dodoi{10.1088/0067-0049/194/2/45}

\bibitem[{{Shen} {et~al.}(2016){Shen}, {Brandt}, {Richards}, {Denney},
  {Greene}, {Grier}, {Ho}, {Peterson}, {Petitjean}, {Schneider}, {Tao}, \&
  {Trump}}]{Shen2016}
{Shen}, Y., {Brandt}, W.~N., {Richards}, G.~T., {et~al.} 2016, \apj, 831, 7,
  \dodoi{10.3847/0004-637X/831/1/7}

\bibitem[{{Shen} {et~al.}(2019{\natexlab{a}}){Shen}, {Wu}, {Jiang},
  {Ba{\~n}ados}, {Fan}, {Ho}, {Riechers}, {Strauss}, {Venemans}, {Vestergaard},
  {Walter}, {Wang}, {Willott}, {Wu}, \& {Yang}}]{Shen2019a}
{Shen}, Y., {Wu}, J., {Jiang}, L., {et~al.} 2019{\natexlab{a}}, \apj, 873, 35,
  \dodoi{10.3847/1538-4357/ab03d9}

\bibitem[{{Shen} {et~al.}(2019{\natexlab{b}}){Shen}, {Hall}, {Horne}, {Zhu},
  {McGreer}, {Simm}, {Trump}, {Kinemuchi}, {Brandt}, {Green}, {Grier}, {Guo},
  {Ho}, {Homayouni}, {Jiang}, {I-Hsiu Li}, {Morganson}, {Petitjean},
  {Richards}, {Schneider}, {Starkey}, {Wang}, {Chambers}, {Kaiser},
  {Kudritzki}, {Magnier}, \& {Waters}}]{Shen2019b}
{Shen}, Y., {Hall}, P.~B., {Horne}, K., {et~al.} 2019{\natexlab{b}}, \apjs,
  241, 34, \dodoi{10.3847/1538-4365/ab074f}

\bibitem[{{Shin} {et~al.}(2019){Shin}, {Nagao}, {Woo}, \& {Le}}]{Shin2019}
{Shin}, J., {Nagao}, T., {Woo}, J.-H., \& {Le}, H. A.~N. 2019, \apj, 874, 22,
  \dodoi{10.3847/1538-4357/ab05da}

\bibitem[{{Stoughton} {et~al.}(2002){Stoughton}, {Lupton}, {Bernardi},
  {Blanton}, {Burles}, {Castander}, {Connolly}, {Eisenstein}, {Frieman},
  {Hennessy}, {Hindsley}, {Ivezi{\'c}}, {Kent}, {Kunszt}, {Lee}, {Meiksin},
  {Munn}, {Newberg}, {Nichol}, {Nicinski}, {Pier}, {Richards}, {Richmond},
  {Schlegel}, {Smith}, {Strauss}, {SubbaRao}, {Szalay}, {Thakar}, {Tucker},
  {Vanden Berk}, {Yanny}, {Adelman}, {Anderson}, {Anderson}, {Annis},
  {Bahcall}, {Bakken}, {Bartelmann}, {Bastian}, {Bauer}, {Berman},
  {B{\"o}hringer}, {Boroski}, {Bracker}, {Briegel}, {Briggs}, {Brinkmann},
  {Brunner}, {Carey}, {Carr}, {Chen}, {Christian}, {Colestock}, {Crocker},
  {Csabai}, {Czarapata}, {Dalcanton}, {Davidsen}, {Davis}, {Dehnen},
  {Dodelson}, {Doi}, {Dombeck}, {Donahue}, {Ellman}, {Elms}, {Evans}, {Eyer},
  {Fan}, {Federwitz}, {Friedman}, {Fukugita}, {Gal}, {Gillespie}, {Glazebrook},
  {Gray}, {Grebel}, {Greenawalt}, {Greene}, {Gunn}, {de Haas}, {Haiman},
  {Haldeman}, {Hall}, {Hamabe}, {Hansen}, {Harris}, {Harris}, {Harvanek},
  {Hawley}, {Hayes}, {Heckman}, {Helmi}, {Henden}, {Hogan}, {Hogg}, {Holmgren},
  {Holtzman}, {Huang}, {Hull}, {Ichikawa}, {Ichikawa}, {Johnston}, {Kauffmann},
  {Kim}, {Kimball}, {Kinney}, {Klaene}, {Kleinman}, {Klypin}, {Knapp},
  {Korienek}, {Krolik}, {Kron}, {Krzesi{\'n}ski}, {Lamb}, {Leger},
  {Limmongkol}, {Lindenmeyer}, {Long}, {Loomis}, {Loveday}, {MacKinnon},
  {Mannery}, {Mantsch}, {Margon}, {McGehee}, {McKay}, {McLean}, {Menou},
  {Merelli}, {Mo}, {Monet}, {Nakamura}, {Narayanan}, {Nash}, {Neilsen},
  {Newman}, {Nitta}, {Odenkirchen}, {Okada}, {Okamura}, {Ostriker}, {Owen},
  {Pauls}, {Peoples}, {Peterson}, {Petravick}, {Pope}, {Pordes}, {Postman},
  {Prosapio}, {Quinn}, {Rechenmacher}, {Rivetta}, {Rix}, {Rockosi}, {Rosner},
  {Ruthmansdorfer}, {Sandford}, {Schneider}, {Scranton}, {Sekiguchi}, {Sergey},
  {Sheth}, {Shimasaku}, {Smee}, {Snedden}, {Stebbins}, {Stubbs}, {Szapudi},
  {Szkody}, {Szokoly}, {Tabachnik}, {Tsvetanov}, {Uomoto}, {Vogeley}, {Voges},
  {Waddell}, {Walterbos}, {Wang}, {Watanabe}, {Weinberg}, {White}, {White},
  {Wilhite}, {Wolfe}, {Yasuda}, {York}, {Zehavi}, \& {Zheng}}]{Stoughton2002}
{Stoughton}, C., {Lupton}, R.~H., {Bernardi}, M., {et~al.} 2002, \aj, 123, 485,
  \dodoi{10.1086/324741}

\bibitem[{{Sulentic} {et~al.}(2007){Sulentic}, {Bachev}, {Marziani}, {Negrete},
  \& {Dultzin}}]{Sulentic2007}
{Sulentic}, J.~W., {Bachev}, R., {Marziani}, P., {Negrete}, C.~A., \&
  {Dultzin}, D. 2007, \apj, 666, 757, \dodoi{10.1086/519916}

\bibitem[{{Trump} {et~al.}(2006){Trump}, {Hall}, {Reichard}, {Richards},
  {Schneider}, {Vanden Berk}, {Knapp}, {Anderson}, {Fan}, {Brinkman},
  {Kleinman}, \& {Nitta}}]{Trump2006}
{Trump}, J.~R., {Hall}, P.~B., {Reichard}, T.~A., {et~al.} 2006, \apjs, 165, 1,
  \dodoi{10.1086/503834}

\bibitem[{{Tsuzuki} {et~al.}(2006){Tsuzuki}, {Kawara}, {Yoshii}, {Oyabu},
  {Tanab{\'e}}, \& {Matsuoka}}]{Tsuzuki2006}
{Tsuzuki}, Y., {Kawara}, K., {Yoshii}, Y., {et~al.} 2006, \apj, 650, 57,
  \dodoi{10.1086/506376}

\bibitem[{{Tytler} \& {Fan}(1992)}]{Tytler1992}
{Tytler}, D., \& {Fan}, X.-M. 1992, \apjs, 79, 1, \dodoi{10.1086/191642}

\bibitem[{{van der Walt} {et~al.}(2011){van der Walt}, {Colbert}, \&
  {Varoquaux}}]{numpy}
{van der Walt}, S., {Colbert}, S.~C., \& {Varoquaux}, G. 2011, Computing in
  Science Engineering, 13, 22

\bibitem[{{Vanden Berk} {et~al.}(2001){Vanden Berk}, {Richards}, {Bauer},
  {Strauss}, {Schneider}, {Heckman}, {York}, {Hall}, {Fan}, {Knapp},
  {Anderson}, {Annis}, {Bahcall}, {Bernardi}, {Briggs}, {Brinkmann}, {Brunner},
  {Burles}, {Carey}, {Castander}, {Connolly}, {Crocker}, {Csabai}, {Doi},
  {Finkbeiner}, {Friedman}, {Frieman}, {Fukugita}, {Gunn}, {Hennessy},
  {Ivezi{\'c}}, {Kent}, {Kunszt}, {Lamb}, {Leger}, {Long}, {Loveday}, {Lupton},
  {Meiksin}, {Merelli}, {Munn}, {Newberg}, {Newcomb}, {Nichol}, {Owen}, {Pier},
  {Pope}, {Rockosi}, {Schlegel}, {Siegmund}, {Smee}, {Snir}, {Stoughton},
  {Stubbs}, {SubbaRao}, {Szalay}, {Szokoly}, {Tremonti}, {Uomoto}, {Waddell},
  {Yanny}, \& {Zheng}}]{VandenBerk2001}
{Vanden Berk}, D.~E., {Richards}, G.~T., {Bauer}, A., {et~al.} 2001, \aj, 122,
  549, \dodoi{10.1086/321167}

\bibitem[{{Venemans}(2020)}]{Venemans2020}
{Venemans}, B.~P. 2020, in prep

\bibitem[{{Venemans} {et~al.}(2019){Venemans}, {Neeleman}, {Walter}, {Novak},
  {Decarli}, {Hennawi}, \& {Rix}}]{Venemans2019}
{Venemans}, B.~P., {Neeleman}, M., {Walter}, F., {et~al.} 2019, \apjl, 874,
  L30, \dodoi{10.3847/2041-8213/ab11cc}

\bibitem[{{Venemans} {et~al.}(2016){Venemans}, {Walter}, {Zschaechner},
  {Decarli}, {De Rosa}, {Findlay}, {McMahon}, \& {Sutherland}}]{Venemans2016}
{Venemans}, B.~P., {Walter}, F., {Zschaechner}, L., {et~al.} 2016, \apj, 816,
  37, \dodoi{10.3847/0004-637X/816/1/37}

\bibitem[{{Venemans} {et~al.}(2012){Venemans}, {McMahon}, {Walter}, {Decarli},
  {Cox}, {Neri}, {Hewett}, {Mortlock}, {Simpson}, \& {Warren}}]{Venemans2012}
{Venemans}, B.~P., {McMahon}, R.~G., {Walter}, F., {et~al.} 2012, \apj, 751,
  L25, \dodoi{10.1088/2041-8205/751/2/L25}

\bibitem[{{Venemans} {et~al.}(2013){Venemans}, {Findlay}, {Sutherland}, {De
  Rosa}, {McMahon}, {Simcoe}, {Gonz{\'a}lez-Solares}, {Kuijken}, \&
  {Lewis}}]{Venemans2013}
{Venemans}, B.~P., {Findlay}, J.~R., {Sutherland}, W.~J., {et~al.} 2013, \apj,
  779, 24, \dodoi{10.1088/0004-637X/779/1/24}

\bibitem[{{Venemans} {et~al.}(2015){Venemans}, {Ba{\~n}ados}, {Decarli},
  {Farina}, {Walter}, {Chambers}, {Fan}, {Rix}, {Schlafly}, {McMahon},
  {Simcoe}, {Stern}, {Burgett}, {Draper}, {Flewelling}, {Hodapp}, {Kaiser},
  {Magnier}, {Metcalfe}, {Morgan}, {Price}, {Tonry}, {Waters}, {AlSayyad},
  {Banerji}, {Chen}, {Gonz{\'a}lez-Solares}, {Greiner}, {Mazzucchelli},
  {McGreer}, {Miller}, {Reed}, \& {Sullivan}}]{Venemans2015a}
{Venemans}, B.~P., {Ba{\~n}ados}, E., {Decarli}, R., {et~al.} 2015, \apj, 801,
  L11, \dodoi{10.1088/2041-8205/801/1/L11}

\bibitem[{{Venemans} {et~al.}(2017){Venemans}, {Walter}, {Decarli},
  {Ba{\~n}ados}, {Carilli}, {Winters}, {Schuster}, {da Cunha}, {Fan}, {Farina},
  {Mazzucchelli}, {Rix}, \& {Weiss}}]{Venemans2017c}
{Venemans}, B.~P., {Walter}, F., {Decarli}, R., {et~al.} 2017, \apjl, 851, L8,
  \dodoi{10.3847/2041-8213/aa943a}

\bibitem[{{Verner} {et~al.}(2003){Verner}, {Bruhweiler}, {Verner}, {Johansson},
  \& {Gull}}]{Verner2003}
{Verner}, E., {Bruhweiler}, F., {Verner}, D., {Johansson}, S., \& {Gull}, T.
  2003, \apjl, 592, L59, \dodoi{10.1086/377571}

\bibitem[{{Vernet} {et~al.}(2011){Vernet}, {Dekker}, {D'Odorico}, {Kaper},
  {Kjaergaard}, {Hammer}, {Randich}, {Zerbi}, {Groot}, {Hjorth}, {Guinouard},
  {Navarro}, {Adolfse}, {Albers}, {Amans}, {Andersen}, {Andersen}, {Binetruy},
  {Bristow}, {Castillo}, {Chemla}, {Christensen}, {Conconi}, {Conzelmann},
  {Dam}, {de Caprio}, {de Ugarte Postigo}, {Delabre}, {di Marcantonio},
  {Downing}, {Elswijk}, {Finger}, {Fischer}, {Flores}, {Fran{\c{c}}ois},
  {Goldoni}, {Guglielmi}, {Haigron}, {Hanenburg}, {Hendriks}, {Horrobin},
  {Horville}, {Jessen}, {Kerber}, {Kern}, {Kiekebusch}, {Kleszcz}, {Klougart},
  {Kragt}, {Larsen}, {Lizon}, {Lucuix}, {Mainieri}, {Manuputy}, {Martayan},
  {Mason}, {Mazzoleni}, {Michaelsen}, {Modigliani}, {Moehler}, {M{\o}ller},
  {Norup S{\o}rensen}, {N{\o}rregaard}, {P{\'e}roux}, {Patat}, {Pena}, {Pragt},
  {Reinero}, {Rigal}, {Riva}, {Roelfsema}, {Royer}, {Sacco}, {Santin},
  {Schoenmaker}, {Spano}, {Sweers}, {Ter Horst}, {Tintori}, {Tromp}, {van
  Dael}, {van der Vliet}, {Venema}, {Vidali}, {Vinther}, {Vola}, {Winters},
  {Wistisen}, {Wulterkens}, \& {Zacchei}}]{Vernet2011}
{Vernet}, J., {Dekker}, H., {D'Odorico}, S., {et~al.} 2011, \aap, 536, A105,
  \dodoi{10.1051/0004-6361/201117752}

\bibitem[{{Vestergaard} \& {Osmer}(2009)}]{Vestergaard2009}
{Vestergaard}, M., \& {Osmer}, P.~S. 2009, \apj, 699, 800,
  \dodoi{10.1088/0004-637X/699/1/800}

\bibitem[{{Vestergaard} \& {Peterson}(2006)}]{Vestergaard2006}
{Vestergaard}, M., \& {Peterson}, B.~M. 2006, \apj, 641, 689,
  \dodoi{10.1086/500572}

\bibitem[{{Vestergaard} \& {Wilkes}(2001)}]{Vestergaard2001}
{Vestergaard}, M., \& {Wilkes}, B.~J. 2001, \apjs, 134, 1,
  \dodoi{10.1086/320357}

\bibitem[{{Vietri} {et~al.}(2018){Vietri}, {Piconcelli}, {Bischetti}, {Duras},
  {Martocchia}, {Bongiorno}, {Marconi}, {Zappacosta}, {Bisogni}, {Bruni},
  {Brusa}, {Comastri}, {Cresci}, {Feruglio}, {Giallongo}, {La Franca},
  {Mainieri}, {Mannucci}, {Ricci}, {Sani}, {Testa}, {Tombesi}, {Vignali}, \&
  {Fiore}}]{Vietri2018}
{Vietri}, G., {Piconcelli}, E., {Bischetti}, M., {et~al.} 2018, \aap, 617, A81,
  \dodoi{10.1051/0004-6361/201732335}

\bibitem[{{Virtanen} {et~al.}(2020){Virtanen}, {Gommers}, {Oliphant},
  {Haberland}, {Reddy}, {Cournapeau}, {Burovski}, {Peterson}, {Weckesser},
  {Bright}, {van der Walt}, {Brett}, {Wilson}, {Jarrod Millman}, {Mayorov},
  {Nelson}, {Jones}, {Kern}, {Larson}, {Carey}, {Polat}, {Feng}, {Moore}, {Vand
  erPlas}, {Laxalde}, {Perktold}, {Cimrman}, {Henriksen}, {Quintero}, {Harris},
  {Archibald}, {Ribeiro}, {Pedregosa}, {van Mulbregt}, \&
  {Contributors}}]{scipy}
{Virtanen}, P., {Gommers}, R., {Oliphant}, T.~E., {et~al.} 2020, Nature
  Methods, 17, 261, \dodoi{https://doi.org/10.1038/s41592-019-0686-2}

\bibitem[{{Volonteri}(2012)}]{Volonteri2012}
{Volonteri}, M. 2012, Science, 337, 544, \dodoi{10.1126/science.1220843}

\bibitem[{{Walter} {et~al.}(2009){Walter}, {Riechers}, {Cox}, {Neri},
  {Carilli}, {Bertoldi}, {Weiss}, \& {Maiolino}}]{Walter2009}
{Walter}, F., {Riechers}, D., {Cox}, P., {et~al.} 2009, \nat, 457, 699,
  \dodoi{10.1038/nature07681}

\bibitem[{{Wang} {et~al.}(2016{\natexlab{a}}){Wang}, {Wu}, {Fan}, {Yang}, {Yi},
  {Bian}, {McGreer}, {Yang}, {Ai}, {Dong}, {Zuo}, {Jiang}, {Green}, {Wang},
  {Cai}, {Wang}, \& {Yue}}]{WangFeige2016}
{Wang}, F., {Wu}, X.-B., {Fan}, X., {et~al.} 2016{\natexlab{a}}, \apj, 819, 24,
  \dodoi{10.3847/0004-637X/819/1/24}

\bibitem[{{Wang} {et~al.}(2017){Wang}, {Fan}, {Yang}, {Wu}, {Yang}, {Bian},
  {McGreer}, {Li}, {Li}, {Ding}, {Dey}, {Dye}, {Findlay}, {Green}, {James},
  {Jiang}, {Lang}, {Lawrence}, {Myers}, {Ross}, {Schlegel}, \&
  {Shanks}}]{WangFeige2017}
{Wang}, F., {Fan}, X., {Yang}, J., {et~al.} 2017, \apj, 839, 27,
  \dodoi{10.3847/1538-4357/aa689f}

\bibitem[{{Wang} {et~al.}(2018){Wang}, {Yang}, {Fan}, {Yue}, {Wu}, {Schindler},
  {Bian}, {Li}, {Farina}, {Ba{\~n}ados}, {Davies}, {Decarli}, {Green}, {Jiang},
  {Hennawi}, {Huang}, {Mazzucchelli}, {McGreer}, {Venemans}, {Walter}, \&
  {Beletsky}}]{WangFeige2018}
{Wang}, F., {Yang}, J., {Fan}, X., {et~al.} 2018, \apjl, 869, L9,
  \dodoi{10.3847/2041-8213/aaf1d2}

\bibitem[{{Wang} {et~al.}(2019){Wang}, {Yang}, {Fan}, {Wu}, {Yue}, {Li},
  {Bian}, {Jiang}, {Ba{\~n}ados}, {Schindler}, {Findlay}, {Davies}, {Decarli},
  {Farina}, {Green}, {Hennawi}, {Huang}, {Mazzuccheli}, {McGreer}, {Venemans},
  {Walter}, {Dye}, {Lyke}, {Myers}, \& {Haze Nunez}}]{WangFeige2019}
---. 2019, \apj, 884, 30, \dodoi{10.3847/1538-4357/ab2be5}

\bibitem[{{Wang} {et~al.}(2013){Wang}, {Wagg}, {Carilli}, {Walter}, {Lentati},
  {Fan}, {Riechers}, {Bertoldi}, {Narayanan}, {Strauss}, {Cox}, {Omont},
  {Menten}, {Knudsen}, {Neri}, \& {Jiang}}]{Wang2013}
{Wang}, R., {Wagg}, J., {Carilli}, C.~L., {et~al.} 2013, \apj, 773, 44,
  \dodoi{10.1088/0004-637X/773/1/44}

\bibitem[{{Wang} {et~al.}(2016{\natexlab{b}}){Wang}, {Wu}, {Neri}, {Fan},
  {Walter}, {Carilli}, {Momjian}, {Bertoldi}, {Strauss}, {Li}, {Wang},
  {Riechers}, {Jiang}, {Omont}, {Wagg}, \& {Cox}}]{WangRan2016}
{Wang}, R., {Wu}, X.-B., {Neri}, R., {et~al.} 2016{\natexlab{b}}, \apj, 830,
  53, \dodoi{10.3847/0004-637X/830/1/53}

\bibitem[{{W}es {M}c{K}inney(2010)}]{pandas_paper}
{W}es {M}c{K}inney. 2010, in {P}roceedings of the 9th {P}ython in {S}cience
  {C}onference, ed. {S}t\'efan van~der {W}alt \& {J}arrod {M}illman, 56 -- 61,
  \dodoi{10.25080/Majora-92bf1922-00a}

\bibitem[{{Wild} \& {Hewett}(2005)}]{Wild2005}
{Wild}, V., \& {Hewett}, P.~C. 2005, \mnras, 358, 1083,
  \dodoi{10.1111/j.1365-2966.2005.08844.x}

\bibitem[{{Willott} {et~al.}(2015){Willott}, {Bergeron}, \&
  {Omont}}]{Willott2015}
{Willott}, C.~J., {Bergeron}, J., \& {Omont}, A. 2015, \apj, 801, 123,
  \dodoi{10.1088/0004-637X/801/2/123}

\bibitem[{{Willott} {et~al.}(2017){Willott}, {Bergeron}, \&
  {Omont}}]{Willott2017}
---. 2017, \apj, 850, 108, \dodoi{10.3847/1538-4357/aa921b}

\bibitem[{{Willott} {et~al.}(2013){Willott}, {Omont}, \&
  {Bergeron}}]{Willott2013}
{Willott}, C.~J., {Omont}, A., \& {Bergeron}, J. 2013, \apj, 770, 13,
  \dodoi{10.1088/0004-637X/770/1/13}

\bibitem[{{Willott} {et~al.}(2007){Willott}, {Delorme}, {Omont}, {Bergeron},
  {Delfosse}, {Forveille}, {Albert}, {Reyl{\'e}}, {Hill}, {Gully-Santiago},
  {Vinten}, {Crampton}, {Hutchings}, {Schade}, {Simard}, {Sawicki}, {Beelen},
  \& {Cox}}]{Willott2007}
{Willott}, C.~J., {Delorme}, P., {Omont}, A., {et~al.} 2007, \aj, 134, 2435,
  \dodoi{10.1086/522962}

\bibitem[{{Willott} {et~al.}(2010){Willott}, {Delorme}, {Reyl{\'e}}, {Albert},
  {Bergeron}, {Crampton}, {Delfosse}, {Forveille}, {Hutchings}, {McLure},
  {Omont}, \& {Schade}}]{Willott2010a}
{Willott}, C.~J., {Delorme}, P., {Reyl{\'e}}, C., {et~al.} 2010, \aj, 139, 906,
  \dodoi{10.1088/0004-6256/139/3/906}

\bibitem[{{Woo} {et~al.}(2018){Woo}, {Le}, {Karouzos}, {Park}, {Park},
  {Malkan}, {Treu}, \& {Bennert}}]{Woo2018}
{Woo}, J.-H., {Le}, H. A.~N., {Karouzos}, M., {et~al.} 2018, \apj, 859, 138,
  \dodoi{10.3847/1538-4357/aabf3e}

\bibitem[{{Wu} {et~al.}(2015){Wu}, {Wang}, {Fan}, {Yi}, {Zuo}, {Bian}, {Jiang},
  {McGreer}, {Wang}, {Yang}, {Yang}, {Thompson}, \& {Beletsky}}]{Wu2015}
{Wu}, X.-B., {Wang}, F., {Fan}, X., {et~al.} 2015, \nat, 518, 512,
  \dodoi{10.1038/nature14241}

\bibitem[{{Yang} {et~al.}(2019){Yang}, {Wang}, {Fan}, {Yue}, {Wu}, {Li},
  {Bian}, {Jiang}, {Ba{\~n}ados}, \& {Beletsky}}]{YangJinyi2019}
{Yang}, J., {Wang}, F., {Fan}, X., {et~al.} 2019, \aj, 157, 236,
  \dodoi{10.3847/1538-3881/ab1be1}

\bibitem[{{Yang} {et~al.}(2020){Yang}, {Wang}, {Fan}, {Hennawi}, {Davies},
  {Yue}, {Banados}, {Wu}, {Venemans}, {Barth}, {Bian}, {Boutsia}, {Decarli},
  {Farina}, {Green}, {Jiang}, {Li}, {Mazzucchelli}, \&
  {Walter}}]{YangJinyi2020}
---. 2020, \apjl, 897, L14, \dodoi{10.3847/2041-8213/ab9c26}

\bibitem[{{Yong} {et~al.}(2020){Yong}, {Webster}, {King}, {Bate}, {Labrie}, \&
  {O'Dowd}}]{Yong2020}
{Yong}, S.~Y., {Webster}, R.~L., {King}, A.~L., {et~al.} 2020, \mnras, 491,
  1320, \dodoi{10.1093/mnras/stz3074}

\bibitem[{{Zakamska} {et~al.}(2016){Zakamska}, {Hamann}, {P{\^a}ris}, {Brandt},
  {Greene}, {Strauss}, {Villforth}, {Wylezalek}, {Alexand roff}, \&
  {Ross}}]{Zakamska2016}
{Zakamska}, N.~L., {Hamann}, F., {P{\^a}ris}, I., {et~al.} 2016, \mnras, 459,
  3144, \dodoi{10.1093/mnras/stw718}

\bibitem[{{Zuo} {et~al.}(2020){Zuo}, {Wu}, {Fan}, {Green}, {Yi}, {Schulze},
  {Wang}, \& {Bian}}]{ZuoWenwen2020}
{Zuo}, W., {Wu}, X.-B., {Fan}, X., {et~al.} 2020, \apj, 896, 40,
  \dodoi{10.3847/1538-4357/ab91a7}

\end{thebibliography}
\bibliographystyle{aasjournal}

\end{document}